\documentclass[mnsc,nonblindrev]{informs3_hide}

\OneAndAHalfSpacedXI %

\usepackage{mathtools}
    \allowdisplaybreaks
\usepackage{bm}
\usepackage[mathscr]{euscript}
\usepackage{enumitem}
\usepackage{xcolor}
\usepackage{graphicx}
\usepackage{multirow}
\usepackage{bigstrut}
\usepackage{booktabs}
\usepackage{microtype}
\usepackage{fix-cm}
\usepackage{xfrac}
\usepackage{float}
\usepackage{tabularx}
\usepackage{bibunits}

\newcommand{\pr}{{\bf\sf P}}
\newcommand{\ex}{{\bf\sf E}}               %
\newcommand{\var}{{\bf\sf Var}}
\newcommand{\cov}{{\bf\sf Cov}}

\newcommand{\bA}{{\bf A}}               %

\newcommand{\bq}{\mathbf{q}}
\newcommand{\bQ}{\mathbf{Q}}
\newcommand{\bP}{\mathbf{P}}
\newcommand{\bR}{\mathbf{R}}
\newcommand{\ba}{\mathbf{a}}

\newcommand{\bB}{\mathbf{B}}
\newcommand{\bC}{\mathbf{C}}
\newcommand{\bSigma}{\boldsymbol{\Sigma}}
\newcommand{\bsigma}{\boldsymbol{\sigma}}
\newcommand{\bdelta}{\boldsymbol{\delta}}
\newcommand{\bGamma}{\boldsymbol{\Gamma}}

\newcommand{\bzero}{\mathbf{0}}

\newcommand{\calg}{{\cal G}}

\newcommand{\cals}{{\cal S}}

\newcommand{\cala}{{\cal A}}
\newcommand{\calf}{{\cal F}}

\newcommand{\g}{\lambda}               %

\newcommand{\calm}{{\cal M}}

\newcommand{\startb}{\parindent0pt\bf}  %

\newcommand{\nvq}{Q^{\rm {\scriptsize NV}}}

\newcommand{\nvmp}{P^{\rm {\scriptsize NV(M)}}}
\newcommand{\nvmr}{R^{\rm {\scriptsize NV(M)}}}

\newcommand{\bgamma}{\boldsymbol{\gamma}}

\newcommand{\chirm}{\chi^{\rm rm}}
\newcommand{\chiiv}{\chi^{\rm iv}}
\newcommand{\HTh}{H_T^{\rm h}}
\newcommand{\HTfin}{H_T^{\rm fin}}
\newcommand{\HTu}{H_T^{\rm u}}

\newcommand{\nvp}{P^{\rm NV}}
\newcommand{\nvr}{R^{\rm NV}}

\newcommand{\ind}{{\bf 1}}

\usepackage{natbib}
 \bibpunct[, ]{(}{)}{,}{a}{}{,}%
 \def\bibfont{\small}%

\TheoremsNumberedThrough     %
\ECRepeatTheorems

\EquationsNumberedThrough    %

\MANUSCRIPTNO{MS-FIN-22-00115.R2} %

\begin{document}

\RUNAUTHOR{Wang, Yao, and Zhang}

\RUNTITLE{How Does Risk Hedging Impact Operations?}

\TITLE{How Does Risk Hedging Impact Operations? Insights from a Price-Setting Newsvendor Model}

\ARTICLEAUTHORS{%
\AUTHOR{Liao Wang, Jin Yao}
\AFF{Faculty of Business and Economics, The University of Hong Kong,
Pofuklam Road, Hong Kong SAR,
 \EMAIL{lwang98@hku.hk}, \EMAIL{u3002692@connect.hku.hk}} %
\AUTHOR{Xiaowei Zhang}
\AFF{Department of Industrial Engineering and Decision Analytics, The Hong Kong University of Science and Technology, Clear Water Bay, Hong Kong SAR, \EMAIL{xiaoweiz@ust.hk}}
} %

\ABSTRACT{If a financial asset's price movement impacts a firm's product demand, the firm can respond to the impact by adjusting its operational decisions. For example, in the automotive industry, car makers decrease the selling prices of fuel-inefficient cars when the oil price rises. Meanwhile, the firm can implement a risk-hedging strategy using the financial asset jointly with its operational decisions. Motivated by this, we develop and solve a general risk-management model integrating risk hedging into a price-setting newsvendor. The optimal hedging strategy is calculated analytically, which leads to an explicit objective function for optimizing price and ``virtual production quantity'' (VPQ). (The latter determines the service level, i.e., the demand fulfillment probability.)
We find that hedging generally reduces the optimal price {when the firm sets the target mean return as its production-only maximum expected profit.
With the same condition on the target mean return}, hedging also reduces the optimal VPQ when the asset price trend positively impacts product demand; meanwhile, it may increase the VPQ by a small margin when the impact is negative.
We construct the return-risk efficient frontier that characterizes the optimal return-risk trade-off.
Our numerical study using data from a prominent automotive manufacturer shows that the markdowns in price and reduction in VPQ are small under our model and that the hedging strategy substantially reduces risk without materially reducing operational profit.
}


\maketitle

\section{Introduction}\label{sec:intro}

Effective risk management for firms facing substantial volatility in product demand must account for various factors that impact demand. An important {\it exogenous} factor is the price of some tradable financial asset (e.g., stock market index, commodity, and foreign currency). The product demand may fluctuate significantly in response to the financial asset price movement. For example, the ``Big Three'' in the automotive industry (General Motors, Fiat Chrysler Automobiles, and Ford Motor Company) all recognize in their annual reports that oil price is a major risk factor affecting their sales volumes (General Motors, 2020 10-K Filing; Fiat Chrysler Automobiles, 2020 10-K Filing; Ford Motor Company, 2020 10-K Filing). Both economic theory and empirical evidence support this observation. Theoretically, it is known that the demand for a product is affected by the price of a complementary good \citep {mankiw2014principles}. The reason is that, in the case of automakers, a higher oil price leads to a higher gasoline price,
directly increasing the expense of using a car and encouraging customers to switch from gas guzzlers to more fuel-efficient vehicles. Empirical studies that support this theory include those of \cite{busse2009pain}, \cite{klier2010price,busse2013consumers}, and \cite{langer2013automakers}. One can also find examples of financial asset prices impacting product demand in other industries. John Deere, the largest farming equipment manufacturer in the United States, discloses that prices of agricultural commodities (e.g., corn) significantly impact their sales, as these prices directly affect the revenues of their customers (e.g., farmers). Another example is Caterpillar Inc., the largest industrial equipment manufacturer, recognizing metal prices as demand risk factors. Our analysis in this paper is not limited to any specific industry.

A major {\it endogenous} factor impacting demand is the product price.
Basic economic theory implies that a higher price leads to lower demand, and the slope of the demand curve describes how demand changes with the product price.
Unlike the financial asset price, the product price is controllable.
The firm may adjust the product price proactively in response to changes in the financial asset price to mitigate its impact on the product demand.
For example, automakers often respond to the fuel price's impact on vehicle demand by setting prices strategically
\citep{busse2009pain,langer2013automakers,allcott2014gasoline}.
In the scenario of rising oil prices, demand for fuel-inefficient cars decreases while demand for fuel-efficient cars increases.
Thus, automakers can reduce the price of fuel-inefficient cars to offset the rising energy costs' negative impact on product demand. At the same time, they can raise the price of fuel-efficient cars in response to the increased demand.

We now demonstrate our results using data from a prominent automotive manufacturer, specifically on two of their popular car models: a sport utility vehicle (SUV) and a compact car.
For the sake of brevity, we will refer to manufacturer and the car models as \texttt{AutoMFR}, \texttt{Sport}, and \texttt{Compact}, respectively.
\texttt{Sport} is categorized as fuel-inefficient, having low miles per gallon (MPG), while \texttt{Compact} is fuel-efficient.
The relevant financial asset is crude oil, West Texas Intermediate (WTI). For each car model, we regress the monthly sales volume and the selling price, respectively, on the monthly average WTI price.
The regression results are summarized in Table~\ref{tab:empirical_result}.
\begin{table}[ht] \def\arraystretch{1}
\TABLE{Regression Results. \label{tab:empirical_result}}
{
    \begin{tabular}{lc lc cc c}
        \toprule
        \multirow{2}{*}{Car} & & \multirow{2}{*}{Estimate} & & \multicolumn{3}{c}{Regression Equation} \\
        \cmidrule{5-7}
        && && Price $\sim$ WTI && Sales Volume $\sim$ WTI \\
        \cmidrule{1-3} \cmidrule{5-5} \cmidrule{7-7}
        \multirow{2}{*}{\texttt{Sport}} && slope && -40.16 (3.60) &&  -168.55 (23.53) \\
        && p-value && $6.6 \times 10^{-20}$ && $1.7\times 10^{-11}$ \\
        \cmidrule{1-3} \cmidrule{5-5} \cmidrule{7-7}
        \multirow{2}{*}{\texttt{Compact}} && slope &&  9.32 (1.90) && 132.99 (25.20) \\
        && p-value && $4.2 \times 10^{-6}$ &&  $2.2 \times 10^{-7}$ \\
	    \bottomrule
    \end{tabular}
}
{{\it Note.}   The numbers in parentheses represent the standard errors. }
\end{table}
We find that the selling price of \texttt{Sport} has a significant negative correlation with the oil price,
while \texttt{Compact} has a significant positive correlation.
The same patterns hold for the sales volume.
These are consistent with our discussions above and existing empirical findings \citep{klier2010price,busse2013consumers,allcott2014gasoline}.

The discussion and data analysis above convey two crucial ideas.
First, financial asset price movement may significantly impact product demand.
Second, decisions regarding the product price may interact with the financial asset price.
Namely, the firm can adjust the product price for risk management in response to the financial asset price's impact on the product demand (as in the practice of automakers discussed above).
The firm may also manage risk by adopting a {\it risk-hedging} strategy of trading this financial asset.
Since operational decisions (pricing and production) and financial decision (risk hedging) can both affect the firm's risk, risk-management strategies can be improved substantially by {\it jointly} optimizing these two types of decisions.
Doing so accounts for endogenous and exogenous factors that impact demand simultaneously.
Despite the rich literature on pricing strategies and risk management using financial assets, these two types of decisions are usually studied separately.
The literature on how to jointly optimize pricing and risk hedging is scarce.
An example analyzed by \cite{guiotto2022combined} is a rare exception.
To the best of our knowledge, our paper is the first study on how risk hedging impacts pricing in a price-setting newsvendor model.

\smallskip
Motivated by the above discussion, we pose the following  research questions:
\begin{enumerate}[label=(\roman*)]
    \item  How can a risk-management strategy be developed by jointly optimizing pricing, production, and hedging decisions using relevant financial assets?

    \item  How does risk hedging affect the pricing decision compared with the no-hedging case?

\end{enumerate}
\smallskip

To answer these questions, we start by building a demand model that incorporates impacts from both the product pricing decision and the financial asset price.
The model allows a general relationship between product demand and asset price.
Based on this demand model, we formulate the risk-management problem
following Markowitz's mean-variance optimization framework \citep{markowitz1952}.
Fixing production and pricing decisions, we first analytically solve for the optimal risk-hedging strategy, which yields an explicit risk objective function for the operational decisions.
We then minimize this objective function to find the optimal production and pricing decisions, which leads to a complete characterization of the efficient frontier.
We demonstrate the effectiveness of the risk-management model using real data about \texttt{Sport} and \texttt{Compact}.

\subsection{Main Results and Contributions}
\label{sec:contribution}
Our main results are Theorems~\ref{thm:hedgingsol}, \ref{thm:benefitpr} and \ref{thm:hurtpr}.
Given a pair of price and VPQ (defined in Section~\ref{sec:NV},
which directly determines the service level),
Theorem~\ref{thm:hedgingsol} characterizes the optimal hedging strategy and the associated variance of the total terminal return.
Risk of a return is quantified by its standard deviation, so the associated variance is the squared risk.
The optimal hedging strategy combines a risk-mitigation position and an investment position.
The variance associated with this hedging strategy is the sum of squared investment and unhedgeable risks
(i.e., the kind of risk that is irrelevant to the financial asset and thus cannot be hedged by it).
This variance, as a function of the price and VPQ, provides an explicit objective to be further minimized.

Assuming the asset price trend positively impacts demand
(e.g., rising oil price boosts the demand for fuel-efficient cars),
Theorem~\ref{thm:benefitpr} characterizes the optimal price and VPQ in the presence of hedging,
with the target mean  return set as the newsvendor's maximum expected profit.
We find that both the optimal price and VPQ are {\it lower} in the presence of hedging than those without hedging.
The reason is that hedging {\it offsets} the positive effect of the asset price trend on product demand, leading to a hypothetically {\it smaller} market than that without hedging.
Adapting to the smaller market size, the optimization of the operational decision %
decreases the price and VPQ while reducing unhedgeable risk.
Moreover, this main result holds for general operational payoff functions with certain structural properties
(refer to Theorem~\ref{thm:genpay} in the e-companion).

Theorem~\ref{thm:hurtpr} assumes
the asset price trend negatively impacts demand
(e.g., rising oil price decreases the demand for fuel-inefficient cars),
the opposite scenario of the one considered in Theorem~\ref{thm:benefitpr}.
Interestingly, the conclusions are similar: the optimal price is lower with hedging than without hedging, and the optimal VPQ with hedging is lower than (or at most exceeds by a small margin) the optimal VPQ without hedging.
We interpret this counter-intuitive result as follows.
{The risk-mitigation position of the hedging strategy}
offsets the negative effect of the asset price trend on product demand, leading to a hypothetically {\it larger} market size.
{
Hence, with the target mean return set as the maximum expected profit of the base model
(i.e., the profit-maximizing price-setting newsvendor without hedging),
the corresponding operational decision of the base model (``NV's decision'') combined with hedging induces a larger expected payoff than the target level.
Thus, to attain this target mean return, with the NV's decision unchanged,
the investment payoff from the hedging strategy
induces a {\it negative} expected payoff while bearing {\it positive} contribution to the total risk.
In this case, the joint operational and risk hedging decision can not be optimal regarding the return-risk trade-off because part of the terminal return has a negative return and positive risk contribution, thereby highly inefficient.
Therefore, to avoid a negative expected investment payoff from hedging with the enlarged market size, the optimal decision has to
{\it leave leeway} in the target mean return for the investment payoff to fill.
As a result, the pricing level needs to be adjusted downward
to suppress the expected hedged production payoff.}
In addition, regarding the optimal VPQ, we identify a critical condition: the negative effect of the asset price is sufficiently strong.
Under this condition, the optimal VPQ with hedging
does not exceed that without hedging (Theorem~\ref{thm:hurtpr}(ii)).
When this condition does not hold, it may exceed that without hedging but the excess amount is small (Theorem~\ref{thm:hurtpr}(iii)).

Both technical and managerial contributions are made.
Technically, we develop and solve a general risk-management model that integrates pricing, production, and risk-hedging decisions.
This model adopts a general diffusion process for the asset price dynamics and does not assume any specific functional relationship between demand and asset price; thus, it can incorporate various application-specific data analytics.
In particular, we explicitly solve the model with the asset price following an exponential Ornstein--Uhlenbeck (EOU) process, a standard oil price model.
We remark that the general asset price process (including EOU) has a {\it stochastic} market price of risk (MPR) process,
rendering the hedging problem much more challenging than those with a constant MPR \citep{wang2017production} or a deterministic MPR \citep{caldentey2006optimal}.

Concerning managerial insights,
the leading message of our paper is that hedging reduces the pricing level.
It points to the insight that hedging not only reduces risk but also enhances a firm's competitiveness in the market.
In addition, we show that the VPQ in the presence of hedging is either lower than or exceeds by a small margin the VPQ without hedging.
We conduct a comprehensive numerical study using data from \texttt{AutoMFR} to demonstrate and complement our theoretical results.
The analysis shows that hedging performs well: it can reduce risk by as much as 40\%.
In particular, markdowns for both the price and VPQ are small---not exceeding 1.03\% for price reduction and 1.70\% for VPQ reduction.
This result is desirable, as firms usually do not want to reduce the price or service level excessively due to concerns about branding or maintaining market share.
It also indicates that, while substantially reducing risk, hedging will not materially reduce a firm's operational levels.

\subsection{Literature Review }
\label{sec:litreview}

This paper falls within the scope of integrated operations and financial risk management. Two main streams of this literature are related to our work. One stream is joint pricing and production/inventory management.
\cite{whitin1955inventory} is the first to study the fundamental connection between pricing and inventory control theory for the single-period setting.
\cite{petruzzi1999pricing} consider a risk-neutral firm that faces a price-dependent random demand, and they examine how the stocking quantity decision interacts with the selling price decision.
\cite{agrawal2000impact} study how a risk-averse retailer facing a price-dependent random demand makes order quantity and pricing decisions to maximize an expected utility.
\cite{chen2009risk} consider a risk-averse decision-maker similar to that analyzed by \cite{agrawal2000impact} but with a different risk objective.
Extensions include multi-period settings \citep{federgruen1999combined,chen2004coordinatingor,chen2004coordinatingmor} and multi-product settings \citep{aydin2008joint,zhu2009coordination}.

The other stream of literature concerns incorporating risk hedging by trading financial assets in operational risk management.
The paper most relevant to ours is that by \cite{wang2017production}.
They study a newsvendor model in which a financial asset's price partially drives product demand. Our paper differs by including product price as a decision variable and adopting a more general model for product demand and asset price.
In particular,
when solving a quadratic hedging problem, they assume the asset price follows a geometric Brownian motion, which has a constant MPR process.
By contrast, we work with a general diffusion process, which has a stochastic MPR process.
This general assumption renders the hedging problem much more technically challenging. Meanwhile, it encompasses a much broader scope of applications than that paper.

\cite{gaur2005hedging} study how to construct an optimal trading strategy to hedge inventory risk using financial market instruments. A study closely related to ours is that of \cite{caldentey2006optimal}.
They formulate a general modeling framework incorporating risk hedging into operations with a quadratic utility function. However, due to the generality of the framework, they do not examine the interaction of risk hedging and any specific operational decision.
\cite{caldentey2009supply} consider a supply chain contracting problem with a supplier and a retailer engaging in a Stackelberg game.
In their setting, the product market size depends on the price of some financial asset.
\cite{ding2007integration} consider an international firm that sells to both domestic and foreign markets and uses operational and financial hedging to manage currency exchange risk.
\cite{kouvelis2019integrated} study a newsvendor problem with correlated operational and financial risks under value-at-risk constraints.

None of the papers reviewed above include product price as a decision variable. Two exceptions are \cite{chen2007risk} and \cite{guiotto2022combined}. \cite{chen2007risk} study joint dynamic inventory control and pricing strategies for a risk-averse decision-maker in a multi-period setting. In the extension part of their study, they assume that the model parameters are correlated with some financial asset price and then include risk hedging.
Our paper differs by considering a mean-variance optimization criterion instead of an expected utility objective.
In contrast to deriving the structure of the dynamic programming problem involved in pricing and inventory control, our focus is to examine how financial risk hedging affects pricing and service levels.
\cite{guiotto2022combined} develop a general framework of combined custom hedging for a risk-averse firm exposed to claimable and non-claimable risks.
The authors consider the pricing decision in a specific example of electricity retailers. Our paper differs from this study in two aspects. First, our setup is a price-setting newsvendor, which also involves the quantity decision, while the example in \cite{guiotto2022combined} is essentially a make-to-order setup. Second, \cite{guiotto2022combined} analyze the interaction of pricing and hedging via additional operational flexibility gained by optimal hedging. By contrast, we study the impact of hedging on pricing by directly analyzing the difference in the pricing levels with and without hedging.

The rest of the paper is organized as follows.
In Section~\ref{sec:NV}, we discuss the price-setting newsvendor model.
In Section~\ref{sec:demandasset}, we develop the demand-asset model and formulate the risk-management problem.
We solve the hedging problem in Section~\ref{sec:hedging} and discuss the optimal price and VPQ in the presence of hedging in Section~\ref{sec:production}.
In Section~\ref{sec:numerical}, we apply the methodology to real data from a prominent automotive manufacturer.
Concluding remarks are collected in Section~\ref{sec:conclusion}.
The e-companion includes lists of notation, supplementary analytical results (part I), and technical proofs (part II).

\section{Price-Setting Newsvendor}
\label{sec:NV}

Consider the price-setting newsvendor (NV) model, which has been
studied extensively in the literature \citep{petruzzi1999pricing,agrawal2000impact};
see also \cite{deyong2020survey} for a recent survey.
Consider a newsvendor over a selling period $[0, T]$.
At time $0$, the newsvendor needs to decide a unit selling price  $P$ and a production quantity $Q$ before a stochastic demand $D_T = D_T(P)$ that depends on $P$ is realized at time $T$.
This setup is suitable for products that do not have frequent price adjustments (e.g., the automobiles considered in this paper) but not for those  that allow dynamic pricing (e.g., retail goods on e-commerce platforms).
With unit production cost $c$ and unit salvage value $s$, the newsvendor's payoff is $P(Q\wedge D_T) + s(Q-D_T)^+ - cQ$,
where $a \wedge b =\min\{a,b\}$, $(x)^+ = \max\{x, 0\}$, and the three terms represent the revenue from sales, the salvage value, and the production cost, respectively.
Because $a \wedge b =  a - (a-b)^+$, the production payoff function can be expressed as:
$H_{T}(P, Q)
=(P-c) Q - (P-s)\left(Q-D_{T}\right)^{+}$.

We model the demand function as $D_T(P) = A_T - bP$,
where $A_T$
is the (random) market size independent of $P$, and
$b$ is a positive parameter capturing demand's sensitivity to price.
We assume that $A_T$ has a continuous distribution and let $f(a)$, $F(a)$, and
$r(a) = f(a)/(1-F(a))$ denote its density function,  distribution function, and hazard rate function,
respectively.

\subsection{{Base Model: Profit Maximization}}
Our \emph{base model} assumes that the newsvendor aims to maximize the expected production payoff:
\begin{equation}
\label{nv}
(\nvp, \nvq) := \argmax_{P, Q} \ex[H_T(P,Q)]\quad {\rm s.t.}\quad P \ge c, \quad Q \ge 0.
\end{equation}

{
We define $R:= Q + bP$ and interpret it as follows.
Note that $R - A_T = Q - D_T$ represents overproduction, so $\pr(A_T \leq R) = \pr(D_T \leq Q)$ is the {\it service level} (i.e., the demand fulfillment probability).
In other words, $R$ serves $A_T$ just like $Q$ serves $D_T$.
Moreover, $R$ and $Q$ have the same unit measuring the number of products. Therefore, $R$ is of similar nature as $Q$, and we call it the {\it virtual production quantity (VPQ)}.
}
In the following,
we will use $(P, R)$ instead of $(P, Q)$ to facilitate theoretical analysis.
This change of variable is standard in the literature \citep{petruzzi1999pricing,agrawal2000impact}.
In particular, problem (\ref{nv}) is equivalent to
\begin{equation}
\label{nv1}
(\nvp, \nvr) := \argmax_{P, R} \ex[H_T(P,R)] \quad {\rm s.t.}\quad P \ge c, \quad R - bP \ge 0,
\end{equation}
where
\begin{equation}
\label{EHTR}
\ex[H_{T}(P, R)]=(P-c) (R-bP) - (P-s)\ex[\left(R-A_{T}\right)^{+}].
\end{equation}
It can be shown that
$\ex[H_T(P,R)]$ is concave in $R$ for a fixed $P$ while concave in $P$ for a fixed $R$.
Moreover,
$\ex[H_T(P,R)]$ is supermodular in $(P,R)$.

We make the following assumptions throughout the paper.\footnote{{Although not imposed on the primitive parameters of the model, Assumptions~\ref{assumption:nvm} and \ref{assumption:Abc} are easy to be verified or disproved numerically.
See Section~\ref{appendix:suffassumption23} of the e-companion (particularly Proposition~\ref{pro:a23suffcond}) for sufficient conditions on the model primitive parameters for these two assumptions to hold.}}
\begin{assumption}
\label{assumption:r}
$2r(a)^2 + r'(a) > 0$ for all $a$.
In addition, $[1-F(a)]^2/r(a) \to 0$ as $a \to \infty$ and $F^2(a)/r(a) \to 0$ as $a \to -\infty$.
Further, the three conditions also hold for distribution of $A_T$ under the probability measure $\pr^M$ defined in (\ref{genZt}) below.
\end{assumption}

\begin{assumption}
\label{assumption:nvm}
$\nvp > c$, $\nvq = \nvr - b\nvp >0$ and $\ex[H_T(\nvp, \nvr)] > 0$.
\end{assumption}

\begin{assumption}
\label{assumption:Abc}
$\ex[(bc - A_T)^+] \leq [2b(c-s)] \wedge \sqrt{4b\ex[H_T(\nvp, \nvr)]}.$
\end{assumption}

Assumption~\ref{assumption:r} is standard in the literature \citep{petruzzi1999pricing}.
It ensures the existence of a unique solution to problem~(\ref{nv1}).
Assumption~\ref{assumption:nvm} excludes the trivial case that the optimal production payoff is nonpositive.
Intuitively, this case only occurs when the market size $A_T$ is very low, which is unlikely for commonly demanded products such as automobiles (the case considered in this paper).
To understand Assumption \ref{assumption:Abc},
note that for any $P \ge c$,
$A_T - bc \geq A_T - bP = D_T$ is an upper bound of the product demand, and we expect it to have a small negative part
(i.e., $\ex[(bc - A_T)^+]$ is small),
which is expressed by Assumption \ref{assumption:Abc}.
In Section~\ref{sec:numerical}, we numerically validate that Assumptions~\ref{assumption:r}--\ref{assumption:Abc} are satisfied by the demand model and parameters calibrated from an automaker's data sets.

We characterize the optimal solution to problem~(\ref{nv1}) in the following proposition.
Analogous results are also known in the literature \citep{petruzzi1999pricing}.
\begin{proposition}
\label{pro:nvsol}
Under Assumptions \ref{assumption:r} and \ref{assumption:nvm}, the profit-maximization problem in (\ref{nv1}) has a unique solution characterized by the following optimality equations:
\begin{equation}
\label{nvopteqns}
2b\nvp - bc = \ex(\nvr \wedge A_T),
\qquad \nvr = F^{-1}\Big(\frac{\nvp - c}{\nvp - s}\Big).
\end{equation}
\end{proposition}

The solution to the base model specified in Proposition~\ref{pro:nvsol} represents a risk-neutral decision: the expected return (throughout the paper, ``return'' stands for profit or loss in dollar amounts rather than the rate of return) is maximized without accounting for the associated risk.
In this study, our ultimate goal is to develop an effective risk-management strategy.

\subsection{{Risk Minimization}}
We follow the common practice of measuring risk by {\it standard deviation}.
Then, minimizing the risk is equivalent to minimizing the variance (i.e., squared risk)
$\var[H_T(P,R)]$.
Note that $\var[H_T(P,R)] = (P-s)^2\var[(R - A_T)^+]$ is increasing in both $P$ and $R$.
A higher $P$ leads to a higher payoff risk because it strengthens the positive (or negative) impact on the payoff from each sold (or unsold) product, increasing the payoff's volatility.
Meanwhile, a higher $R$
{(i.e., VPQ)}
represents a higher service level and thus increases the exposure to demand volatility, amplifying the payoff risk.
Managing risk without considering the associated return makes little economic sense, and the focal issue here is to consider the trade-off between the two.

Of particular interest are two optimization problems that examine the return-risk trade-off induced by the pricing ($P$) or VPQ ($R$) decisions.
For the former, for $P \ge c$, we define
\begin{equation}
\label{NVP}
\nvr(P) := \argmax_R \ex[H_T(P, R)], \;\; m(P) := \ex[H_T(P, \nvr(P))],\;\; v(P) := \var[H_T(P, \nvr(P))].
\end{equation}
In parallel, for $R \ge bc$, we define
\begin{equation}
\label{NVR}
P^{\rm NV}(R) := \argmax_P \ex[H_T(P, R)], \;\; m(R) := \ex[H_T(P^{\rm NV}(R), R)], \;\; v(R) := \var[H_T(\nvr(R), R)].
\end{equation}

Further, we formulate a risk-management problem under the \emph{conditional minimum variance (mV)}  criterion\footnote{There exist alternative, mathematically equivalent formulations; see Section~\ref{appendix:mveq} of the e-companion for a discussion.} to explicitly express the return-risk trade-off:
\begin{eqnarray}
\label{nvmv}
(\nvp_m, \nvr_m):= \argmin_{P, R} \var[H_T(P,R)] \quad {\rm s.t.}\quad \ex[H_T(P,R)] = m,
\end{eqnarray}
where $m \in (0, \ex[H_T(\nvp, \nvr)]]$ is the target mean return.
{We call \eqref{nvmv} the \emph{no-hedging model} to differentiate with the risk-minimization model with risk hedging in Section~\ref{sec:demandasset}. }
With $P = c$ or $R = bP$,
the production payoff is at most $-(c-s)\ex[(bc - A_T)^{{+}}] \leq 0$.
Thus, with $m > 0$, neither $P \ge c$ nor $R \ge bP$ can be binding and thus can be removed from the formulation in (\ref{nvmv}).
The same applies to all relevant settings throughout the paper.

{
Before proceeding, we review the concept of the \emph{efficient frontier}
\citep{markowitz1952}.
For a random return $W(x)$ induced by a decision $x$,
the corresponding mV formulation is: $v(m):=\min_x \var(W(x))$ subject to $\ex[W(x)] = m$, for which $m$ is an exogenously set target mean level of return.
(In our NV's setup, $x = (P,R)$ and $W(x) = H_T(P,R)$.)
If $v(m)$ is an increasing function in $m$, then $(m, \sqrt{v(m)})$ is said to be an efficient frontier.
It means that with all decisions being optimized, every increment of return is accompanied by an increment of risk.
Economically, the efficient frontier is a complete characterization of the best possible (i.e., efficient) trade-off between return and risk.
}

Now, we characterize these three optimization problems in the following proposition.
\begin{proposition}
\label{pro:nvfrontier}
Suppose Assumptions~\ref{assumption:r}--\ref{assumption:Abc} hold.
\begin{enumerate}[label=(\roman*)]
    \item  For problem~(\ref{NVP}), $R^{\rm NV}(P)$ increases in $P$.
Both $m(P)$ and $v(P)$ increase in $P \in [\underline{P}, P^{\rm NV}]$,
where $\underline{P}$ is the smallest root of $m(P) =0$,
 so $(m(P), \, \sqrt{v(P)})$ constitutes an efficient frontier.
    \item For problem~(\ref{NVR}), $P^{\rm NV}(R)$ increases in $R$.
Both $m(R)$ and $v(R)$ increase in $R \in [\underline{R}, R^{\rm NV}]$,
where $\underline{R}$ is the smallest root of $m(R) =0$, so $(m(R), \, \sqrt{v(R)})$ constitutes an efficient frontier.
 \item For problem~(\ref{nvmv}), $\var[H_T(\nvp_m, \nvr_m)]$ increases in $m$, so $(m, \sqrt{\var[H_T(\nvp_m, \nvr_m)]})$ constitutes an efficient frontier for $m \in (0, \ex[H_T(\nvp, \nvr)]]$.
In particular, $\frac{d}{dm}\sqrt{\var[H_T(\nvp_m, \nvr_m)]} = \infty$ at $m = \ex[H_T(\nvp, \nvr)]]$.
\end{enumerate}
\end{proposition}
Proposition~\ref{pro:nvfrontier} offers insights into how operational decisions impact return-risk trade-offs.
Part (i) indicates that a higher pricing level increases returns and risk simultaneously,
and analogously, Part (ii) considers the impact of the VPQ.\footnote{We expect $\underline{P}$ and $\underline{R}$ to be close to $c$ and  $bc$, respectively. This is confirmed in the data analysis in Section~\ref{sec:calibration}.}
Part (iii) mirrors the last two parts: higher returns induce higher risk (after $P$ and $R$ are optimized).
It also indicates that the profit-maximizing solution to the base model is just one point on the efficient frontier, inducing both the maximum risk and the maximum  {\it incremental} risk.
As the return approaches the newsvendor's maximum profit in the base model, the additional risk affected by a slight increase in return is enormous. In other words, the solution to the base model induces an extremely risky payoff. This paper proposes a risk-hedging model that improves the efficient frontier via substantial risk reductions from the no-hedging model.

Parts (i) and (ii) of Proposition~\ref{pro:nvfrontier} are symmetrical:
a higher $P$ induces a higher $R$ and vice versa.
This property confirms the {\it supermodularity} of $\ex[H_T(P,R)]$ in $(P,R)$, which holds because the pricing level's marginal value,
$\frac{\partial \ex[H_T(P,R)]} {\partial P} = -2bP + bc + R - \ex[(R-A_T)^+]$,
is  increasing in $R$.
Economically, this means that a higher VPQ (hence also a higher service level) increases the pricing level's marginal value, and we interpret it as follows.
Increasing the VPQ has opposing effects on the pricing level's marginal value. On the one hand, it increases revenue by capitalizing on the higher price from the increased VPQ (i.e., the term $R$ in $\frac{\partial \ex[H_T(P,R)]} {\partial P}$ above).
On the other hand, the higher the service level
{(induced by the higher VPQ value)
},
the greater the expected loss due to overproduction
(i.e., the term $\ex[(R-A_T)^+]$ in $\frac{\partial\ex[H_T(P,R)]}{ \partial P}$ above.
The positive effect outweighs the negative one, as the probability of selling a unit of product is always nonzero.
In a nutshell, a higher VPQ induces a higher pricing level due to its positive impact on the pricing level's marginal value.
This insight is crucial in proving our main results, Theorems \ref{thm:benefitpr} and \ref{thm:hurtpr}, in Section~\ref{sec:production}.

\section{Price-Setting Newsvendor with Risk Hedging}
\label{sec:demandasset}
We first introduce a general continuous-time model for financial asset price and product demand (Section~\ref{sec:demand}), and then formulate a risk-minimization problem  in the presence of risk hedging (Section~\ref{sec:mvhedging}).

\subsection{Financial Asset Price and Product Demand}
\label{sec:demand}
We fix a filtered probability space $\{\Omega, \{\calf_t\}_{t\ge 0}, \pr \}$ and define two independent standard Brownian motions
$B_t$ and $\tilde B_t$.
Let
$\calg_t$ and $\calf_t$ denote the filtrations
generated by $B_t$ and $(B_t, \tilde B_t)$, respectively.\footnote{Throughout the paper,
unless otherwise stated, adapted processes, martingales, and local martingales are defined with respect to $\calf_t$.}
Clearly, $\calg_t \subseteq \calf_t$.
Suppose the asset price $X_t$ is a continuous diffusion process driven by $B_t$:
\begin{equation}
\label{Xt}
dX_t = X_t (\mu_t dt + \sigma_t dB_t),
\end{equation}
where $\mu_t$ and $\sigma_t > 0$ are continuous  processes  adapted to $\calg_t$.
Suppose also that $(X_t, \mu_t, \sigma_t)$ is Markovian
under $\pr$.
We make the following integrability assumption.
\begin{assumption}
\label{assumption:X}
$\int_0^T \ex[(X_t\sigma_t)^2] dt < \infty.$
\end{assumption}

Let $D_t$ be the cumulative demand up to time $t$. Suppose
\begin{equation}
\label{Dt}
dD_t = dA_t - \frac{b}{T} P dt
\quad{\rm with}\quad dA_t = dC_t + \tilde\sigma d\tilde B_t,
\end{equation}
where $\tilde\sigma>0$ is a constant and $C_t$ is a continuous process adapted to $\calg_t$.
We further assume that for all $u \leq v $, $C_v - C_u$ is
adapted to the sigma-algebra generated by $\{X_t, u \leq t \leq v\}$; that is, the asset price fluctuation over any time interval
impacts the demand accumulated over the same period, which is consistent with the discussion in Section~\ref{sec:intro}.
Note that for all $t \leq u$, $A_u = A_t + (C_u - C_t) + \tilde\sigma(\tilde B_u - \tilde B_t)$.
Given $\calf_t$, the distribution of $A_u$ under $\pr$ is determined by $(X_t, \mu_t, \sigma_t, A_t)$ because of the assumed Markovian property of $(X_t, \mu_t,\sigma_t)$
and the independence of $\tilde B_t$.

Three factors determine the instantaneous incremental demand $d D_t$ in equation~\eqref{Dt}.
That is,
$dC_t$ (representing the impact of the asset price fluctuation),
$\tilde\sigma d\tilde B_t$ (capturing the demand's inherent volatility),
and $-(b/T)Pdt$ (incorporating the sensitivity of the demand rate with respect to the selling price).
Without loss of generality, we assume that $C_0 = 0$.
Thus, the base demand $D_0 = 0$.
In light of the demand function $D_T(P) = A_T - bP$,
equation~(\ref{Dt}) models the market size $A_T$
as
\begin{equation}
\label{AT1}
A_T = C_T + \tilde\sigma \tilde B_T.
\end{equation}

The demand model above is {\it general}.
We do not assume any specific functional form of $C_t$ in $\{X_s: 0\leq s\leq t\}$, allowing the model to be adapted to various specific application scenarios.
A simple example is $C_t = g(X_t)$ for some function $g$; see, e.g., \cite{gaur2005hedging}.
One can extend this simple model to permit dependence on the path of asset prices.
For example, $C_T = \int_0^T \tilde\mu(X_t)dt$, where $\tilde\mu$ is a function of the asset price and models the demand rate (see Section~\ref{sec:numerical}).

Assuming the interest rate is zero,
the associated MPR process
is
\begin{equation}
\label{eta}
\eta_t := \frac{\mu_t}{\sigma_t}.
\end{equation}
It captures the risk-adjusted {\it trend} of the asset price which may positively or negatively impact product demand.
For example, with rising oil prices,
the demand for fuel-efficient cars (usually sedans) increases while the demand for fuel-inefficient cars decreases.

We introduce several technical definitions to formalize the impact of the asset price trend on product demand.
Define the risk-neutral measure $\pr^M$ for the asset price process via the Radon--Nikodym derivative $Z_T$. That is,
\begin{equation}
\label{genZt}
\frac{d\pr^M}{d\pr} := Z_T,
\quad Z_t := e^{-\int_0^t \eta_s dB_s - \frac{1}{2}\int_0^t \eta_s^2 ds},\; t\in[0,T].
\end{equation}
Suppose  $(X_t, \sigma_t)$ is Markovian under $\pr^M$.
Then, given $\calf_t$, the distribution of $A_u$ under $\pr^M$ is determined by \ $(X_t, \sigma_t, A_t)$ for all $t \leq u$.
As will become evident later in the paper,  $Z_t$ is crucial in determining the optimal hedging strategy.
We impose a mild technical assumption on it.
\begin{assumption}
\label{assumption:Z}
$Z_t$ is a square-integrable martingale over $[0, T]$ under $\pr$.
\end{assumption}

Under the probability measure $\pr^M$, the asset price process $X_t$, which is driven by $B_t$, is {\it de-trended}, becoming a local martingale.
By contrast, the change of measure does not affect the non-financial noise $\tilde B_t$.
Moreover, let $C_T^M$ be a random variable whose distribution under $\pr$ is identical to that of $C_T$ under $\pr^M$; that is, $\pr(C_T^M \leq x) = \pr^M(C_T \leq x)$ for all $x$.\footnote{{
For example, suppose $C_T = \int_0^T \tilde\mu(X_t)dt$ for some function $\tilde\mu$.
Then, $C_T^M = \int_0^T \tilde\mu(X_t^M)dt$, where $\{X_t^M:t \in [0, T]\}$ is a stochastic process whose distribution under $\pr$ is identical to that of $\{X_t:t\in[0, T]\}$ under $\pr^M$.
In Section~\ref{sec:numerical}, we assume $X_t$ is an EOU process under $\pr$, following the dynamics (\ref{expOUX}).
Then, it is easy to show that the distribution of  $X_t$ under $\pr^M$ equals $X_0e^{-(1/2)\sigma^2t + \sigma B_t^M}$, where $B_t^M$ is a Brownian motion under $\pr^M$.
Hence, the distribution of $X_t^M$ under $\pr$ equals $X_0e^{-(1/2)\sigma^2t + \sigma B_t}$, where $B_t$ is a Brownian motion under $\pr$.}}
We can interpret $C_T^M$ as the version of the financial component of the demand that incorporates no asset price trend.

We also need the notion of {\it stochastic order} to compare random variables \citep{rossbook}.
Recall that for two random variables $X$ and $Y$,
$X$ is said to be {\it stochastically larger}  than $Y$, denoted by $X \succeq Y$ or
$Y \preceq X$,
if $\pr(X \leq a) \leq \pr(Y \leq a)$ for any real  $a$.
An immediate result following this definition is that,
$X \succeq Y$ if and only if $\ex[f(X)] \ge \ex[f(Y)]$ for any increasing function $f$.

{To illustrate this order relation between two stochastic processes,
consider two geometric Brownian motions (GBMs): $X^{(1)}_t = X^{(1)}_0\exp{((\mu_1 - 0.5\sigma^2)t + \sigma B^{(1)}_t)}$ and $X^{(2)}_t = X^{(2)}_0\exp{(- 0.5\sigma^2t + \sigma B^{(2)}_t)}$, where $B^{(1)}_t$ and $B^{(2)}_t$ are two Brownian motions.
If $X^{(1)}_0= X^{(2)}_0$ and $\mu_1 \ge 0$, then the two GBMs have the same initial value and volatility, but $X^{(1)}_t$ has a positive trend while $X^{(2)}_t$ does not.
It is easy to verify that
$X^{(1)}_t \succeq X^{(2)}_t$ for any $t \ge 0$.
Note that in the special case that these two Brownian motions are identical in the path-wise sense, we have that $X^{(1)}_t \ge X^{(2)}_t$ path-wise, which implies the stochastic ordering between them.
Furthermore, in this special case, if we assume in the market size model~\eqref{AT1} that $A^{(1)}_T = \int_0^T \tilde\mu(X^{(1)}_t)dt + \tilde\sigma \tilde B_T$ and $A^{(2)}_T = \int_0^T \tilde\mu(X^{(2)}_t)dt + \tilde\sigma \tilde B_T$,
then $A^{(1)}_T \succeq$ $A^{(2)}_T$ for an increasing function $\tilde\mu$ while $A^{(1)}_T \preceq$ $A^{(2)}_T$ for a decreasing function $\tilde\mu$.
In Section~\ref{sec:numerical}, we assume that the asset price follows an EOU process, and the demand rate function $\tilde{\mu}$ is linear. In this example, it is analytically intractable to establish a stochastic order between $C_T$ and $C_T^M$. Instead, we resort to a hypothesis-testing approach, applying the Mann--Whitney U-test \citep{MW1947} to simulated samples of these two quantities with parameters estimated from real world data; see Section~\ref{appendix:sdtest} of the e-companion.}

Now, we formally define the positive/negative effect of the asset price trend on product demand.
\begin{itemize}
\item The asset price trend is said to {\it positively impact} demand, if $C_T \succeq C_T^M$.
In this case,
$\ex[H_T(P,R)] \ge \ex^M[H_T(P,R)]$ for all $(P,R)$;
refer to Proposition~\ref{pro:prx0-positive} in Section~\ref{sec:benefitpr}.
\item The asset price trend is said to {\it negatively impact} demand, if $C_T \preceq C_T^M$.
In this case, $\ex[H_T(P,R)] \leq \ex^M[H_T(P,R)]$ for all $(P,R)$;
refer to Proposition~\ref{pro:prx0-negative} in Section~\ref{sec:hurtpr}.
\end{itemize}
In other words, the positive effect occurs if
the demand incorporating the asset price trend is stochastically larger than the demand impacted by the de-trended asset price.
We can interpret the negative effect likewise.

\begin{remark}
\label{rem:ATorder}
{
The positive/negative impact of asset price trend on product demand defined above directly translates to the same stochastic order relation between the market sizes under the real-world and risk-neutral probability measures. That is, $C_T \succeq C_T^M$ implies $A_T \succeq A_T^M$, and likewise $C_T \preceq C_T^M$  implies $A_T \preceq A_T^M$.
This can be easily shown using the definitions of $A_T$ and $A_T^M$ and the fact that stochastic ordering is preserved under convolutions of probability distributions; see \citet[page~6]{StochOrderBook}.
}
\end{remark}

In Section~\ref{sec:production}, we study the impact of risk hedging on the operational decisions,
which entails characterizing how the stochastic order of $C_T$ between $C_T^M$ influences the operational decisions.
We show that a (stochastically) larger market induces a higher profit-maximizing price and a higher VPQ; see Propositions~\ref{pro:prx0-positive} and \ref{pro:prx0-negative}  in Section~\ref{sec:production}.

\subsection{Mean-Variance Risk-Hedging Model}
\label{sec:mvhedging}
Having built the impact of asset price into the demand model, one can design a real-time hedging strategy using the financial asset to mitigate production payoff risk.
Specifically, in problem~(\ref{nvmv}), we include a risk-hedging strategy as a third decision variable in addition to $P$ and $R$:
\begin{equation}
\label{mvprob0}
\begin{aligned}
&\min_{P,\,R,\, \vartheta := \{\theta_t, t\in[0,T]\} \in \cala_X}\var[H_T(P,R) + \chi_T(\vartheta)] \\
&{ s.t.}\quad \chi_t(\vartheta):= \int_0^t \theta_s dX_s, \quad \ex[H_T(P,R) + \chi_T(\vartheta)] = m,
\end{aligned}
\end{equation}
where $\theta_t$ is the number of shares held in the asset at time $t$,
$\chi_t$ is the cumulative profit/loss from the hedging strategy  until time $t$, and
$\cala_X$ is the set of all \emph{admissible} strategies that satisfy certain conditions.\footnote{For example,
$\theta_t$ is an adapted process (so the strategy does not look into the future)
and $\chi_T(\vartheta)$ has a finite variance;
other conditions are described in Section~EC.7.2.1 of the e-companion.}
{We call \eqref{mvprob0} the \emph{hedging model} in contrast to the no-hedging model \eqref{nvmv}.
We remark that the hedging model  applies to large and financially unconstrained manufacturers; hence there is no budget limit on the execution of the hedging strategy.}

To solve problem~(\ref{mvprob0}), we first optimize the hedging strategy with fixed $P$ and $R$ (see Section~\ref{sec:hedging}):
\begin{equation}
\label{mvprob}
\begin{aligned}
B(m, P, R) &:= \min_{\vartheta = \{\theta_t, t\in[0,T]\} \in \cala_X}\var[H_T(P,R) + \chi_T(\vartheta)] \\
           &{ s.t.}\quad \chi_t(\vartheta):= \int_0^t \theta_s dX_s, \quad \ex[H_T(P,R) + \chi_T(\vartheta)] = m.
\end{aligned}
\end{equation}
Then, we minimize the optimal value of this hedging problem over $P$ and $R$ to further reduce risk:
\begin{equation}
\label{prprob}
(P^h_m, R^h_m):= \argmin_{P, R} B(m, P, R).
\end{equation}
Moreover, we examine how hedging affects optimal operational decisions compared to the no-hedging case and completely characterize the return-risk efficient frontier
in Section~\ref{sec:production}.

{
Last, we emphasize that a primary goal of this paper (see the research question (ii) in Section~\ref{sec:intro}) is to compare the price and VPQ under  the no-hedging model \eqref{nvmv} and those under the hedging model \eqref{mvprob0}.
Both models are formulated as a risk minimization problem subject to an exogenously set target mean return (i.e., $m$), following the classical Markowitz's model \citep{markowitz1952}.
To make a fair comparison between the two models, the values of $m$ are \emph{identical} in both models.
(Note that $m$ is the sum of the mean operational and hedging payoffs.)}

\begin{remark}
 \label{remark:MV}
{The risk-minimization objective of the problem (\ref{mvprob0}) may lead to a reduction in the price and VPQ (see Sections~\ref{sec:benefitpr} and \ref{sec:hurtpr}).
However, the firm may be incentivized to expand operational activities and use risk hedging to improve the return-risk trade-off instead of minimizing risk with a \emph{fixed} target mean return.
Our model can readily capture this different attitude of the firm as follows.
Suppose the firm wants to expand production and prescribes a target production quantity.
Then, the only change needed in our model is to fix $Q$ at this prescribed level in the variance function $B(m, P, R)$ in (\ref{mvprob}) and optimize the price $P$.
Alternatively, one may adopt the \emph{mean-variance utility maximization} (MV) formulation to encompass the objective of expanding operational activities.
It is straightforward to show that the MV formulation is mathematically equivalent to our mV formulation~(\ref{mvprob0}), meaning they lead to the same efficient frontier.
However, because the mV formulation fixes the target mean return whereas the MV formulation does not, the impact of hedging on the operational decisions may be substantially different. See more discussion in Section~\ref{appendix:mveq} of the e-companion.
}

\end{remark}

\section{Solving the Hedging Problem}
\label{sec:hedging}
We apply the quadratic hedging technique \citep{gourieroux1998mean} to solve the hedging problem~(\ref{mvprob}).
First, define the following martingale
under $\pr^M$:
\begin{equation}
\label{genZtM}
Z_t^M:= \ex^M(Z_T \,|\, \calf_t), \quad {\rm with} \quad
dZ_t^M =\zeta_t dX_t,
\end{equation}
where the dynamics of $Z_t^M$ (i.e., $dZ_t^M$) follows directly from Martingale Representation Theorem,
and $\zeta_t$ is a stochastic process adapted to $\calg_t$
(refer to Corollary~\ref{cor:calfg} in the e-companion).
The solution approach essentially relies on the {\it projected} production payoff process, defined as
\begin{equation}
\label{Vt}
V_t(P,R) := \ex^M[H_T(P,R) \,|\, \calf_t], \quad 0\leq t \leq T.
\end{equation}
Clearly, $V_t$ is a martingale under $\pr^M$, and its martingale representation (summarized as the lemma below) plays a crucial role in determining the optimal hedging strategy.
\begin{lemma}
\label{lem:Vt}
$V_t(P,R)$ defined in (\ref{Vt}) has the following representation:
\begin{equation}
\label{GKW}
V_t(P,R) = V_0(P,R) + \int_0^t \xi_s(P,R) dX_s + \int_0^t \delta_s(P,R)d\tilde B_s,
\end{equation}
where $V_0(P,R) = \ex^M[H_T(P,R)]$ and
$\xi_t$ and $\delta_t$ are adapted processes.
In particular,
\begin{equation}
\label{delta}
\delta_t(P,R) = \tilde\sigma(P-s)\pr^M(A_T \leq R \,|\, \calf_t ),
\end{equation}
which increases in both $P$ and $R$, where $A_T$ is defined in (\ref{AT1}) and $s$ is the unit salvage price.
\end{lemma}

\subsection{Optimal Hedging Strategy and Minimal Variance}
\label{sec:strategy}
\begin{theorem}
\label{thm:hedgingsol}
Suppose Assumptions \ref{assumption:X}--\ref{assumption:Z} hold.
\begin{enumerate}[label=(\roman*)]
\item The optimal solution to the problem in (\ref{mvprob}) is
\begin{equation}\label{opttheta}
    \theta^*_t(P,R) = \underbrace{-\xi_t(P,R)}_{\text{risk-mitigation position}} + \underbrace{\iota_t(P,R)}_{\text{investment position}},
\end{equation}
where, with the terms $Z_t^M$, $\zeta_t$, and $\xi_t$ being defined in  (\ref{genZtM}) and (\ref{GKW}),
\[
\iota_t(P,R) = -\frac{\zeta_t}{Z_t^M}\biggl[\underbrace{\frac{mZ_0^M - V_0(P,R)}{Z_0^M - 1}}_{\g_m} - V_t(P,R) - \underbrace{\int_0^t \theta^*_s(P,R)dX_s}_{\chi_t^*}\biggr].
\]

\item
The optimal objective function value in (\ref{mvprob}) has the following expression:
\begin{equation}
\label{bmpr}
B(m, P, R) =\frac{[m - V_0(P,R)]^2}{Z_0^M-1} + \int_0^T \ex\Big[\frac{Z_t}{Z_t^M}\delta_t^2(P,R)\Big]dt,
\end{equation}
where $\delta_t(P,R)$ is defined in (\ref{delta}).
In particular, $Z_0^M = \ex(Z_T^2) > 1$ and $0\leq Z_t / Z_t^M \leq 1$.

\item
{
Define the following terms:
\begin{equation}
\label{chidecomp}
\chirm_T(P,R) := \int_0^T (-\xi_t(P,R))dX_t, \quad \chiiv_T(P,R) := \int_0^T \iota_t(P,R) dX_t,
\end{equation}
and
\begin{equation}
 \label{hedgedprodpayoff}
 \HTh(P, R):= H_T(P,R) + \chirm_T(P,R).
\end{equation}
Then, $\var(\chiiv_T) + \cov(\chiiv_T, \HTh) \geq 0$, $\var(\HTh) + \cov(\chiiv_T, \HTh) \geq 0$, and
\begin{gather}
\underbrace{\sqrt{\var(\chiiv_T) + \cov(\chiiv_T, \HTh)}}_{\text{investment risk}}
= \Biggl[\underbrace{\frac{[m-V_0(P,R)]^2}{Z_0^M-1}}_{\text{1st term of $B(m, P, R)$}}\Biggr]^{1/2}, \label{invrisk} \\
\underbrace{\sqrt{\var(\HTh) + \cov(\chiiv_T, \HTh)}}_{\text{unhedgeable risk}}
= \Biggl[\underbrace{\int_0^T \ex\Big[\frac{Z_t}{Z_t^M}\delta_t^2(P,R)\Big]dt}_{\text{2nd term of $B(m, P, R)$}}\Biggr]^{1/2}.   \label{unhedgeablerisk}
\end{gather}
}
\end{enumerate}
\end{theorem}

Part (i) of Theorem~\ref{thm:hedgingsol} is intuitively appealing.
The structure in (\ref{opttheta}) reveals two aspects of the optimal hedging strategy.
First, the term $\xi_t(P,R)$, which also appears in (\ref{GKW}), captures the impact of asset price movement (i.e., $dX_t$) on production payoff.
By taking the opposite position, $-\xi_t(P,R)$, the hedging strategy offsets such impact to mitigate risk.
The second term in (\ref{opttheta}) involves the difference between $\g_m$ (a proxy of the target mean return $m$) and $V_t(P,R) + \chi_t^*$ (the projected total return at time $t$).
This term functions as an {\it investment} to close this gap between the target and  the total return by taking a position in the asset.
\cite{wang2017production} also note these two aspects but with specific asset price (GBM) and demand models.

Part (ii) of Theorem~\ref{thm:hedgingsol} presents the minimal variance of the hedging problem~(\ref{mvprob}).
Its structure is closely related to that of the optimal hedging strategy and can be interpreted from the perspective of risk decomposition, clarified in Part (iii).
{By definition, the total return can be rewritten as
\begin{equation}\label{eq:total_wealth}
    H_T + \chi_T^* = \HTh + \chiiv_T,
\end{equation}
where $\chiiv_T$ is the payoff from the investment position $\iota_t$ of the optimal hedging strategy $\theta_t^*$, and $\HTh$ represents the \emph{hedged production payoff} for the following reason.
Note that
\begin{equation}\label{eq:hedged-prod-payoff}
    \HTh = H_T + \chirm_T = V_0 + \int_0^T \xi_t dX_t + \int_0^T \delta_t d\tilde B_t + \int_0^T (-\xi_t)dX_t =V_0 + \int_0^T \delta_t d\tilde B_t,
\end{equation}
where the first equality is the definition \eqref{hedgedprodpayoff}, the second equality holds because of the fact that $H_T = V_T$, Lemma~\ref{lem:Vt}, and the definition of $\chirm_T$ in \eqref{chidecomp}.
The calculations in \eqref{eq:hedged-prod-payoff} reveals that $\chirm_T$ (which is the payoff from
the risk-mitigation payoff of the optimal hedging strategy)
offsets $\int_0^T \xi_t dX_t$ (which is the part of production payoff $H_T$ that is driven by the asset price movement).
Furthermore, taking expectations in \eqref{eq:hedged-prod-payoff} yields $V_0 = \ex(\HTh)$, so $V_0$ carries the economic meaning of {\it expected hedged production payoff}.
}

{We now provide a risk-decomposition interpretation of \eqref{bmpr}.
According to the definition of $B(m, P, R)$ in \eqref{mvprob} and the total return's representation in \eqref{eq:total_wealth}, we have
\[B(m, P, R)= \var(\HTh + \chiiv_T)
= \left[\var(\chiiv_T) + \cov(\chiiv_T, \HTh)\right] + \left[\var(\HTh) + \cov(\chiiv_T, \HTh)\right].\]
Namely, we may decompose the variance of the total return into two parts,\footnote{{This kind of variance decomposition is also used in Risk Parity, a widely applied portfolio management strategy in the financial industry; see \cite{AFP2012} and the white paper by Wealthfront, LLC. (\url{https://research.wealthfront.com/whitepapers/risk-parity}).}}
one associated with the investment payoff $\chiiv_T$ and the other with the hedged production payoff $\HTh$.
In particular, we call $\sqrt{\var(\chiiv_T) + \cov(\chiiv_T, \HTh)}$ {\it investment risk} and call $\sqrt{\var(\HTh) + \cov(\HTh, \chiiv)}$  {\it unhedgeable risk}.
}

{
Furthermore, (\ref{invrisk}) and \eqref{unhedgeablerisk} show that these two types of risk directly translate to the two terms of the decomposition of the total variance $B(m, P, R)$ in (\ref{bmpr}).
Their expressions are intuitive and revealing.
First,
note that the expected investment payoff needs to fill in the gap between the target mean return $m$ and the expected hedged production payoff $V_0$, that is,
\begin{equation}
\label{invfillgap}
\ex(\chiiv_T) = m - V_0.
\end{equation}
Hence, the investment risk is proportional to
$m - V_0$.
(Proposition~\ref{pro:prbound-part1} in Section~\ref{sec:production} implies that
$m - V_0 > 0$ for the optimal $(P, R)$ in the presence of hedging.)
This implies that a larger gap (i.e., $m-V_0$) needs a higher investment payoff, inducing higher investment risk.
Second,
regarding the unhedgeable risk,
because it does not involve $m$ and increases in the demand noise $\tilde\sigma$ via $\delta_t$ in (\ref{delta}),
it can not be reduced by trading financial asset.
Moreover, the unhedgeable risk increases in both $P$ and $R$, meaning that a higher operational level increases the exposure to the demand uncertainty that is unhedgeable.
Assuming the product price is exogenously given, \cite{wang2017production} and \cite{wang2021mean} note results similar to \eqref{bmpr} under particular asset price models.
However, these two papers do not establish the kind of relationship like (\ref{invrisk}) and \eqref{unhedgeablerisk} between the risk decomposition and the two terms of the minimum variance \eqref{bmpr}; they  do not recognize $V_0$ as the expected hedged production payoff from an economic perspective.
}

In the presence of hedging, the optimal operational decisions minimizing $B(m, P, R)$ essentially assumes a market of size $A_T^M$,
the version of $A_T$ under the risk-neutral probability measure $\pr^M$
(formally, $A_T^M$ is defined via $\pr(A_T^M \leq a) = \pr^M(A_T \leq a)$ for all $a$),
because the operations-related terms in (\ref{bmpr}) (i.e.,  $V_0(P,R)$ and $\delta_t(P,R)$) only depend on the distribution of $A_T$ under $\pr^M$.
Hedging---particularly the risk mitigating term $-\xi_t$ in (\ref{opttheta})---offsets any impact of the asset price trend on production payoff (via the impact on the market size $A_T$).
Consider the following  examples.
\begin{itemize}
    \item The asset is a stock index, reflecting the general economic condition.
    Suppose its price $X_t$ follows a GBM, and a higher asset price boosts the product demand.
Then, the asset price trend's positive effect on the distribution of the market size $A_T$ under $\pr$ can be captured by an upward trend of $X_t$ (i.e., $\mu > 0$).
In the presence of hedging, $A_T$ is evaluated under $\pr^M$, which corresponds to a GBM with zero drift (i.e., the asset price under $\pr^M$ has no trend).
\item The asset is a commodity (e.g., crude oil) with mean-reverting prices.
The price tends to return to its long-term mean when deviating from the latter, impacting product demand.
For instance, when the oil price rises towards its long-term mean, the demand for fuel-inefficient cars will fall.
Suppose the price $X_t$ follows an EOU process.
Then, the asset price negatively impacts the demand if the initial price $X_0$ is below the long-term mean (see Sections~\ref{sec:numerical}).
Similar to the last example, the distribution of $A_T$ under $\pr^M$ also corresponds to a GBM with zero drift.
\end{itemize}

In both examples, hedging removes the asset price trend's positive/negative effect on production payoff,
and the firm optimizes operational decisions assuming that the market size will be as if the asset price has no trend, remaining at the current level in expectation.
In other words, when the optimization of the operational decision $(P,R)$ is concerned,
hedging effectively replaces the actual market size $A_T$ in the real world
(i.e., under $\pr$) with a market size $A_T^M$ in the risk-neutral world (i.e., under $\pr^M$) where the asset price has no trend.
This insight will be crucial in analyzing optimal price and VPQ in Section~\ref{sec:production}.
\begin{remark}
We solve the hedging problem~(\ref{mvprob}) for a general continuous diffusion process having a {\it stochastic} MPR process.
Our solution procedure is significantly more challenging than that of \cite{wang2017production}---where the asset price is assumed to follow a GBM, inducing a \emph{constant} MPR process---mainly concerning the characterization of $\zeta_t$ defined in (\ref{genZtM}).
Specifically,
$\zeta_t$ must be such that the optimal hedging strategy, $\theta_t^*$ in (\ref{opttheta}), is admissible (see the discussion below equation~(\ref{mvprob0})).
While it is straightforward to verify this property for GBMs,
doing so for general diffusion processes is considerably more difficult (see Lemma~\ref{lem:genztmmg} in the e-companion).
Due to the generality, our model encompasses a much broader scope of applications than that of \cite{wang2017production}.
For example, in Section~\ref{sec:expou},  we apply Theorem~\ref{thm:hedgingsol} to an EOU process capturing asset prices' mean-reverting property, whereas the corresponding result of \cite{wang2017production} does not cover mean-reverting processes.
In particular, verifying Assumption~\ref{assumption:Z} for EOU processes is nontrivial because of the difficulty of calculating critical terms including $\zeta_t/Z_t^M$ in the hedging strategy (\ref{opttheta}) and $Z_t/Z_t^M$ in the variance function (\ref{bmpr}) (see Section~\ref{appendix:hedgingsolEOU} of the e-companion).
By contrast, Assumption~\ref{assumption:Z} trivially holds for GBMs.
Again, the difference stems from the nature of the asset price's MPR process:
it is stochastic for EOU processes, whereas constant for GBMs.
\end{remark}

\subsection{Example: EOU Asset Price Model}
\label{sec:expou}

To align with the automakers' example in Section~\ref{sec:intro},
we assume the financial asset is crude oil, and its price $X_t$ follows an EOU process, commonly used to model the mean-reverting feature of commodity prices \citep{schwartz1997stochastic}.
Suppose $X_t = e^{Y_t}$ with $dY_t = \kappa(\alpha-Y_t)dt +\sigma dB_t$. Then,
\begin{equation}
\label{expOUX}
dX_t = \kappa\Big(\alpha+\frac{\sigma^2}{2\kappa}-\log X_t\Big)X_tdt+ \sigma X_t dB_t,
\end{equation}
where $\kappa$, $\alpha$, and $\sigma$ are all positive constants and the dynamics of the asset price $X_t$ directly follows from It\^{o}'s Lemma.
In particular, $\alpha$ represents the long-term mean of $Y_t$ and $\kappa$ is the mean-reversion coefficient.
The MPR $\eta_t$ in (\ref{eta}) corresponding to the EOU process is
\begin{equation}
\label{etaeOU}
\eta_t =\frac{\kappa}{\sigma}\Big(\alpha+\frac{\sigma^2}{2\kappa}-\log X_t\Big).
\end{equation}
When the mean-reversion coefficient $\kappa$ approaches zero,
$Y_t$ in (\ref{expOUX}) reduces to $\sigma B_t$, $X_t$ in (\ref{expOUX}) reduces to a GBM, and $\eta_t$ in (\ref{etaeOU}) becomes $\sigma / 2$.
All the results in this section hold for this limiting case.

To apply Theorem~\ref{thm:hedgingsol} to the EOU model,
we need to verify Assumptions~\ref{assumption:X} and \ref{assumption:Z}.
The former holds trivially, and for the latter, we impose the following sufficient condition
(see Lemma~\ref{lem:OUztsqmg} in
the e-companion):
\begin{equation}
\label{kappa}
\kappa T<\frac{\pi}{4}.
\end{equation}

\begin{proposition}
\label{pro:hedgingsolEOU}
Suppose that $X_t$ follows the EOU process (\ref{expOUX}) and that the condition~(\ref{kappa}) holds.
\begin{enumerate}[label=(\roman*)]
    \item The optimal solution to the problem in (\ref{mvprob}) is
		\begin{equation*}
		\label{optthetaEOU}
		\theta^*_t(P,R) = -\xi_t(P,R) - \frac{a(T-t)+\frac{\kappa}{\sigma^2}(\alpha-Y_t)b(T-t)}{X_t}[\g_m - V_t(P,R) - \chi_t^*],
		\end{equation*}
		where $\chi_t^* = \int_0^t \theta^*_s(P,R)dX_s$, $\xi_t(P, R)$ is defined in (\ref{GKW}),
		$\g_m$ is defined in (\ref{opttheta}), and $a$ and $b$ are functions defined as   \begin{equation}
       \label{abtau}
        a(\tau) = \frac{1}{2} + \frac{1}{\cos{\kappa \tau}-\sin{\kappa \tau}}, \quad b(\tau) =  \frac{\cos{\kappa \tau}+\sin{\kappa \tau}}{\cos{\kappa \tau}-\sin{\kappa \tau}},\quad \tau\in[0,T].
       \end{equation}

    \item The optimal objective function value of the problem in (\ref{mvprob}) is
		\begin{align}
		\label{EOUbmpr}
		B(m, P,  R) = \frac{[m-V_0(P, R)]^2}{Z_0^M-1}+ \int_{0}^T\ex [e^{-f_0(T-t)-f_1(T-t)Y_t-f_2(T-t)Y_t^2 } \delta^2_t(P, R)]dt,
		\end{align}
		where $\delta_t(P, R)$ is defined in (\ref{GKW}), and $f_0$, $f_1$, and $f_2$ are functions defined as
		\begin{equation}
		\label{OUf}
		\begin{aligned}
		& f_0(\tau) = -\alpha -(\frac{1}{2}\kappa +\frac{1}{4}\sigma^2)\tau -\frac{1}{2}\log [\cos{\kappa \tau}-\sin{\kappa \tau}] + \frac{\alpha+(\frac{\alpha^2\kappa}{\sigma^2}+\frac{\sigma^2}{2\kappa})\sin{\kappa \tau}}{\cos{\kappa \tau}-\sin{\kappa \tau}},\\
		& f_1(\tau) = \frac{-1+\cos{\kappa \tau} - (\frac{2\kappa\alpha}{\sigma^2}+1)\sin{\kappa \tau} }{\cos{\kappa \tau}-\sin{\kappa \tau}},\quad
		 f_2(\tau ) = \frac{\kappa}{\sigma^2} \frac{\sin{\kappa \tau}}{\cos{\kappa \tau}-\sin{\kappa \tau}},\quad \tau\in[0,T].
		\end{aligned}
		\end{equation}
\end{enumerate}
\end{proposition}

Functions in (\ref{abtau}) and (\ref{OUf}) are well defined under the condition (\ref{kappa}).
By further specifying the market size $A_T$ in (\ref{AT1}),
we can derive $\xi_t(P,R)$ and $\delta_t(P,R)$ explicitly (see Section~\ref{appendix:eoudeltaxi} of the e-companion).

\section{Optimal Price and VPQ in the Presence of Hedging}
\label{sec:production}
In this section, we focus on the operational task and solve problem~(\ref{prprob}) to find the jointly optimal price and VPQ in the presence of hedging.\footnote{{For a given $m\geq 0$, numerically minimizing $B(m, P, R)$ over $(P, R)$ can be reduced to a line search over a finite interval—instead of a two-dimensional grid search—because $B(m, P, R)$ is convex with respect to $P$ within a specific range; see Proposition~\ref{pro:rcvx} in Section~\ref{sec:numerical-procedure} of the e-companion.}}
We present the main results of this paper regarding optimal price and VPQ in the presence of hedging, discussing their properties depending on whether the asset price trend positively (Section~\ref{sec:benefitpr}) or negatively (Section~\ref{sec:hurtpr}) impacts the demand.
We also fully characterize the efficient frontier for the hedging model (Section~\ref{sec:hedgingeff}).

Sections~\ref{sec:benefitpr} and \ref{sec:hurtpr} focus on the case that the target mean return is set as $m = \ex[H_T(\nvp, \nvr)]$.
This return level is an important benchmark for decision-making as it is the maximum expected profit that the firm can attain (without hedging) in the base model \eqref{nv}.
It is economically sensible for a non-financial firm (e.g., a manufacturer) to target its earnings around this level.
Setting a substantially lower target is certainly inconsistent with the profit-generating goal of the firm.
On the other hand, targeting at a significantly higher level than $m$ will alter the non-financial nature of the firm since the extra income can only come from trading the financial asset, which does not align with the firm's primary operations.
{
With $m = \ex[H_T(\nvp, \nvr)]$, it is straightforward to see that the risk-minimizing price and VPQ in the no-hedging model \eqref{nvmv} are identical to the profit-maximizing decisions in the base model \eqref{nv}, both equal to  $(\nvp, \nvr)$.
Hence, Theorems~\ref{thm:benefitpr} and \ref{thm:hurtpr} compare $(\nvp, \nvr)$ with the risk-minimizing decisions $(P_m^h, R_m^h)$, and their conclusions are based on the condition $m = \ex[H_T(\nvp, \nvr)]$.
}

\begin{remark}
{There exist situations in which other choices of $m$ may better reflect the firm's attitude.
For example, if it anticipates the product demand is low (hence a low value of $\ex[H_T(\nvp, \nvr)]$), the firm may want to set $m$
significantly higher than $\ex[H_T(\nvp, \nvr)]$ and use hedging to fill in the gap between the operational payoff and the target.
In other words, the firm views hedging as a tool for profit improvement instead of risk minimization.
For another example, if its risk tolerance is low, the firm may prefer a target $m$ that is substantially below $\ex[H_T(\nvp, \nvr)]$ to reduce risk.
When $m = \ex[H_T(\nvp, \nvr)]$ does not reasonably reflect the firm's return or risk preferences, or when it is not straightforward to select a proper $m$, the mean-variance utility maximization (MV) formulation may be a better choice, provided that the firm's risk-aversion coefficient is known or easy to estimate.
In this case, the efficient frontier generated by our mV formulation can be used to solve the MV formulation; see Section~\ref{appendix:mveq} of the e-companion.
}
\end{remark}

\subsection{Asset Price Trend Positively Impacting Demand}
\label{sec:benefitpr}
We first examine the case of the asset price trend positively impacting demand (i.e., $C_T \succeq C_T^M$). A leading example is automakers: when oil prices exhibit a downward trend, demand for fuel-inefficient vehicles such as SUVs or pickup trucks will increase.
{
We compare the risk-minimizing operational decisions in the hedging and no-hedging models, that is, $(P_m^h, R_m^h)$ versus $(P^{\rm NV}, R^{\rm NV})$ for $m = \ex[H_T(\nvp, \nvr)]$.
To this end, we introduce an intermediate set of operational decisions.
Define the profit-maximizing price and VPQ under the risk-neutral measure $\pr^M$ as
\begin{equation*}
(P^{\rm NV(M)},\, R^{\rm NV(M)}) := \argmax_{P,R} V_0(P, R).
\end{equation*}
Recall that, by definition, $V_0(P, R) = \ex^M[H_T(P, R)]$ is the expected production payoff under the risk-neutral measure, so $(\nvmp, \nvmr)$ maximizes the expected production payoff without considering the asset price trend.
In the following, we present two sets of comparisons: one comparing $(P^{\rm NV}, R^{\rm NV})$ with $(P^{\rm NV(M)},\, R^{\rm NV(M)})$, the other comparing $(P^{\rm NV(M)},\, R^{\rm NV(M)})$ with $(P_m^h, R_m^h)$.
}

\begin{proposition}
\label{pro:prx0-positive}
{
If $C_T \succeq C_T^M$, then (i) $P^{\rm NV} \geq P^{\rm NV(M)} $ and $R^{\rm NV} \geq R^{\rm NV(M)}$; (ii) $\ex[H_T(P,R)] \geq  V_0(P,R)$ for all $(P,R)$.
}
\end{proposition}

{Part (i) of Proposition~\ref{pro:prx0-positive} follows from an analysis on the marginal effect of the price and VPQ on the expected production payoff $\ex[H_{T}(P, R)]$.
In particular,
$\frac{\partial \ex(H_{T})}{\partial P} =  R-2bP+bc- \ex[\left(R-A_{T}\right)^{+}]$
and
$\frac{\partial \ex(H_{T})}{\partial R} = (P-c) -(P-s)\pr(A_T \leq R)$ are both increasing in $A_T$, indicating that a larger market size induces more significant marginal benefits from increases in $P$ and $R$.
As a result, the optimal price and VPQ are higher in the presence of a stochastically larger market size.
The economic intuition is as follows.
A larger market size provides more buffer to cope with the negative impact of a higher price; hence, the firm has more room to increase the price and thus generate a higher maximum profit.
In addition to increasing the price, the firm can simultaneously increase the production quantity $Q$ to capture the greater demand.
The VPQ (recall, $R = Q + bP$) integrates the combined effect of pricing and production decisions and thus increases with a larger market size.
We know from Remark~\ref{rem:ATorder} that if $C_T \succeq C_T^M$, then $A_T \succeq A_T^M$, so the profit-maximizing price and VPQ are lower with a market size $A_T^M$ than those with a market size $A_T$. By definition, the former decisions are $(P^{\rm NV(M)}, R^{\rm NV(M)})$ and the latter are $(P^{\rm NV}, R^{\rm NV})$.
}

{Regarding Part (ii) of Proposition~\ref{pro:prx0-positive}, we know from its expression in \eqref{EHTR} that $\ex[H_{T}(P, R)]$ is increasing in $A_T$ for any price and VPQ.
Hence, with $A_T \succeq A_T^M$, the expected production payoff is greater in the real-world than in the risk-neutral world.
By definition, the former is $\ex[H_{T}(P, R)]$ and the latter is $\ex^M[H_T(P,R)] = V_0(P, R)$.
}

\begin{proposition}
\label{pro:prbound-part2}
{
$P_m^h \leq P^{\rm NV(M)}$ and $R_m^h \leq R^{\rm NV(M)}$ for all $m\geq 0$.
}
\end{proposition}

{
Proposition~\ref{pro:prbound-part2} holds regardless of the relationship between $C_T$ and $C_T^M$.
It asserts that the risk-minimizing price and VPQ  in the presence of hedging are bounded above by the profit-maximizing price and VPQ under the risk-neutral measure, respectively.
From an economic perspective, among the set of operational decisions $(P, R)$ that attain the same expected hedged production payoff $V_0(P, R)$,
those with smaller values are preferred as they induce lower exposure to unhedgeable uncertainties.
Hence, values higher than $(\nvmp,\nvmr)$, which maximizes $V_0(P, R)$, correspond to over-pricing and over-stocking, thereby not return-risk efficient.
}

{To further clarify the intuition behind Proposition~\ref{pro:prbound-part2}, consider the particular case in which $C_T \succeq C_T^M$ and $m = \ex[H_T(\nvp, \nvr)]$ (i.e., the setting of Theorem~\ref{thm:benefitpr}).
In this case, we can show\footnote{{First, because $(\nvp, \nvr)$ maximizes $\ex[H_T(P, R)]$, we have $m \geq \ex[H_T(P, R)] $ for all $(P, R)$.
Next, by Proposition~\ref{pro:prx0-positive}, if $C_T \succeq C_T^M$, $\ex[H_T(P, R)] \geq V_0(P, R)$ for all $(P, R)$.}} that $V_0(P, R) \leq m$ for all $(P, R)$.
Because $(\nvmp, \nvmr)$ maximizes $V_0(P, R)$,
increasing $(P, R)$ beyond $(\nvmp, \nvmr)$ would make $V_0(P, R)$ begin to decrease,
which increases the value of $|m - V_0(P, R)| = m-V_0(P,R)$ and thus increases the investment risk defined in (\ref{invrisk}).
Meanwhile, the unhedgeable risk defined in  (\ref{unhedgeablerisk}) is always increasing in both $P$ and $R$.
Thus, increasing $(P, R)$ beyond $(\nvmp, \nvmr)$ would make the total risk measured by $B(m, P, R)$ in \eqref{bmpr} begin to increase, so $(P_m^h, R_m^h)$, which minimizes $B(m, P, R)$, must not exceed $(\nvmp, \nvmr)$.
}

\begin{theorem}
\label{thm:benefitpr}
Suppose Assumptions~\ref{assumption:r}--\ref{assumption:Z} hold and the asset price trend positively impacts demand (i.e., $C_T \succeq C_T^M$).
Let $m = \ex[H_T(\nvp, \nvr)]$.
Then, $P^h_m \leq P^{\rm NV} $ and $R^h_m \leq \nvr$.
\end{theorem}

{Theorem~\ref{thm:benefitpr} is an immediate result of Propositions~\ref{pro:prx0-positive} and \ref{pro:prbound-part2}.
It indicates that the risk-minimizing operational activities in the presence of hedging are lower
than the profit-maximizing operational activities without hedging, when the asset price positively impacts demand.
On the one hand,
as the discussion following Theorem~\ref{thm:hedgingsol} reveals, with risk hedging, the firm operates as if in the risk-neutral world with a market size $A_T^M$,
which is (stochastically) smaller than the market size $A_T$ in the real world when $C_T \succeq C_T^M$.
This implies that the profit-maximizing price and VPQ are lower in the presence of hedging than  without hedging;
that is, $P^{\rm NV} \geq P^{\rm NV(M)} $ and $R^{\rm NV} \geq R^{\rm NV(M)}$ (Proposition~\ref{pro:prx0-positive}).
On the other hand, through the risk decomposition in Theorem~\ref{thm:hedgingsol}, we know that increasing the price and VPQ beyond $(\nvmp, \nvmr)$ would increase both the investment risk and the unhedgeable risk, so the risk-minimizing price and VPQ in the presence of hedging must not exceed  $(\nvmp, \nvmr)$, meaning $P_m^h \leq P^{\rm NV(M)}$ and $R_m^h \leq R^{\rm NV(M)}$ (Proposition~\ref{pro:prbound-part2}).
Therefore, the risk-minimizing operational activities must be at most the profit-maximizing levels of the base model to avoid unnecessary risk that does not bring any additional return.
}

\begin{remark}
{
The insight of Theorem~\ref{thm:benefitpr}---that is, within our risk-minimization framework,  hedging reduces operational activities when the asset price positively impacts demand---can be generalized beyond the price-setting newsvendor problem.
Indeed, it holds for general operational payoff functions with certain structural properties;
see Section~\ref{sec:generalization} of the e-companion.
}
\end{remark}

\subsection{Asset Price Trend Negatively Impacting Demand}
\label{sec:hurtpr}
Now we consider the case of the asset price trend negatively impacting demand (i.e., $C_T \preceq C_T^M$).
{
In this case, hedging can offset the negative impact on the production payoff.
For example, when oil prices exhibit an upward trend, the demand for fuel-inefficient cars such as SUVs decreases.
The firm can enter a long position\footnote{{The risk mitigation component of the hedging strategy, i.e., the term $-\xi_t$ in (\ref{opttheta}), essentially reverses the effect of asset price on the production payoff.}} for oil shares to hedge against the oil price's negative impact on the demand for SUVs.
By doing so, the firm receives a positive return from the hedging, making the expected hedged production payoff higher than the expected production payoff (without hedging).
}

\begin{proposition}
\label{pro:prx0-negative}
{
If $C_T \preceq C_T^M$, then (i) $P^{\rm NV} \leq P^{\rm NV(M)} $ and $R^{\rm NV} \leq R^{\rm NV(M)} $;
 $\ex[H_T(P,R)] \leq V_0(P,R)  $ for all $(P,R)$.
}
\end{proposition}

{Proposition~\ref{pro:prx0-negative} is analogous to Proposition~\ref{pro:prx0-positive} and follows from the same argument based on analyzing the marginal effect of the price and VPQ on the expected production payoff. }

\begin{proposition}
\label{pro:prbound-part1}
{
$V_0(P^h_m, R^h_m) \leq m$ for all $m\geq 0$.
}
\end{proposition}

{
Proposition~\ref{pro:prbound-part1} is crucial for establishing Theorem~\ref{thm:hurtpr}, which analyzes the optimal operational decisions in the presence of hedging when the asset price trend negatively impacts the demand.
Recall from the discussion following Theorem~\ref{thm:hedgingsol} that
$V_0(P_m^h, R_m^h)$ represents the {\it expected hedged production payoff}.
Thus, if $V_0(P_m^h, R_m^h) > m$, we can {\it increase} the target mean return from $m$ to $m' = V_0(P_m^h, R_m^h)$ while retaining the same operational decisions $(P_m^h, R_m^h)$.
By doing so, the investment risk defined in \eqref{invrisk} is reduced to zero, and the unhedgeable risk,
 which is defined in (\ref{unhedgeablerisk}) and only depends on the operational decisions, remains the same,
implying $B(m', P_m^h, R_m^h) < B(m, P_m^h, R_m^h)$.
In other words, when the target mean return is $m' = V_0(P_m^h, R_m^h)$,
the operational decisions $(P_m^h, R_m^h)$ along with the hedging strategy $\vartheta^*(m',P_m^h, R_m^h)$ lead to an expected payoff $m' > m$  and
and a variance $B(m', P_m^h, R_m^h) < B(m, P_m^h, R_m^h)$.
Therefore, if $V_0(P_m^h, R_m^h) > m$, the set of integrated operational and hedging decisions $(P_m^h, R_m^h, \vartheta^*(m,P_m^h, R_m^h))$ would not be efficient, because it is dominated by another set of decisions $(P_m^h, R_m^h, \vartheta^*(m',P_m^h, R_m^h))$ which induces a higher return and lower risk.}

{
To better understand why such inefficiency would arise, note that $\ex[\chiiv_T(P_m^h, R_m^h)] = m - V_0(P_m^h, R_m^h)$ represents the {\it expected investment payoff} from hedging (see  (\ref{chidecomp}) and (\ref{invfillgap})), and it would become \emph{negative} if $V_0(P_m^h, R_m^h) > m$.
While the investment payoff's contribution to the total risk is always nonnegative,
a negative return from the investment
is clearly detrimental to the return-risk trade-off of the total return, which
can be expressed as $\HTh + \chiiv_T$, the sum of the {hedged} production payoff and the investment payoff; see \eqref{eq:total_wealth}.
Consequently,
the joint decisions $(P_m^h, R_m^h, \vartheta^*(P_m^h, R_m^h))$ cannot be return-risk efficient because part of the payoff (i.e., the investment payoff $\chiiv_T$) is highly inefficient as it bears a negative return and a positive contribution to the total risk.
\cite{wang2017production} and \cite{wang2021mean} also note that the condition $V_0(P_m^h, R_m^h) \leq m$ is necessary for the optimality based on the observation that increasing the target mean return from $m$ to $V_0$ would induce a better return-risk trade-off. However, these two papers do not provide an economic interpretation of this result as we do here,
due to their lack of the risk decomposition presented in Parts (ii) and (iii) of Theorem~\ref{thm:hedgingsol}.
}

\begin{theorem}
\label{thm:hurtpr}
Suppose Assumptions \ref{assumption:r}--\ref{assumption:Z} hold and the asset price trend negatively impacts demand  (i.e., $C_T \preceq C_T^M$).
Let $m = \ex[H_T(\nvp, \nvr)]$ and let $\bar P = P - s$ for any $P$. Define
    \begin{equation}
    \label{PCirc}
       P^\circ := P^{\rm NV(M)}(\nvr) \quad \mbox{and}\quad r^\circ := \frac{\bar P^\circ}{\bar P^{\rm NV}}.
    \end{equation}
(Note: $r^\circ \ge 1$.) Then, the following properties hold.
\begin{enumerate}[label=(\roman*)]
\label{hurtpr}
    \item $P^h_m \leq \nvp$.
    \item $R^h_m \leq \nvr$, if
    \begin{equation}
    \label{hurtrcond}
   \bigg[\frac{\bar P^{\rm NV}}{r^\circ + \sqrt{(r^\circ)^2 - 1}}\bigg]\cdot\pr^M(A_T \ge \nvr)  \leq c-s.
    \end{equation}
    \item $R^h_m \leq R^\circ$, if (\ref{hurtrcond}) does not hold,
    where $R^\circ$ (which exists uniquely) is determined by
    \begin{equation*}
    \label{hurtcond1}
    \frac{\ex^M\big[A_T \ind\{A_T \leq R^\circ\}\big]}{\pr^M(A_T \ge R^\circ)}
    = \frac{\bar P^{\rm NV}(2b\bar P^\circ + 2bs - bc)}{(c-s)(r^\circ + \sqrt{(r^\circ)^2-1})} - \nvr.
     \end{equation*}
     In particular, $\nvr \leq R^\circ \leq R^{\rm NV(M)}$.
\end{enumerate}
\end{theorem}

{Part (i) of Theorem~\ref{thm:hurtpr} indicates that hedging reduces the pricing level when the asset price negatively impacts demand.
This is counter-intuitive because the assumption here ($C_T \preceq C_T^M$) is opposite to the assumption of Theorem~\ref{thm:benefitpr} ($C_T \succeq C_T^M$), and yet the conclusions regarding the risk-minimizing price are the same.
Nevertheless, the economic intuition here is different from that of Theorem~\ref{thm:benefitpr}.
}

{
When the asset price negatively impacts demand, with risk hedging, the firm operates as if in the risk-neutral world where the market size $A_T^M$ is (stochastically) larger than the actual market size $A_T$ in the real world.
In this case, the expected production payoff is greater in the risk-neutral world than in the real world regardless of the operational decisions; that is, $V_0(P, R) \geq \ex[H_T(P, R)]$ for all $(P, R)$ (Proposition~\ref{pro:prx0-negative}).
In particular, we have $V_0(\nvp, \nvr) \geq \ex[H_T(\nvp, \nvr)] = m$,
so the expected investment payoff (defined in \eqref{chidecomp}) associated with $(\nvp, \nvr)$ would be $\ex[\chiiv_T(\nvp, \nvr)] = m - V_0(\nvp, \nvr) \leq 0$.
However, by Proposition~\ref{pro:prbound-part1}, the expected investment payoff associated with the risk-minimizing operational decisions must satisfy $\ex[\chiiv_T(P^h_m, R^h_m)] = m - V_0(P^h_m, R^h_m) \geq  0$.
In other words, the profit-maximizing operational decisions $(\nvp, \nvr)$ are overly aggressive for risk minimization, because they generate more production payoff (in the risk-neutral world) than necessary to meet the target $m$, resulting in a negative expected investment payoff and introducing unnecessary risk.
Instead, the firm should decrease the price from $\nvp$ to reduce the expected hedged production payoff $V_0$.
Doing so would \emph{leave leeway} in the target mean return for the hedging to provide a positive expected investment payoff while simultaneously reducing the investment and unhedgeable risks.
}

Both Parts (ii) and (iii) of Theorem~\ref{thm:hurtpr} characterize the optimal VPQ in the presence of hedging.
However, they differ in the assumption regarding the magnitude of the asset price trend's negative effect, particularly whether the condition (\ref{hurtrcond}) holds.
This condition is interpreted as follows.
Suppose the asset price trend's negative effect on product demand strengthens. (For example, if we fix the initial oil price below the long-term average level, any increment in this level will negatively affect the demand for fuel-inefficient cars more due to the upward oil price trend.)
Then, the market size under the risk-neutral measure (i.e., $A_T^M$) is not affected since it does not involve the oil price trend while the actual market size (i.e., $A_T$) decreases, which induces a decreased VPQ $\nvr$.
Then, because
It then follows from the fact that $\nvp =[\ex(\nvr \wedge A_T) + bc]/(2b)$ and $P^\circ = [\ex(\nvr \wedge A_T^M) + bc]/(2b)$  that
both $\nvp$ and $P^\circ$ decrease.
Because $A_T$ decreases while $A_T^M$ is not affected,
we expect $\nvp$ to decrease more than $P^\circ$ does,
so $r^\circ$, defined in (\ref{PCirc}), will increase as the negative effect strengthens.
Thus, the term in the bracket on the left-hand side of the inequality~(\ref{hurtrcond}), which is the main factor controlling the magnitude of the left-hand side (since the probability term never exceeds 1), decreases.

In a nutshell,
(\ref{hurtrcond}) means that the asset price trend's negative effect on product demand is strong.
Part (ii) states that under this circumstance,
the optimal VPQ is lower in the presence of hedging than without hedging.
This is consistent with the insight from Part (i).
When the negative effect of asset price on product demand is strong,
the expected hedged production payoff $V_0(P,R)$ is substantially larger than the mean of actual production payoff ($\ex [H_T(P, R)]$) because hedging contributes significant return in this scenario.
Consequently, to prevent the hedged production payoff from exceeding $m$ (i.e., the benchmark production payoff without hedging), the exposure to asset price needs to be contained to reduce the exposure to the product market.
This is achieved by---in addition to the price markdown---further suppressing the VPQ.

On the other hand, if the asset price trends' negative effect is not strong enough, Part (iii) states that we can upper bound the optimal VPQ $R^h_m$ by a value lower than $R^{\rm NV(M)}$, the optimal VPQ under the risk-neutral measure.
In this case, the market size under the risk-neutral measure is not substantially larger than the actual one due to the mild negative effect on product demand, and we expect $\nvr$ to be not far below $\nvmr$.
Therefore, even if $R_m^h$ exceeds $\nvr$, we expect the exceeding amount to be moderate.

In summary, Parts (ii) and (iii) of Theorem~\ref{thm:hurtpr} indicate that, in the presence of hedging, either the optimal VPQ is lower than $\nvr$ when the negative effect is strong, or otherwise it exceeds $\nvr$ by a moderate amount (refer to Section~\ref{app:numerical_hurt} of the e-companion for a numerical illustration).

\subsection{Efficient Frontier}
\label{sec:hedgingeff}
\begin{proposition}
\label{pro:optfrontier}
$B(m, P_m^h, R_m^h)$ increases in $m$; thus, $(m, B(m, P_m^h, R_m^h))$ is an efficient frontier.
\end{proposition}

Proposition~\ref{pro:optfrontier} indicates that, at optimality, higher risk accompanies a higher return.
This efficient frontier lies lower than the one specified in Part (iii) of Proposition~\ref{pro:nvfrontier}.
That is,  $B(m, P^h_m, R^h_m)$ is smaller than $\var(H_T(P^{\rm NV}_m, R^{\rm NV}_m))$ for any target mean return $m \ge 0$ and the gap between the two represents the risk reduction achieved by the hedging model from the no-hedging model.
In addition to characterizing the return-risk trade-off,
we can use this efficient frontier to optimize other risk objectives, such as mean-variance or risk-averse utility functions; see Section~\ref{appendix:mveq} of the e-companion.

\section{Numerical Study}
\label{sec:numerical}
We now implement the hedging model developed in this paper using real-world financial and automotive sales/price data sets.
With the asset and demand models calibrated from data, we conduct a comprehensive numerical study\footnote{There are two crucial differences between our numerical study and that by \cite{wang2017production}. First, we use real-world data sets for model calibration, whereas they only use simulated data.
Second, we use the EOU process to model oil prices, whereas they use GBM, which is more suitable for modeling stock prices.} to illustrate various aspects of the analytical results derived in Sections~\ref{sec:hedging} and \ref{sec:production}.
We briefly introduce the data and model calibration in Sections~\ref{sec:data} and \ref{sec:calibration}, respectively, and relegate the details to the e-companion.
We examine the risk-minimizing operational decisions in the hedging model when the target return is the newsvendor's maximum profit in the base model, and compare them with the counterparts in the no-hedging model (Section~\ref{sec:Numerical_Implementation}).
Then, we examine the hedging performance and compare the efficient frontiers for the hedging and no-hedging models (Section~\ref{sec:Numerical_Implementation_varying}).

\subsection{Data Description}
\label{sec:data}

Two sets of data are used: one is financial data, and the other is operational data from \texttt{AutoMFR}, a prominent automotive manufacturer. The financial asset in our context is the WTI crude oil, a primary global
oil benchmark. The data source is the Federal Reserve Bank of St. Louis database. The financial data includes {\it daily} spot prices from 2010 to 2019. \texttt{AutoMFR}'s operational data, including {\it monthly} sales volumes and manufacturer-suggested retail prices (MSRPs), was purchased from a commercial vendor specializing in automotive business data. The operational data includes brands, models, versions, MSRPs, and combined MPG. We focus on two popular models manufactured by \texttt{AutoMFR}---\texttt{Sport} (fuel-inefficient, low MPG) and \texttt{Compact} (fuel-efficient, high MPG).
The data for \texttt{Sport}  ranges from January 2011 to December 2019, and the data for \texttt{Compact}  ranges from January 2010 to May 2018. We deseasonalize the sales data using the standard STL method \citep{CCMT1990}. In addition, we adjust the MSRP for inflation using monthly Consumer Price Index (CPI) data ranging from January 2010 to December 2019 for all urban consumers from the Federal Reserve Bank of St. Louis database. Specifically, the adjusted price of a given month is the original price divided by the CPI index of that month, then multiplied by the CPI index
of December 2019.
In other words, each price is adjusted to December 2019 by accounting for the inflation rate prevailing from that month to December 2019. The sales volume for each model is aggregated from sales across various versions and the price is the weighted (by sales volume) average price of different versions.
{The summary statistics of these two car models can be found in Section~\ref{sec:parameters} of the e-companion .}

\subsection{Model Calibration}
\label{sec:calibration}
In the automaker's case under consideration, one selling period is one month (i.e., $T = 1/12$).
We assume that the price of WTI---the relevant financial set---follows an EOU process.
We also adopt the specification of \cite{wang2017production} to model the market size $A_T$ defined in (\ref{AT1}) as
\begin{equation}
    \label{specdemand}
    A_T = \int_0^T \tilde\mu(X_t)dt + \tilde\sigma \tilde B_T = \int_0^T (\mu_0 + \mu_1 X_t)dt + \tilde\sigma \tilde B_T,
\end{equation}
where $\tilde\mu(x)$ is the {\it demand rate} function and we assume it to be linear in the asset price $X_t$.

Although, in reality, the unsold units of each month can be carried over to subsequent months, we do not model this inventory carryover in this numerical study.
Instead, we assume for simplicity that the firm makes a myopic decision each month without considering the sales revenue generated by inventory carried over to and sold in later months.
With this assumption, we set the salvage value to zero (i.e., $s=0$) to reflect that units sold in later months are not accounted for in the current month's payoff (i.e., $H_T$).
{Furthermore, the monthly production quantity is unavailable in the data set, so we can not match its model-implied values to the real ones.
Instead, the calibrating procedure matches, besides the model-implied and real prices, the model-implied average monthly sales volumes and the real ones.
Despite the compromises made in the calibrating procedure, the calibrated model does capture qualitative characteristics of the real-world scenario.
For example, the price sensitivity $b$ is positive,
and $\mu_1$ in (\ref{specdemand}) is positive (resp., negative) for fuel-efficient (resp., fuel-inefficient) car model, i.e., \texttt{Compact} (resp., \texttt{Sport}).
In particular, the estimated average gross profit margins match the real-world value. }
The calibrating procedure for the asset price and demand models and the estimated parameters are detailed in Section~\ref{sec:parameters} of the e-companion.\footnote{We have also verified that Assumptions \ref{assumption:r}--\ref{assumption:Abc}
and Assumption \ref{assumption:Z}
(to be exact, its sufficient condition (\ref{kappa})) hold for the estimated parameters and the initial oil prices used in Sections~\ref{sec:Numerical_Implementation} and \ref{sec:Numerical_Implementation_varying}.
In addition, for $\underline{P}$ and $\underline{R}$ defined in Proposition \ref{pro:nvfrontier}, we have $(\underline{P}-c)/c = 0.01\%$ and $(\underline{R}-bc)/(bc) = 0.1\%$; thus, $\underline{P}$ and $\underline{R}$ are close to $c$ and $bc$, respectively.}

\subsection{Assessing the Hedging Model: Fixed Target Return}
\label{sec:Numerical_Implementation}
Suppose the target return is the newsvendor's maximum profit in the base model (i.e., $m = \ex[H_T(\nvp, \nvr)$).
We consider three initial oil prices: $X_0=40$, $70$, and $100$.
The estimated long-run average oil price is around $\$ 70$
(see Section~\ref{sec:WTI} of the e-companion).
Thus, the case of $X_0 = 40$ represents an upward asset price trend; in this case,
the trend positively impacts the demand for \texttt{Compact} but negatively affects the demand for \texttt{Sport}.
By contrast, the case of $X_0 = 100$ represents a downward asset price trend; in this case, the effects of the trend on the demands for these two models are reversed.
The positive/negative effect (formally $C_T \succeq C_T^M$ or $C_T \preceq C_T^M$) is confirmed by the Mann--Whitney U-test applied to the simulated samples of $C_T$ and $C_T^M$ with the calibrated parameters (see Section~\ref{appendix:sdtest} of the e-companion).
The case of $X_{0} = 70$ represents a negligible trend scenario as it is very close to the long-run average.
With these three initial oil prices,
we compute the optimal pricing/VPQ/production decisions for both \texttt{Sport} and \texttt{Compact}.
The results are summarized in Table~\ref{tab:PQR_Sport} and we discuss them below.

\begin{table}[ht] \def\arraystretch{1}
\TABLE{Production ($Q$), Price ($P$), VPQ ($R$), Risks and Return with ${\bf m=\ex[H_T(\nvp,\nvr)]}$.\label{tab:PQR_Sport}}
{
	\begin{tabular}{c l c c c c c c c}
	\toprule
	Car & 	$X_{0}$	&  Model &  $Q$  &    $P$  & $R$  & Risk & {Unhedgeable Risk}  & Return $(m)$  \\
        \midrule
	\multirow{6.7}{*}{\texttt{Sport}} &	\multirow{2}{*}{40}   & NV     & 15,548  &  42,079        & 100,549   &    40.86  &  {40.52}   & 103.41      \\
	&	       & Hedging&  (-5.28\%) &  (-1.04\%)    &   (-1.69\%)  &   (-36.73\%) &  {(-43.73\%)}  & (96.38\%)     \\
	\cmidrule{2-9}
	&	\multirow{2}{*}{70}       &  NV    &   12,552      &  40,627        & 94,618              &   34.09  & {33.72}  & 65.58     \\
	 &              & Hedging &  (-1.99\%)&  (-0.34\%)  &  (-0.56\%)  &  (-10.18\%)  & {(-13.88\%)}  & (99.49\%)          \\
	 \cmidrule{2-9}
	&	\multirow{2}{*}{100}     &  NV     &    9,516       &  39,155          & 88,608    &    28.72   &  {28.05} & 35.97                   \\
	&	           & Hedging & (-6.19\%)  &   (-0.78\%)&  (-1.36\%)     &  (-23.24\%) & {(-32.20\%)}  & (96.55\%)    \\
		 \midrule
	\multirow{6.7}{*}{\texttt{Compact}} &	 \multirow{2}{*}{40} &  NV  & 9,826 & 21,946   &   154,511    &    7.99  &  {7.97}   & 12.90      \\
 	 &       & Hedging  & (-6.90\%) &  (-0.48\%)   &   (-0.89\%)  & (-39.86\%)  &  {(-51.09\%)}  & (94.19\%)          \\ %
 	 \cmidrule{2-9}
	&	 \multirow{2}{*}{70} &  NV      & 12,263    &  22,313    & 159,364         & 9.30  & {9.19}   &  20.49           \\
     &       & Hedging  &   (-1.42\%)  &   (-0.13\%)    & (-0.23\%)   &  (-8.47\%)  & {(-13.94\%)}  &(99.63\%)           \\
     \cmidrule{2-9}
	&	\multirow{2}{*}{100} &  NV      & 14,646  &  22,670              &   164,105         & 11.09  & {10.67}   & 29.55             \\
     &       & Hedging  &     (-3.23\%) &  (-0.33\%)    & (-0.59\%)    & (-32.44\%)  &  {(-36.27\%)} &(98.08\%)         \\
		\bottomrule
	\end{tabular}
	}
{ {\it Note.} The percentages in the columns labeled ``$Q$'', ``$P$'', ``$R$'', `Risk'', and ``Unhedgeable Risk'' represent the reduction relative to the no-hedging model;
{see Section~\ref{appendix:baseriskdecomp} for the formal definitions of risk and unhedgeable risk for the no-hedging model.}
The percentages in the column labeled ``Return'' represent the contribution from the production payoff.
{The units of ``Risk'', ``Unhedgeable Risk'', and ``Return'' are all one million dollars.}}
\end{table}

First, when the initial oil price for a particular month significantly deviates from the long-run average (i.e., $X_0 = 40$ or $100$), the asset price exhibits a prominent trend towards its long-run level due to the mean-reverting effect.
For such cases, the risk reduction achieved by the hedging model is significantly more than the risk reductions achieved for the case of $X_0 = 70$ (which is close to the long-run average).
{The key reason is that a more salient asset price trend leads to a higher return-to-risk ratio of the investment, meaning subject to the same level of investment risk, one earns a higher investment payoff $\ex(\chiiv_T)$ and thus requires a lower production payoff $V_0$ to achieve the target for the total expected payoff $m$. (Recall from \eqref{invfillgap} that $\ex(\chiiv_T) + V_0 = m$.)
Therefore, the higher return-to-risk ratio of the investment payoff induces more reductions in the operational levels in the presence of hedging, thereby reducing the unhedgeable risk (and thus the total risk) by a more significant amount. }

{We now explain why the case of $X_0 = 40$ or $100$ leads to a higher return-to-risk ratio of the investment than the case of $X_0=70$.
By the expressions for the expected investment payoff and the investment risk (see \eqref{invfillgap} and \eqref{invrisk}, respectively), we know that the return-to-risk ratio of the investment is $\sqrt{Z_0^M-1}$.
In the example that $X_t$ follows an EOU process, we can show---using the fact that $Z_0^M = \ex(Z_T^2)$ and Jensen's inequality---that
$Z_0^M \ge \exp(\frac{1}{2}\int_0^T\ex^M(\eta_t^2)dt)$,
where $\eta_t = \frac{\kappa}{\sigma}(\alpha + \frac{\sigma^2}{2\kappa}- \log X_t)$ is the market price of risk  (see (\ref{etaeOU})).
Hence, when $X_0 = 40$ or $100$, $\log X_t$ deviates significantly from the long-run average $\alpha$,\footnote{{It is easy to verify that $\log X_t = -\frac{1}{2}\sigma^2t + \sigma B_t^M$.
Then, $\eta_t = \frac{\kappa}{\sigma}(\alpha + \frac{\sigma^2}{2\kappa} + \frac{1}{2}\sigma^2t -  \sigma B_t^M)$.
When $X_0 = 40$, $\log X_t$ is significantly below $\alpha$, and $\eta_t$ tends to become more positive as $t$ increases.
When $X_0 = 100$, $\log X_t$ is above $\alpha$; thus, so long as the gap is significant and $t$ is small
($t \leq \frac{1}{12}$ in our case), $\eta_t$ still tends to take a significantly negative value.}}  leading to a large value of $\eta_t^2$.
Then, the above lower bound suggests a large value of $Z_0^M$, meaning a high return-to-risk ratio of the investment.
}

A second observation from Table~\ref{tab:PQR_Sport} is that, for both car models, the optimal operational decisions (price $P$, VPQ $R$ and production $Q$) in the presence of hedging are adjusted downward relative to those without hedging.
However, the decrements are small, around $7\%$ at most, so the production payoff still contributes most of the total return.
This is reassuring, as hedging does not excessively decrease the operational level.
A particularly desirable trait is that the price markdown is small: while price reduction enhances market competitiveness, automakers are usually reluctant to reduce prices too much due to the concern of protecting their brands' values.
{Moreover, the ``Unhedgeable Risk'' column of Table~\ref{tab:PQR_Sport} shows that even small markdowns in the operational levels can substantially reduce the unhedgeable risk.
The reason is as follows.
The slope of the efficient frontier for the no-hedging model is exploding as $m$ approaches the maximum profit level $\ex[H_T(\nvp, \nvr)]$ (see Proposition \ref{pro:nvfrontier}(iii)).
Thus, decreasing the operational levels even slightly, which reduces the expected production payoff, can substantially reduce the exposure to the noise in the product demand, which is unhedgeable.}

Last, when the asset price trend's impact on product demand is strong ($X_0 = 40$ or $100$), hedging makes more significant adjustments to the operational decision than the weaker impact case ($X_0 = 70$).
This is consistent with the finding that hedging is more effective in the former case.

\subsection{Assessing the Hedging Model: Varying Target Returns}
\label{sec:Numerical_Implementation_varying}

\subsubsection{Efficient Frontiers}
\begin{figure}[ht]
\FIGURE
{\includegraphics[width=0.85\textwidth]{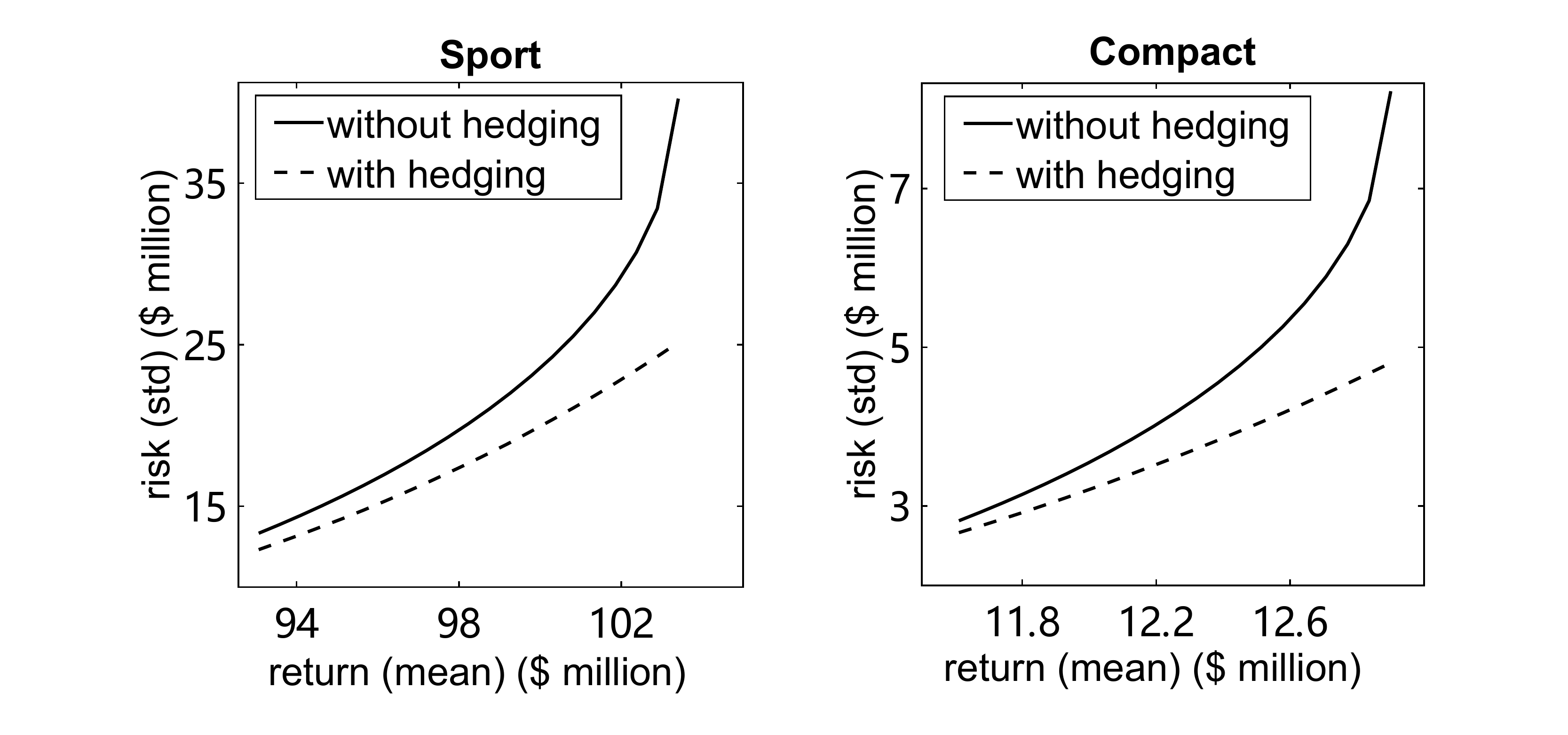}}
{Efficient Frontiers. \label{fig: efficientfrontier}}
{Initial WTI price is $X_{0}=40$.}
\end{figure}

Figure~\ref{fig: efficientfrontier} plots the efficient frontiers for both car models with an initial oil price $X_{0} = 40$.
The target return $m$ ranges from 90\% to 100\% of $\ex[H_T(\nvp, \nvr)]$, the newsvendor's maximum profit in the base model.
The risk is measured by the standard deviation of the terminal return.
(The right ends of the curves correspond to the results in
Table~\ref{tab:PQR_Sport}).
All curves are upward sloping, which is the hallmark of efficient frontiers: when all decisions are optimized, an increase in return is always accompanied by increased risk.

{Figure~\ref{fig: efficientfrontier} clearly shows, as expected, that the hedging model outperforms the no-hedging model: the frontiers for the former lie below the frontiers for the latter.
More significantly, it sheds light on the difference in \emph{incremental risk} (i.e., the slope of a frontier) between the two models.
Specifically, for \texttt{Sport} or \texttt{Compact}, the frontier for the hedging model increases more slowly than that for the no-hedging model, especially towards the right ends of the frontiers.
In other words, the hedging model's incremental risk is smaller than the no-hedging model's, and this phenomenon is more prominent for high-return cases. The reason is that, in the presence of hedging, the operational levels $P$ and $R$ need not grow as fast as those in the no-hedging model (see Figure~\ref{fig: impact_of_hedging_on_pricing} and the related discussion) to reach higher values of the target mean return $m$. Such smaller increments in $P$ and $R$ induce the slower growth of the unhedgeable risk, thereby, the slower growth of the total risk. }

{The smaller incremental risks of the hedging model drive the risk reduction to increase in the target mean return $m$: the gap between the frontiers for the two models widens as $m$ increases, especially towards the frontier's right end.
(Recall from Proposition~\ref{pro:nvfrontier}(iii) that the no-hedging model's incremental risk explodes as $m$ approaches the newsvendor's maximum profit.)
The same pattern also holds in relative terms: the percentage risk reduction increases in $m$.
Hence, the hedging model's advantage over the no-hedging model is more significant when the target mean return is higher.
This also demonstrates the hedging model's effectiveness: on top of its first-order advantage (i.e., risk reduction), it also exhibits a second-order advantage (i.e., reduction in incremental risk).}

\begin{figure}[ht]
\FIGURE
{\includegraphics[width=0.8\textwidth]{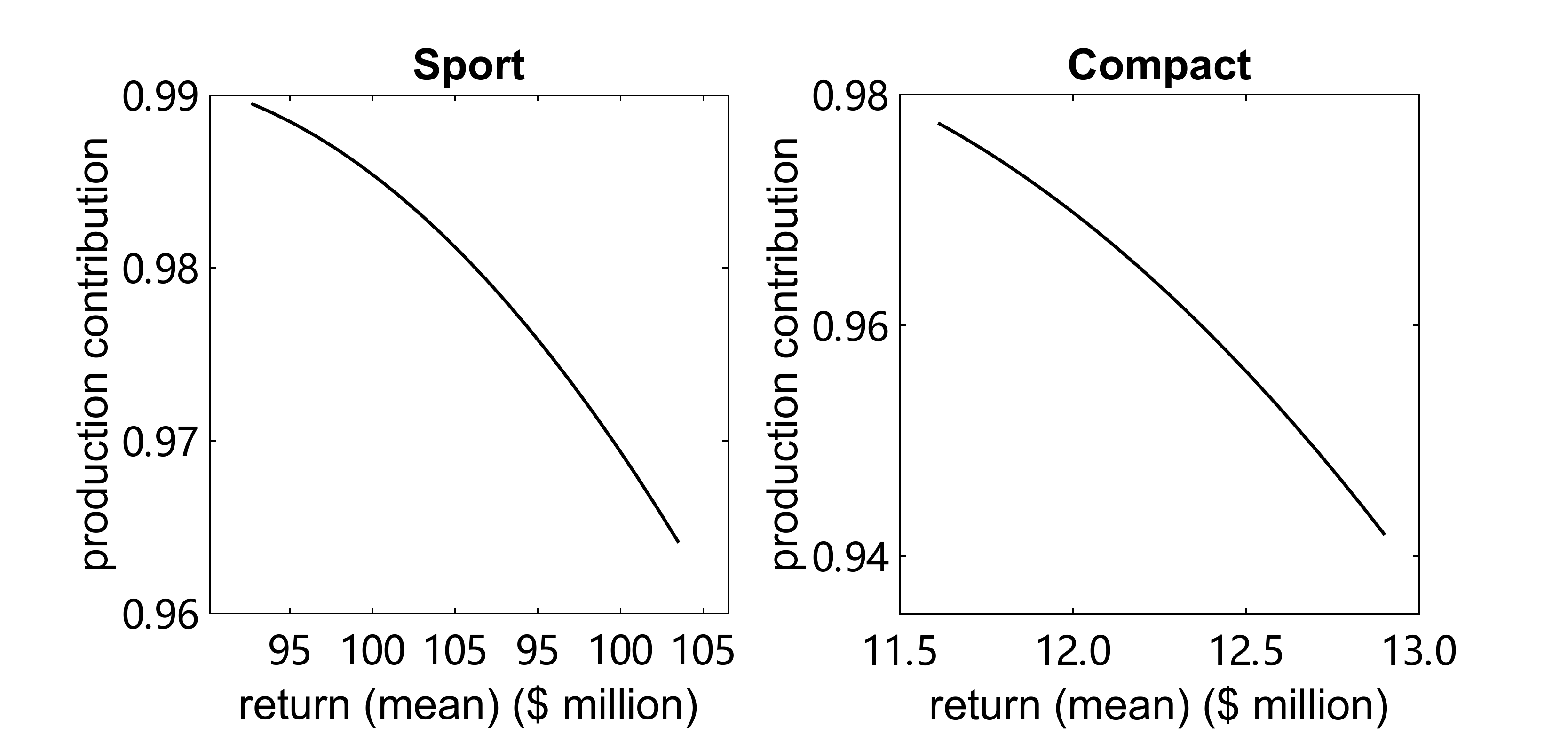} }
{Production Payoff Contribution with Risk Hedging. \label{fig: production_pay_contribution}}
{Initial WTI price is $X_{0}=40$.}
\end{figure}

Figure~\ref{fig: production_pay_contribution} illustrates the contribution of production payoff in (\ref{EHTR}) to total return for varying target returns.
In particular, production payoff accounts for at least $94\%$ of total return in all instances, indicating that optimized operations are the primary source of profit for the automakers.
We can also observe that as the target return increases, the contribution from hedging increases (i.e., the contribution from production decreases).
This is consistent with how hedging reduces risk.
In the no-hedging model, as the return increases, risk increases due to higher price and VPQ; the increment in risk is most prominent when the return approaches the newsvendor's maximum profit (see Proposition~\ref{pro:nvfrontier} (iii)).
By contributing more to the return as it increases,
hedging suppresses the growth of production payoff, which therefore suppresses the growth of unhedgeable risk to control the total risk.
This is reflected by the increased percentage contribution from hedging when the target return is higher.

\subsubsection{Optimal Operational Decisions}
\begin{figure}[ht]
\FIGURE
{\includegraphics[width=1.2\textwidth]{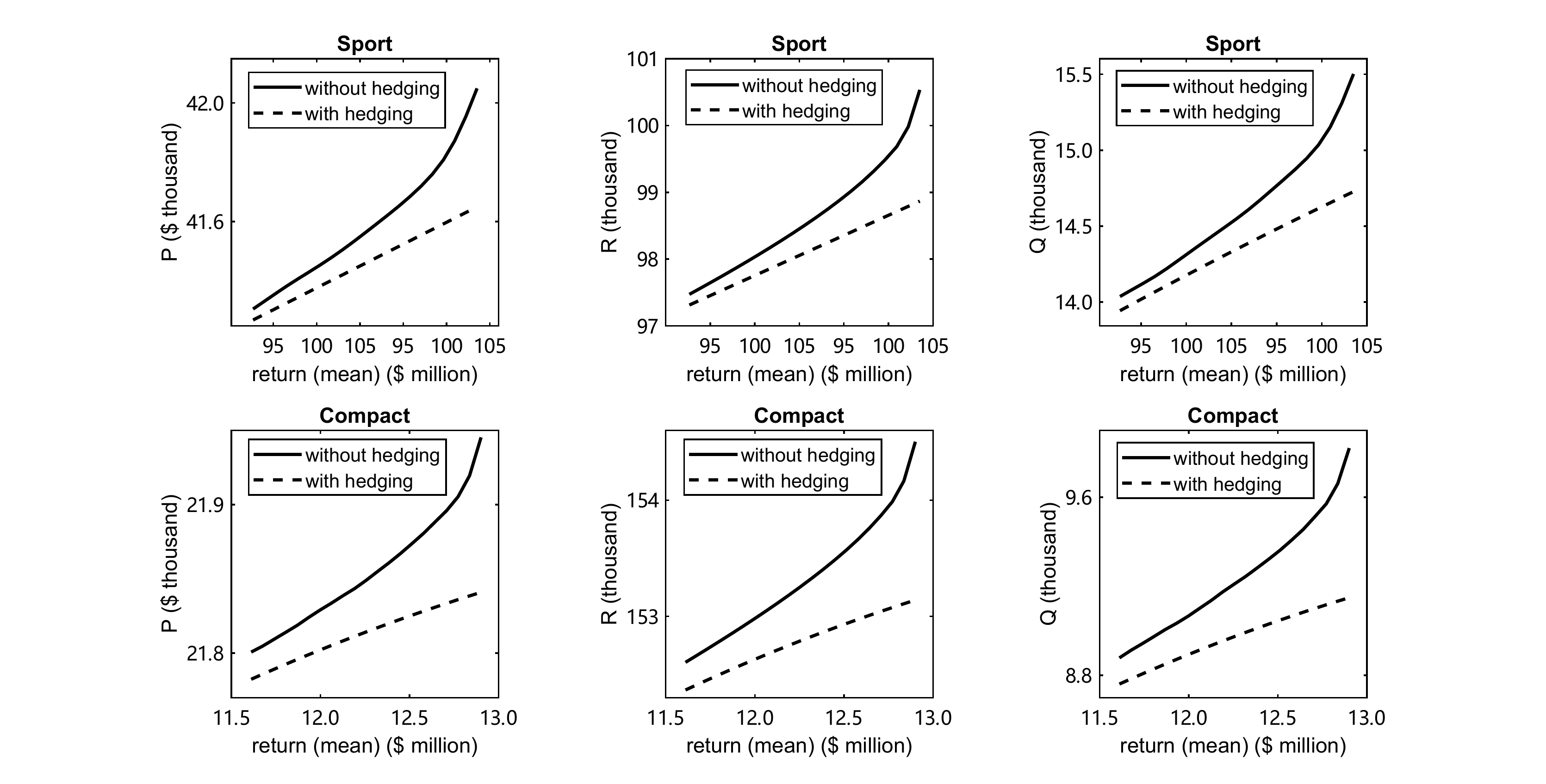} }
{Optimal Price (P), VPQ (R), and Production Quantity (Q). \label{fig: impact_of_hedging_on_pricing}}
{Initial WTI price is $X_{0}=40$.}
\end{figure}

Figure~\ref{fig: impact_of_hedging_on_pricing} plots
the optimal prices ($P$), VPQ ($R$), and production quantities ($Q$) for both the hedging and no-hedging models
(with an initial oil price $X_{0} = 40$) over a range of target returns.
Theorems~\ref{thm:benefitpr} and \ref{thm:hurtpr} show that if the target return equals the newsvendor’s maximum profit in the base model,
the optimal price with hedging is lower than without hedging.
Despite the absence of analytical results for other values of the target return, Figure~\ref{fig: impact_of_hedging_on_pricing} shows that over the given range of target returns, the optimal price with hedging never exceeds the optimal price without hedging, and the price markdown is small.

In addition, as return increases,
the optimal price and VPQ grow more slowly with hedging than without hedging.
In particular,
they increase in the target return $m$ at most linearly in the hedging model (dashed curves),
in contrast to their faster growths without hedging (solid curves),
especially when $m$ is reaching the right ends of the curves.
This can be explained as follows.
By the (partial) concavity of the expected production payoff $\ex[H_T(P,R)]$ in $P$ and $R$, as the target mean return $m$ approaches NV's maximum profit $\ex[H_T(\nvp, \nvr)]$, $P$ and $R$ in the no-hedging model need to increase sharply for the expected production payoff to reach this maximum point.
This is consistent with the sharp increases in $P$ and $R$ around the right ends of the solid curves observed in Figure \ref{fig: impact_of_hedging_on_pricing}.
On the other hand, in the hedging model, part of the target mean return $m$ stems from the hedging payoff. Thus, the operational levels $P$ and $R$ can be adjusted downward relative to the no-hedging model because the production payoff can be lower than the target mean return $m$.
In other words, as $m$ increases, the optimal operational levels of the hedging model need not grow as fast as those in the no-hedging model.

The observation above is also consistent with the concavity of the curves in Figure~\ref{fig: production_pay_contribution}.
Since the operational levels with hedging
increase slowly in $m$, the percentage contribution from the production decreases at an accelerating pace.
Thus, when the firm demands a higher return, the price increment
with hedging need not be as high as that without hedging, which helps the manufacturer to stay competitive in the market. Similar patterns also hold for the optimal $R$ and $Q$.

\section{Concluding Remarks}
\label{sec:conclusion}
In this study, we propose and solve a general model that integrates pricing, production, and risk hedging using financial assets.
We completely characterize the return-risk efficient frontier.
We find that the pricing level is lower with hedging than without hedging when the asset price trend either positively or negatively impacts product demand.
The price reduction is desirable for firms that operate in a competitive market.
In addition, the service level is lower
(which is induced by a lower VPQ) with hedging than without hedging when the asset price trend has a positive effect on product demand.
However, when the asset price trend negatively impacts demand,
the service level with hedging may exceed the service level without hedging, but only by a small margin.
Our numerical study using data sets of \texttt{AutoMFR} shows that the hedging model performs substantially better than a price-setting newsvendor without hedging.
The markdowns in pricing and service levels are small,
which are appealing because hedging does not materially decrease operational profit while substantially reducing risk.

We conclude with a discussion on potential research directions.
One direction is to consider the setup where the time horizon $[0, T]$ is a {\it production period}, and the pricing decision is made at time $t = T$ instead of at $t = 0$.
Two problems need to be solved in this setup. One is to maximize payoff by determining the price at time $t = T$
after all the uncertainties are resolved;
the other is an mV risk-management problem at time $t = 0$ to determine the production quantity and hedging strategy.
This new setup focuses on the impact of hedging on the production decision because it is through this decision that hedging impacts the pricing.

Another potential research direction is to incorporate dynamic hedging into a dynamic pricing setup.
As pointed out earlier, the model developed in this paper is suitable for products that do not allow frequent price adjustments.
Integrating dynamic hedging with dynamic pricing for risk management is significant for products or services with such price adjustments.
Under this setup, one usually models the product demand as a Poisson process with an intensity decreasing in the selling price \citep{GR1994}.
One may further allow the intensity to depend on the asset price to incorporate the impact of the financial market.
For example, if the intensity is an increasing function in the asset price, then the asset price trend positively impacts product demand.
Incorporating asset price into the intensity turns the demand model into a doubly stochastic Poisson process.
While the hedging methodology specified in this paper can potentially be applied to this dynamic-pricing setup, the key challenge lies in minimizing the variance function to find the optimal pricing policy,
which is now a stochastic process, making the problem analytically intractable.
In this regard, a feasible approach is to introduce heuristics, such as approximating the optimal pricing decision by linear functions in the asset price.

\bibliographystyle{informs2014} %
\bibliography{NV_hedging} %

\ECSwitch

\EquationsNumberedBySection

\ECHead{E-Companion}

\section*{Lists of Notation and Assumptions}
\begin{table}[H]%
\TABLE{Basic Notation.\label{table_notation}}
{
	\begin{tabular}{ll}
	\toprule
	$c$		& unit production cost  \\
    $s$     & unit salvage value   \\
    $D_t$   & demand accumulated over time period $[0,t]$ for $0
    \leq t \leq T$ \\
    $b$     & positive parameter capturing demand’s sensitivity to price   \\
    $m$ & target mean return \\
    $\pr$ & real world probability measure \\
    $\pr^M$ & risk-neutral probability measure  \\
    $\mu_t$ & stochastic return rate of the asset price $X_t$\\
    $\sigma_t$ & stochastic volatility of the asset price $X_t$ \\
    $\eta_t$ & market price of risk, $\eta_t = \mu_t/\sigma_t $ \\
    $A_t$   & market size accumulated over time period $[0,t]$ for $0\leq t \leq T$ \\
    $A^M_t$   & version of $A_t$ under $\pr^M$, i.e., $\pr(A_t^M \leq a) = \pr^M(A_t\leq a), \;\; \forall a$.\\
    $C_t$  & component of $A_t$ that is determined by the financial asset only \\
    $C^M_t$   & version of $C_t$ under $\pr^M$, i.e., $\pr(C_t^M \leq x) = \pr^M(C_t\leq x), \;\; \forall x$.\\
    $\tilde\sigma$ & non-financial noise volatility coefficient of the cumulative demand process\\
    $\tilde\mu(\cdot)$  &  demand rate function\\
    \bottomrule
	\end{tabular}
}
{}
\end{table}
\begin{table}[H] %
\TABLE{Notation for Operational Decisions. }
{
	\begin{tabular}{p{0.5in}p{4in}}
	\toprule
	$P$     & pricing level \\
    $Q$     & production quantity \\
    $R$     &%
    { virtual production quantity (VPQ)
    }, $R:=Q+bP$ \\
    $\nvp$  &  optimal $P$ to maximize expected NV profit  under $\pr$\\
    $P^{\rm NV(M)}$ & optimal $P$ to maximize expected NV profit  under $\pr^M$\\
    $\nvq$   &  optimal $Q$  to maximize expected NV profit  under $\pr$ \\
    $Q^{\rm NV(M)}$   &  optimal $Q$  to maximize expected NV profit  under $\pr^M$ \\
    $\nvr$   &  optimal $R$ to maximize expected NV profit  under $\pr$ \\
    $R^{\rm NV(M)}$   &  optimal $R$ to maximize expected NV profit  under $\pr^M$ \\
    $\nvp_m$  &  optimal $P$ to minimize payoff variance subject to target mean return $m$ in the no-hedging model \\
    $\nvq_m$  &  optimal $Q$ to minimize payoff variance subject to target mean return $m$ in the no-hedging model\\
    $\nvr_m$  &  optimal $R$ to minimize payoff variance subject to target mean return $m$ in the no-hedging model \\
     $P_m^h$  &  optimal $P$  to minimize payoff variance subject to target mean return $m$ in the hedging model\\
    $Q_m^h$  &  optimal $Q$  to minimize payoff variance subject to target mean return $m$ in the hedging model\\
    $R_m^h$  & optimal $R$  to minimize payoff variance subject to target mean return $m$ in the hedging model \\
    \bottomrule
	\end{tabular}
}
{{\it Note.} NV stands for newsvendor.}
\end{table}
\begin{table}[H] %
\TABLE{Assumptions for Main Results\label{table_collectionassumption}}
{
	\begin{tabular}{ll}
	\toprule
	Theoretical Results		                 &    Assumptions Involved \\
 \midrule
    Proposition~\ref{pro:nvsol}                        &  Assumptions 1--2  \\
    Proposition~\ref{pro:nvfrontier}                      &  Assumptions 1--3  \\
    Lemma~\ref{lem:Vt}, Theorem~\ref{thm:hedgingsol} and Proposition~\ref{pro:hedgingsolEOU}    &  Assumptions 4--5  \\
    Propositions~\ref{pro:prx0-positive}--\ref{pro:prbound-part1} and Theorems~\ref{thm:benefitpr}--\ref{thm:hurtpr}      &  Assumptions 1--5 \\
    \bottomrule
	\end{tabular}
}
{}
\end{table}

\begin{table}[H] \def\arraystretch{1.3}
\TABLE{List of Definitions.\label{table_collectiondefinition}}
{
	\begin{tabular}{p{2in}p{4in}}
	\toprule
        service level & $\pr(D_T \leq Q)$, which equals $\pr(A_T \leq R)$   \\
	  mV	        &   minimization of variance subject to an exogenously set target mean return\\
                      & i.e., $\min \var$ subject to $\ex = m$, where $m$ is exogeneously set   \\
        MV            &   mean-variance utility maximization (i.e., $\max \ex - \gamma\var$)        \\
       efficient frontier   & In mV, let $v(m)$ be the minimum variance.
       If $v(m)$ increases in $m$, then $(m, \sqrt{v(m)})$ constitutes an efficient frontier \\
        risk-mitigation payoff ($\chirm_T$)   &  defined as $\chirm_T = \int_0^T (-\xi_t)dt$ in (\ref{chidecomp}) \\
        investment payoff ($\chiiv_T$)            &  defined as $\chiiv_T = \int_0^T \iota_t dX_t$ in (\ref{chidecomp})\\
         hedged production payoff ($\HTh$)  &  defined as $\HTh = H_T +\chirm_T$ in (\ref{hedgedprodpayoff}) \\
         expected hedged production payoff ($V_0$) & $ V_0 := \ex^M(H_T) =\ex(\HTh) $ \\
        investment risk &  defined as $\sqrt{\var(\chiiv_T) + \cov(\chiiv_T, \HTh)}$ in (\ref{invrisk}) \\
         unhedgeable risk &  defined as $\sqrt{\var(\HTh) + \cov(\chiiv_T, \HTh)}$ in (\ref{unhedgeablerisk})  \\
    \bottomrule
	\end{tabular}
}
{}
\end{table}

\vspace{18pt}
\noindent
\begin{center}
\large \textbf{Part I: Supplementary Materials}
\end{center}

\section{Equivalents to the Mean-Variance Formulation}
\label{appendix:mveq}
The {mV} optimization problem~(\ref{nvmv}) adopted throughout the paper has the form of
\begin{equation}\label{eq:general-MV}
    \min_x \var[W(x)] \quad  {\rm s.t.} \quad \ex[W(x)] = m,
\end{equation}
where $W(x)$ represents a random return that depends on a decision $x$.
With the above formulation,  a decision-maker aims to minimize the variance of his return while meeting a pre-specified target on the mean return $m$.
The following two alternative formulations are mathematically equivalent to \eqref{eq:general-MV}.

The decision-maker may choose to maximize the mean return with a constraint on the variance:
\begin{equation}\label{eq:general-VM}
    \max_x \ex[W(x)] \quad {\rm s.t.} \quad \var[W(x)] \leq v.
\end{equation}
It is easy to show that for any $m$, there exists $v$ such that the optimal solution to problem~\eqref{eq:general-MV} is optimal for problem~\eqref{eq:general-VM}, and vice versa.
Thus, the efficient frontier constructed from the {mV} problem by varying the target mean return $m$ is identical to the efficient frontier constructed from the latter problem by varying the risk tolerance coefficient $v$.

Another formulation is the mean-variance utility maximization (MV) :
\begin{equation}\label{eq:general-MVU}
    \max_x\; \ex[W(x)] - \gamma\var[W(x)],
\end{equation}
where $\gamma \ge 0$ is a given risk-aversion coefficient.
{It is straightforward to show that any pair of $(\ex[W(x^*(\gamma))], \var[W(x^*(\gamma))])$ induced by the optimal solution $x^*(\gamma)$ of the problem in (\ref{eq:general-MVU}),
for a given $\gamma$,
must also be on the efficient frontier generated by the mV model, with $\ex[W(x^*(\gamma))]$ being the corresponding target mean return $m$ in the {mV} formulation.
In particular, $x^*(\gamma)$ and the optimal solution of mV model with $m = \ex[W(x^*(\gamma))]$ coincide.
Therefore, the {mV} and {MV} models are mathematically equivalent.
In particular, in our setup, with the efficient frontier built from the {mV} model, the {MV} model (with a given $\gamma$) can be translated to the following problem (recall, $B(m, P, R)$ in (\ref{mvprob}) is the minimum variance of {mV} with target mean return set as $m$):
$$\max_m \quad \ m - \gamma B(m, P_m^h, R_m^h).$$
Accounting for that $(P_m^h, R_m^h)$ minimizes $B(m,P,R)$,
the problem above is clearly equivalent to
$$\max_{m, P, R} \quad \ m - \gamma B(m, P, R).$$
This problem is in general not concave.
Moreover, the evaluation of $B(m,P, R)$ can be computationally expensive as it involves nested simulation.
So, numerical global optimization procedure is needed.
There is an established body of literature on the numerical global optimization with computationally expensive objective function (e.g., \cite{jones1998}) and such methods are implemented in many software packages.
Once the optimal $m$ of the problem above, denoted $m_\gamma$, is found,
the {MV} problem reduces to an {mV} with $m = m_\gamma$.
}

In practice, however, unless the decision-maker has a reasonable estimate of $\gamma$ a priori,
the {mV} formulation has a modeling advantage as the parameter $m$ carries a more straightforward economic meaning.
In addition, a decision-maker usually maximizes a risk-averse {\it utility function}, which can be approximated by its first two moments.
The problem directly translates to selecting points from the efficient frontier generated by the {mV} formulation in such a situation \citep{markowitz1952, markowitz2014}.

\section{{Risk Decomposition for the No-hedging Model}}
\label{appendix:baseriskdecomp}
{In parallel with the risk decomposition of the total return in the hedging model as described in Part (ii) of Theorem \ref{thm:hedgingsol},
in particular the decomposition in (\ref{invrisk}) and \eqref{unhedgeablerisk},
here we define analogous decomposition of risk for the no-hedging model.
Below, all arguments $P$ and $R$ are dropped as they are given.}

{By Lemma~\ref{lem:Vt} and definition of $V_t$ in (\ref{Vt}),
$H_T$ has the following representation:
\begin{equation*}
 H_T= V_T =  V_0 + \int_0^T \xi_t dX_t + \int_0^T \delta_t d\tilde B_t.
\end{equation*}
As discussed in Section~\ref{sec:strategy}, $\xi_t$ captures the impact from the asset price movement on product demand and $\delta_t$ is associated with the intrinsic demand noise that can not be reduced by hedging.
Therefore, in parallel with the decomposition of total return of the hedging model, $H_T + \chi_T^*$, into hedged production payoff and the investment payoff,
here we decompose $H_T$ into a {\it financial} part, $\HTfin$,
and an {\it unhedgeable} part, $\HTu$:
\begin{equation}
\label{HTdecomp}
\HTfin:=  V_0+\int_0^T \xi_t dX_t \quad{\rm and}\quad \HTu:=\int_0^T \delta_t d\tilde B_t.
\end{equation}
Then, we decompose the risk of $H_T$ as:
$${\Big[\underbrace{\sqrt{\var(H_T)}}_{\text{total risk}}}\Big]^2 = \var(\HTfin + \HTu) =
\Big[\underbrace{\sqrt{\var(\HTfin) + \cov(\HTfin,\HTu)}}_{\text{financial risk}}\Big]^2
+ \Big[\underbrace{\sqrt{\var(\HTu) + \cov(\HTfin,\HTu)}}_{\text{unhedgeable risk}}\Big]^2.$$
In the numerical implementation of the model in Section~\ref{sec:numerical},
$\xi_t$ and $\delta_t$ are simulated according to their expressions in (\ref{xiGBM}) and (\ref{deltaGBM}), respectively.
Then, the $\HTfin$ and $\HTu$ are simulated based on the generated sample paths of $\xi_t$ and $\delta_t$,
with which we compute the variances and the covariance to evaluate the financial and unhedgeable risks of $H_T$ defined above.
}

\section{{Sufficient Conditions for Assumptions~\ref{assumption:nvm} and \ref{assumption:Abc}}}
\label{appendix:suffassumption23}
{
Strictly speaking, Assumptions~\ref{assumption:nvm} and \ref{assumption:Abc} are not explicitly on primitive model parameters as they involve $\nvp$, $\nvr$ and $\ex[H_T(\nvp, \nvr)]$, which are functions of the model primitives.
In this section, we derive sufficient conditions explicitly on the model primitives that guarantee Assumptions~\ref{assumption:nvm} and \ref{assumption:Abc} to hold.
Note that it is sufficient to derive conditions under which $\ex[H_T(P, R)] > 0$ for {\it some} $P$ and $R$.
Clearly, under such condition $\ex[H_T(\nvp,\nvr)] > 0$ while $P = c$ or $R = bc$ leads to $\ex[H_T(P, R)] < 0$, hence $\nvp > c$ and $\nvr > bc$ hold also.}

{To this end, define $h(P,R) = \ex[H_T(P,R)]$ and derive
(with $\mu_A:= \ex(A_T)$ and $\sigma_A^2 = \var(A_T)$):
\begin{eqnarray*}
h(P,R) &=& (P-c)(R-bP) - (P-s)\ex[(R- A_T)^+] \\
&\ge& -bP^2 + bcP +(P-c)R - (P-s)(R-\mu_A)^+ - (P-s)\ex[(\mu_A - A_T)^+] \\
&\ge& -bP^2 + bcP +(P-c)R - (P-s)(R-\mu_A)^+ - \frac{1}{2}(P-s)\sigma_A.
\end{eqnarray*}
The first inequality is due to the fact that $(R - A_T)^+ = (R-\mu_A + \mu_A - A_T)^+ \leq
(R - \mu_A)^+ +(\mu_A - A_T)^+$;
the second inequality follows from the fact that for any random variable $X$,
$\ex[(E(X)-X)^+] = (1/2)\ex(|X-\ex(X)|) \leq (1/2) \sqrt{\var(X)}$.
}

{Now, choose $R = \mu_A$, we have
\begin{eqnarray}
\label{fFuncP}
 \ex[H_T(\nvp, \nvr)]  &\ge& h(\mu_A, P) \ge -bP^2 + bcP +(P-c)\mu_A - \frac{1}{2}(P-s)\sigma_A \nonumber\\
 &=&-bP^2+\Big(bc +\mu_A - \frac{1}{2}\sigma_A\Big)P -\Big(c\mu_A-\frac{1}{2}s\sigma_A\Big)
 =:f(P)
\end{eqnarray}
which is a concave quadratic function in $P$.
Denote the discriminant of $f(P)$ by $\Delta(\mu_A)$
(the argument $\mu_A$ is to stress its dependence on $\mu_A$):
$$\Delta(\mu_A) = \Big(bc +\mu_A - \frac{1}{2}\sigma_A\Big)^2 - 4b\Big(c\mu_A-\frac{1}{2}s\sigma_A\Big)
=\mu_A^2 -(\sigma_A + 2bc)\mu_A + \frac{1}{4}\sigma_A^2-bc\sigma_A + b^2c^2+2bs\sigma_A.$$
}
{
For existence of $P$ such that $f(P)>0$, it is necessary that $\Delta(\mu_A) > 0$.
Note that $\Delta(\mu_A)$ is a convex quadratic function in $\mu_A$ with discriminant
$$\Delta' = (\sigma_A + 2bc)^2-\sigma_A^2+4bc\sigma_A-4b^2c^2-8bs\sigma_A = 8b(c-s)\sigma_A>0.$$
Then,
\begin{equation}
\label{muAcond0}
\mu_A  > \frac{\sigma_A+2bc + \sqrt{8b(c-s)\sigma_A}}{2} \Rightarrow \Delta(\mu_A) > 0.
\end{equation}
Then, for $f(P) > 0$ (which implies $g(P) > 0$ hence $\ex[H_T(\nvp, \nvr)]$), it is sufficient that
$$P\in \Big(c\vee \frac{bc +\mu_A - \frac{1}{2}\sigma_A - \sqrt{\Delta(\mu_A)}}{2b}, \frac{bc +\mu_A - \frac{1}{2}\sigma_A + \sqrt{\Delta(\mu_A)}}{2b}\Big).$$
Note that the interval above exists due to:
\begin{eqnarray*}
\frac{bc +\mu_A - \frac{1}{2}\sigma_A + \sqrt{\Delta(\mu_A)}}{2b} - c
&=& \frac{1}{2b}\Big(bc + \mu_A -\frac{1}{2}\sigma_A + \sqrt{\Delta(\mu_A)}-2bc\Big) \\
&=& \frac{1}{2b}\Big(\mu_A -\frac{1}{2}\sigma_A-bc
+ \sqrt{\Delta(\mu_A)}\Big) \\
&>& \frac{1}{2b}\Big(\sqrt{2b(c-s)\sigma_A} + \sqrt{\Delta(\mu_A)}\Big) > 0,
\end{eqnarray*}
where the first inequality is due to the condition in the left hand side of (\ref{muAcond0}).
In summary, under the condition on the left hand side of (\ref{muAcond0}), there exists $P$ and $R$ such that $\ex[H_T(P,R)] > 0$ and thus Assumption~\ref{assumption:nvm} holds.
}

{Now we proceed to deriving sufficient conditions for Assumption~\ref{assumption:Abc}.
To start with, note that
$$\ex[(bc - A_T)^+] = \ex[(bc - A_T^+)^+] + \ex[(-A_T)^+] \leq bc\pr(A_T \leq bc) + \ex[(-A_T)^+].$$
Let $p_0:=\pr(A_T \leq bc)$ and $\epsilon_0 := \ex[(-A_T)^+]$, then the following conditions are clearly sufficient for Assumption~\ref{assumption:Abc}:
$$bcp_0 +\epsilon_0 \leq 2b(c-s) \quad{\rm and}\quad bcp_0+\epsilon_0 \leq 2\sqrt{b\max_P f(P)},$$
where $f(P)$ is defined in (\ref{fFuncP}).
Since $f(P)$ is concave quadratic function,
it is straightforward to derive its maximum value:
\begin{equation}
\label{maxfP}
\begin{aligned}
f^* &:= \max_P f(P) = f(P^*)
= \frac{1}{4b}\Big[\Big(\mu_A-bc-\frac{1}{2}\sigma_A\Big)^2-2b(c-s)\sigma_A\Big]>0, \\
P^* &= \frac{\mu_A-\frac{1}{2}\sigma_A+bc}{2b} > c.
\end{aligned}
\end{equation}
In (\ref{maxfP}) above, both $f^*>0$ and $P^*>c$ follow immediately from the condition on the left hand side of (\ref{muAcond0}).
Clearly, $\ex[H_T(\nvp, \nvr)] > f^*$.}

{Summarizing the analysis above, we have the following result.}
{
\begin{proposition}
\label{pro:a23suffcond}
Recall, $\mu_A:= \ex(A_T)$ and $\sigma_A^2:= \var(A_T)$;
in addition, $p_0:=\pr(A_T \leq bc)$ and $\epsilon_0 := \ex[(-A_T)^+]$.
Then, Assumptions~\ref{assumption:nvm} and \ref{assumption:Abc} are met under the following conditions:
\begin{equation}
\label{a23suffcond}
\mu_A  > \frac{\sigma_A+2bc + \sqrt{8b(c-s)\sigma_A}}{2}
\quad{\rm and}\quad
bcp_0 +\epsilon_0 \leq 2b(c-s) \wedge 2\sqrt{bf^*},
\end{equation}
where the expression of $f^*$ is in (\ref{maxfP}).
\end{proposition}
}

\section{Supplementary Materials for Section~\ref{sec:numerical}}
 \label{appendix:supplementaryS6}
\subsection{Expressions of $\delta_t(P, R)$ and $\xi_t(P, R)$}
\label{appendix:eoudeltaxi}
{Here we give explicit expressions of $\delta_t(P,R)$ and $\xi_t(P,R)$ in
Section~\ref{sec:expou};
recall, these two quantities are involved in the martingale representation of $V_t$ in
(\ref{GKW}).
With the dynamics of $X_t$ specified in (\ref{expOUX}) and the demand model specified in (\ref{specdemand}),
we can obtain more explicit expressions for $\delta_t$ and $\xi_t$ as follows.
}
Define $f(t, x, a) := \ex^M[(R-A_T)^+ \,|\, X_t = x, A_t = a]$.
Note that since in (\ref{expOUX}), $\sigma_t = \sigma$ is a constant,
we can remove $\sigma_t$ from the arguments of the function $f(\cdot)$.
Then,
\begin{eqnarray*}
f(t, x, a) &=& \ex^M\Big[\Big(R - a -
\int_t^T \tilde\mu(xe^{\sigma (B^M_u-B^M_t)-\frac{1}{2}\sigma^2(u-t)})
-\tilde\sigma(\tilde B_T - \tilde B_t)\Big)^+\,\Big{|}\, X_t = x, A_t = a\Big] \\
&=& \ex\Big[\Big(R - a -
\int_t^T \tilde\mu(xe^{\sigma (B_u-B_t)-\frac{1}{2}\sigma^2(u-t)})
-\tilde\sigma(\tilde B_T - \tilde B_t)\Big)^+\Big],
\end{eqnarray*}
where the second equality uses the facts that $B_t^M$ is a Brownian motion under $\pr^M$,
the independence between $\{B_t\}$ and $\{\tilde B_t\}$,
and that both $\{B_u-B_t, t \leq u \leq T\}$ and $\tilde B_T - \tilde B_t$ are independent of $\calf_t$.
Using the fact that $\tilde B_T - \tilde B_t$ follows Normal distribution with zero mean and $\sqrt{T-t}$ as standard deviation,
taking derivative with respect to $a$ leads to
$$%
f_a(t,x,a)= -\ex\Big[\Phi\Big(\frac{R-a-\int_{t}^{T} \tilde{\mu}(x e^{\sigma (B_u-B_t)-\frac{1}{2}\sigma^2(u-t)}) d u}{\tilde{\sigma} \sqrt{T-t}}\Big)\Big], $$
where $\Phi(x)$ is the distribution function of standard normal random variable.
Furthermore, note that the $dX_t$-term in (\ref{GKW}) is produced by $f_x$.
Taking derivative with respect to $x$ leads to
\begin{eqnarray*}
f_x(t,x,a)&=& -\ex \Big[\Phi\Big(\frac{R-a-\int_{t}^{T} \tilde{\mu}(x e^{\sigma (B_u-B_t)-\frac{1}{2}\sigma^2(u-t)}) d u}{\tilde{\sigma} \sqrt{T-t}}\Big)\\
&&\cdot\int_{t}^{T} \tilde{\mu}'(x e^{\sigma (B_u-B_t)-\frac{1}{2}\sigma^2(u-t)})e^{\sigma (B_u-B_t)-\frac{1}{2}\sigma^2(u-t)} d u\Big].
\end{eqnarray*}

Therefore, by Lemma \ref{lem:Vt},
{
\begin{equation}
\label{deltaGBM}
\delta_t =-\tilde\sigma(P-s)f_a(t, X_t,A_t)
=\tilde{\sigma}(P-s) \ex\Big[\Phi\Big(\frac{R-A_t-\int_{t}^{T} \tilde{\mu}(X_t X_{u-t}^{M}) d u}{\tilde{\sigma} \sqrt{T-t}}\Big) \Big{|} X_t, A_t \Big],
\end{equation}
and
\begin{eqnarray}
\label{xiGBM}
\xi_t &=&-(P-s)f_x(t,X_t,A_t) \nonumber\\
&=&(P-s)\ex \Big[\Phi\Big(\frac{R-a-\int_{t}^{T} \tilde{\mu}(X_t X_{u-t}^{M}) d u}{\tilde{\sigma} \sqrt{T-t}}\Big)\int_{t}^{T} \tilde{\mu}'(X_t X_{u-t}^{M})X_{u-t}^{M} d u\Big{|}X_t, A_t\Big].
\end{eqnarray}
Here, $X^M_{u-t} := e^{\sigma (B_u-B_t)-\frac{1}{2}\sigma^2(u-t)}$.
}

\subsection{Parameter Estimation}
\label{sec:parameters}
The summary statistics of the data set is presented in Table~\ref{tab:data_statistics}.

\begin{table}[ht]\def\arraystretch{1}
\TABLE{Summary Statistics. \label{tab:data_statistics}}
{\begin{tabular}{l c c c c }
		\toprule
		Car	& Sample Size & Price (\$)  & Sales Volume  &  MPG \\
		\midrule
		\multirow{2}{*}{\texttt{Sport}} & \multirow{2}{*}{108}  & 40,615  & 15,933 & 16.89  \\
		         & & (1,239)  & (3,516) &  (1.96) \\
	    \cmidrule{1-5}
		\multirow{2}{*}{\texttt{Compact}} & \multirow{2}{*}{101}  & 22,300   & 16,190 & 24.92  \\
		      &   &(444)   & (3,378) & (3.34)  \\
		\bottomrule
	\end{tabular}
}
{{\it Note.}  The numbers in parentheses are standard deviations.}
\end{table}

\subsubsection{WTI Price}
\label{sec:WTI}
To estimate the parameters of the EOU process in (\ref{expOUX}),
we calibrate the model to the log WTI price data by applying Ordinary Least Squares (OLS) method to the discretized equation produced by (\ref{expOUX}).
{Recall, $Y_t =\log X_t$ is modeled in (\ref{expOUX}).
To calibrate this model to WTI prices, we start with discretizing $Y_t$:
\begin{align}
\label{disY}
Y_{t+\nu}=(1- \kappa \nu)Y_{t}+\kappa\alpha\nu+\sigma \sqrt{\nu} \cdot \epsilon_t,
\end{align}
where $\nu = 1/252$ is the time step size  (i.e., one trading day), and $\epsilon_t$ is a standard normal random variable independent of $Y_t$.
Then, (\ref{disY}) leads to the following linear regression model
($n$ stands for the n-th day in the data set):
\begin{align*}
\label{estimateY}
Y_{n+1} = aY_n +b +\epsilon_n,
\end{align*}
where $ a= 1- \kappa \nu $,  $b= \kappa\alpha\nu$, and
$\epsilon_n$'s are i.i.d zero-mean random noises with variance $\sigma^2 \nu$.
Applying OLS produces estimators of $a$, $b$, and standard error of residuals,
which are denoted by
$\hat a$, $\hat b$, and $\hat \sigma^2$, respectively.
Then, the estimators (denoted by corresponding symbols with hats)
for $\kappa$, $\alpha$, and $\sigma^2$ are
\begin{equation*}
\label{estimateX}
\begin{aligned}
&\hat{\kappa} = \frac{1-\hat{a}}{\nu}, \quad
\hat{\alpha} = \frac{\hat{b}}{1-\hat{a}}, \quad {\rm and} \quad
\hat\sigma^2 =  \frac{1}{N\nu} \sum_{n=1}^{N} (Y_{n+1}-(\hat{a}Y_n +\hat{b} ))^2,
\end{aligned}
\end{equation*}
where $N = 2600$ is the number of observations.
}

Here we remark that the regression model above is essentially an autoregressive model
(with lag $1$).
Although for such models, OLS estimators in general are {\it not} unbiased
due to the past errors influencing future values of the process,
they are {\it consistent} estimators;
see \cite{tsbookjames}.
Considering the large sample size ($N = 2600$), we deem that the OLS estimators are appropriate under the current context.

The estimated parameters are as follows:
\begin{equation*}
    \label{wtipara}
    \hat\kappa = 0.5356, \quad \hat\alpha = 4.1847, \quad{\rm and}\quad \hat\sigma = 0.3327.
\end{equation*}
The estimated parameters suggests that the long-run average of WTI is
$$\ex[X_{\infty}] = e^{\alpha + \frac{\sigma^2}{2\kappa}} = 72.82.$$

\subsubsection{Demand Model Calibration}
\label{sec:modelcalibration}
Recall, market size is assumed to take the form shown in (\ref{specdemand}).
To evaluate the integral involved in (\ref{specdemand}) in each period (month),
we apply the discretized numerical integration:
\begin{equation}
\label{demandintegral}
   C_T =  \int_0^T \tilde{\mu}(X_t)dt \;\approx\; T(\mu_0 + \mu_1 \bar{X}) \equiv A +B \bar{X},
\end{equation}
where $\bar{X}$ is the average asset price within one month:
$\bar{X} = \Big(\sum_{j=0}^J X_j\Big) / (J+1)$, and $J = 20$---that is, there are $21$ trading days in a month.
Thus,
$\mu_0 = A/T$, and $\mu_1= B/T$, with $T = 1/12$.

Substituting (\ref{demandintegral}) to (\ref{specdemand}),
we have the following linear regression model:
\begin{equation}
\label{demandmodel}
D = A + B \bar{X} -b P +\tilde\sigma\sqrt{T}\epsilon,
\end{equation}
where $\epsilon$ is {a standard normal error that is independent of $\bar{X} $. }
The pricing level $P$ in (\ref{demandmodel}) and the production quantity $Q$ are determined by the manufacturer,
which we assume to be accord with the newsvendor's profit-maximization problem
(recall, the salvage value $s$ is zero):
\begin{equation}
\label{equiprice}
(P^*, Q^*)  = \argmax_{P, Q} \ex[(P-c)Q - P(Q-D)^+].
\end{equation}
The set of model parameters to be calibrated is $\{A, B, b, c, \tilde{\sigma} \}$.
Given the oil price $X_0$ on the first trading day of month $i$,  the model-implied price and production quantity for the same month are
$$(P_i^*, Q_i^*)  = \argmax_{P, Q} \ex[(P-c)Q - P(Q-D_i)^+],$$
where $D_i$ is simulated given $\{A, B, b, c, \tilde{\sigma} \}$ and $X_0$.
In principle, we need to minimize the distance between the model-implied $P$ and $Q$ and those observed in data.
While $P$ is observable, $Q$ is not available in our data set.
Thus, instead of fitting $Q$, we turn to minimizing the error of the demand model in (\ref{demandmodel}).
Here, we use sales volume to proxy demand.
This is appropriate because the inventory-to-sales ratio in the automotive industry in the U.S. market has been around $2.5$,
which is substantially above $1$,
for decades (\cite{dunn2016inventory}),
suggesting that the inventory held has been able to cover the demand.
Therefore, we minimize the distance between the model-implied expected demand $ A + B \bar{X_i} -b P_i^*$ and observed sales volume $S_i$.
Thus,  our calibration model is
\begin{equation*}
    \label{squarederror}
\min_{A, B, b, c, \tilde{\sigma} }\sum_{i=1}^{n}[(P_i^*-P_i)^2 +(A + B \bar{X_i} -b P_i^*-S_i)^2].
\end{equation*}
Then, we apply off-the-shelf global optimization algorithm to solve the problem above to obtain the solutions. \\
The calibrated parameters are summarized in Table~\ref{tab:calibrated_result}.
The cost $c$ is the average direct production cost that includes raw material and labor/assembly costs.

\begin{table}[ht] \def\arraystretch{1.3}
\TABLE{Calibrated Model Parameters.\label{tab:calibrated_result}}
{
	\begin{tabular}{ l c c c c c }
	\toprule
	Car		& $A\;(=\mu_0 T)$  &$B\;(=\mu_1 T)$ & $b$ & $c$ (cost) & $\tilde{\sigma}$  \\
	\midrule
		\texttt{Sport} & 111,155.66  &  -185.42 & 2.02  &   34,543.91 &  11,577.37 \\
		\texttt{Compact} & 151,887.67 &  157.41 & 6.59 &   20,467.10 &  \phantom{0}8,619.46   \\
		\bottomrule
	\end{tabular}
}
{{\it Note.} $ D_T = \int_0^T ({\mu}_0+ {\mu}_1 X_t)dt + \tilde\sigma \tilde B_T - bP$, and  $T = 1/12$.}
\end{table}
Note that the sign of the parameter $B$ determines how asset price impacts demand.
Specifically, the plus (resp., minus) sign of $B$ represents the positive (resp., negative) impact of oil price on product demand.
That is, the demand of \texttt{Sport} (resp., \texttt{Compact}), the fuel-inefficient (resp., fuel-efficient) model, is negatively (resp., positively) impacted by oil price.
This is consistent with the economic intuition and empirical evidence discussed in Section~\ref{sec:intro}.
Another point worth noting is that the estimated production costs ($c$) for both cars are around $85\%$ of the average selling price,
which matches the profit margin observed in the automotive industry
(Margins by Sector,
Auto \& Truck,
\url{https://pages.stern.nyu.edu/~adamodar/New_Home_Page/datafile/margin.html}).
With the parameters in Table \ref{tab:calibrated_result},
we simulate samples of $C_T$ and $C_T^M$ and
apply Mann--Whitney U-test (\cite{MW1947}) to test the stochastic dominance relationships between these two quantities, respectively for the positive ($C_T \succeq C_T^M$) and negative ($C_T \preceq C_T^M$) cases.
The test results support such relationships for all model instances;
see Section~\ref{appendix:sdtest} for details.

\subsection{Numerical Procedure}\label{sec:numerical-procedure}
{
We apply Monte Carlo simulation to evaluate the expectations involved in the variance function $B(m, P, R)$.
To simulate sample paths,
we set the discretized time step size at $\nu = 1/252$ and assume that each month has $21$ trading days.
To evaluate the first term of $B(m, P, R)$, in particular $V_0$,
Monte Carlo simulation is readily applied with simulated values of
$D_T$,
which is part of the simulated paths $(X_t, D_t)$
(the simulation procedure is described below).
Evaluating the second term of $B(m, P, R)$ in (\ref{EOUbmpr})
involves the following steps: }

\begin{enumerate}[label=(\roman*)]
    \item Use Monte Carlo simulation to generate $N_1$ sample paths of $(Y_t, X_t, A_t)$ according
to (\ref{expOUX}), (\ref{AT1}), and (\ref{demandintegral}).
\item At each time point $t \leq T$, given each realized value of $(Y_t=y, X_t=x, A_t=a)$,  generate $N_2$ paths of $X_u^M$ for $u\in[t, T]$ with initial value $X_0^M = 1$ and evaluate the integral $\int_t^T[\mu_0 + \mu_1\cdot(xX_{u-t}^M)]du$ along each path of $X_u^M:=e^{\sigma (B_u-B_t)-\frac{1}{2}\sigma^2(u-t)}$; the integral is evaluated by trapezoidal rule.
\item Evaluate $\delta_t(P, R)$ in (\ref{deltaGBM}) via simulation using the paths generated in step (ii) for each path  $(X_t=x, A_t=a, Y_t=y)$ generated in step (i).
\item Evaluate the function values in (\ref{OUf}) and use $\delta_t(P, R)$ to evaluate, for $t\in[0, T]$,
\[\ex[e^{-f_0(T-t)-f_1(T-t)Y_t-f_2(T-t)Y_t^2 } \delta^2_t(P, R)]. \]
\item Evaluate the second term in $B(m, P, R)$ via the trapezoidal rule.
\end{enumerate}
To minimize $B(m, P, R)$ over $(P, R)$, we follow two steps.
First, given $R$, we perform a line search for the corresponding optimal $P$ over $[c,\;  P^{\rm NV(M)}]$.
This is fairly efficient, as $B(m, P, R)$ is convex in $P$ given $m$ and $R$ (see Proposition~\ref{pro:rcvx}).
Then, we perform a line search for the optimal $R$ over $[bc, \; R^{\rm NV(M)}]$.

\begin{proposition}
\label{pro:rcvx}
{For a given $R$, $\bar P(R)$ is the smallest root of $V_0(P, R) = m$ if $m \leq \max_P V_0(P, R)$;
otherwise,  let $\bar P(R)=  \argmax_P V_0(P, R)$.
Then,
$\argmin_P B(m, P, R) \leq \bar P(R)$
and $B(m, P, R)$ is convex in $P$ over $[c, \bar P(R)]$.
}
\end{proposition}

\subsection{Stochastic Dominance Tests for $C_T$ and $C^M_T$}
\label{appendix:sdtest}
In this section, we test the stochastic dominance relationship between $C_T$ and $C^M_T$ by applying the classical Mann--Whitney U-test (\cite{MW1947})
on samples of $C_T$ and $C_T^M$ simulated with calibrated parameters
(see Section~\ref{sec:WTI}, Section~\ref{sec:modelcalibration} and Table \ref{tab:calibrated_result}).
As the testing procedures are similar,
in the following,
we detail the test results for calibrated parameters and briefly report the conclusion for the modified parameters.

For the positive (resp.,\ negative) case, we want to test
$$H_0: C_T^M \stackrel{d}{=} C_T \quad ({\rm resp.,\ } \; C_T \stackrel{d}{=} C_T^M )
\quad \text{versus} \quad H_1: C_T^M \preceq C_T \quad ({\rm resp.,\ } \; C_T \preceq C_T^M ).  $$
With the initial oil price at the levels of $X_0 = 40$ and $100$,
and calibrated parameters of WTI price in Section~\ref{sec:WTI},
and those of demand model in Table \ref{tab:calibrated_result},
for each problem instance,
we simulate $100000$ realizations for $C_T$ and $C^M_T$ independently.
The Mann–Whitney U test
(of the null hypothesis of equal distributions against appropriate alternatives)
is implemented in many software packages and we utilize the Python package
(i.e., scipy.stats.mannwhitneyu) to conduct the test.
The p-values are summarized in Table \ref{tab:MW_U_test}.

\begin{table}[ht]
\TABLE{P-values for the Mann–Whitney U test. \label{tab:MW_U_test}}
{\begin{tabular}{l c c c }
		\toprule
		Car	     & $X_0=40$  & $X_0=100$ \\
		\midrule
		\texttt{Sport} & $9.70\times 10^{-4}$   &  $1.72\times 10^{-153}$  \\
        \texttt{Compact}  & 0  & $9.56\times 10^{-173}$\\
		\bottomrule
	\end{tabular}
}
{ }
\end{table}

In this table, two cases of asset price trends positively impacting the demands are reported: \texttt{Compact} ($X_0=40$) and \texttt{Sport} ($X_0=100$).
For these two cases, all the p-values are extremely small,
which
strongly reject the null hypothesises in favor of the alternative ones
(i.e.,  $C_T \succeq C_T^M$).
The other two are the cases of asset price's negative effects on the demands:
\texttt{Sport} ($X_0=40$) and \texttt{Compact} ($X_0=100$).
The p-values of both tests also provide strong evidences to reject the null hypothesises and support the alternative ones, $C^M_T \succeq C_T$.
In the Section~\ref{app:numerical_hurt}, hypothetical model parameters are used.
We conduct the same tests for all problem instances.
We find that all the p-values are extremely small,
so conclude that the tests reject the null hypothesis and support the alternative one, $C^M_T \succeq C_T$
(note that all problem instances in this part are for the case of asset price trend negatively impacting demand).

\subsection{Numerical Illustration of Theorem \ref{thm:hurtpr}} \label{app:numerical_hurt}

Theorem~\ref{thm:hurtpr} characterizes how risk hedging affects the optimal price and VPQ  when the asset price trend negatively impacts demand.
Part (i) of the theorem shows that hedging induces a price markdown.
However, as shown in Parts (ii) and (iii), the effect of hedging on the optimal VPQ is more involved, depending on whether the asset price trend’s negative effect on product demand is sufficiently strong.
We now illustrate Parts (ii) and (iii) using the same demand and asset models as in Section~\ref{sec:numerical}, but with different parameter values.\footnote{To make the asset price trend more prominent to strengthen the negative effect, we amplify the value of the mean-reverting coefficient $\kappa$ in (\ref{expOUX}) by a factor of $10$.
To control for the magnitude of MPR process  (\ref{etaeOU}),
we also increase the volatility $\sigma$ by a factor of $10$.
In addition, we change the parameters in equation~\eqref{specdemand} to make the demand more sensitive to the asset price while keeping $\ex(A_T)$ at the same level as in the calibrated model in  Section~\ref{sec:numerical}.
}

We first define the following metric to measure the negative effect of the asset price trend on the demand,
{which we call ``hurting factor'',
because ``negatively impacting demand'' can be expressed informally as ``hurting demand'':}
\begin{equation*}
    \label{Delta}
    \Delta := c - s - \bigg[\frac{\bar P^{\rm NV}}{r^\circ + \sqrt{(r^\circ)^2 - 1}}\bigg]\cdot\pr^M(A_T \ge \nvr).
\end{equation*}
Note that $\Delta$ is the difference between the right- and left-hand sides of the inequality~(\ref{hurtrcond}).
Recall from the discussion following Theorem~\ref{thm:hurtpr} that $\Delta\geq 0$ indicates that the negative effect is sufficiently strong.
In this case (Part~(ii)), the optimal VPQ with hedging is not higher than that without hedging, that is, $R^h_m \leq \nvr$.
Otherwise (Part~(iii)), $R_m^h$ may exceed $\nvr$, but by only a moderate amount at most.
In the following numerical illustration,
we first show $\Delta$ indeed captures the negative effect and then examine how $R^\circ$, $\nvr$, and $R^{\rm NV(M)}$, quantities involved in Part~(iii),  respond as $\Delta$ varies.

In the EOU model~(\ref{expOUX}), a simple way to vary the magnitude of the asset price's negative effect is to vary the parameter $\alpha$.
For \texttt{Sport}, fixing the initial oil price $X_0=30$, we increase $\alpha$ to increase the long-run average of oil prices, strengthening the negative effect of the oil price trend on product demand.
On the other hand, for \texttt{Compact}, we fix $X_0=250$ to introduce a prominent downward oil price trend and then decrease $\alpha$ to strengthen the negative effect.

\begin{figure}[ht]
\FIGURE{\includegraphics[width=0.85\textwidth]{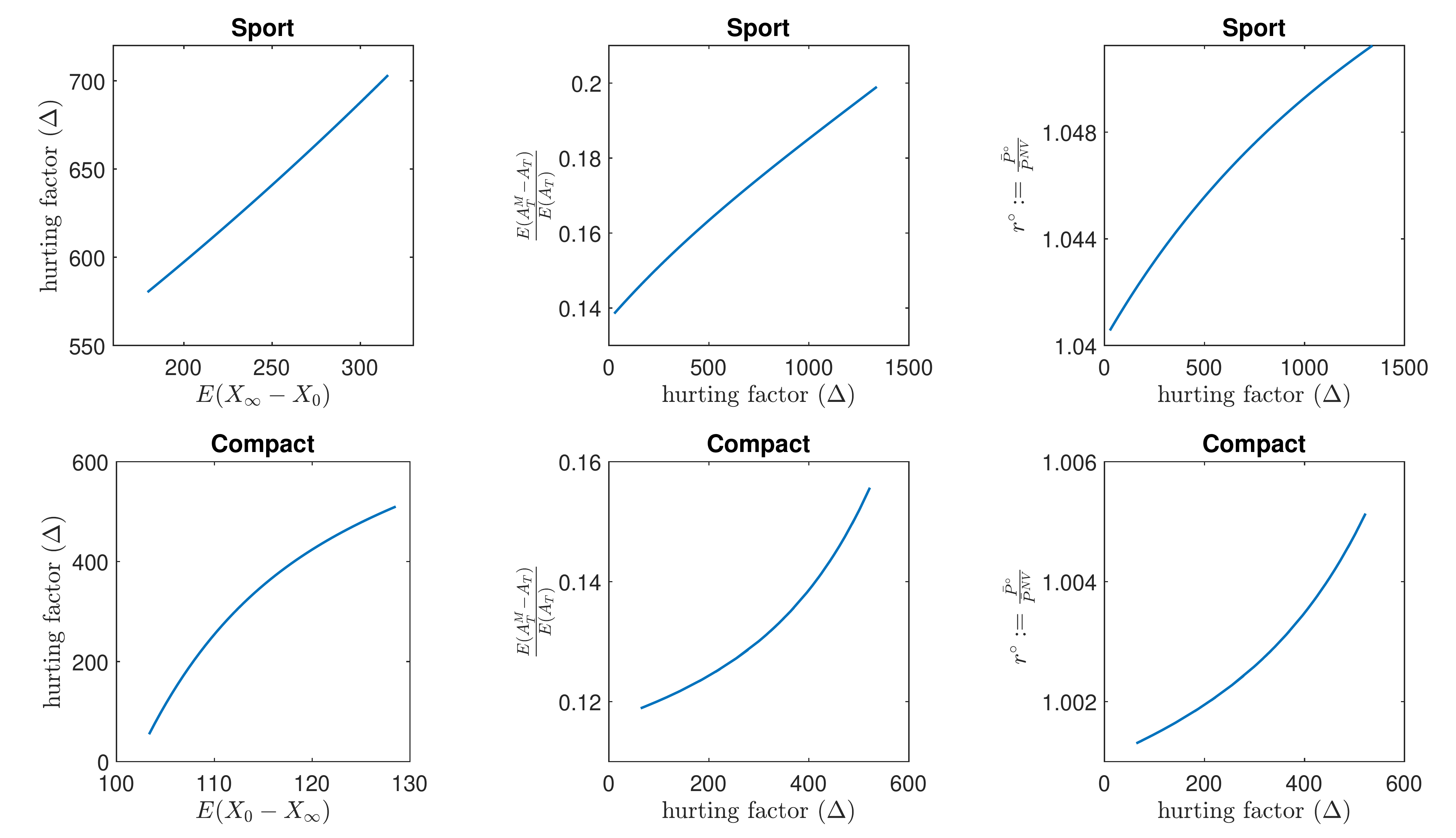}}
{Illustration of Theorem~\ref{thm:hurtpr}(ii). \label{fig:thm3part2}}
{}
\end{figure}

Figure~\ref{fig:thm3part2} illustrates Part (ii) of Theorem~\ref{thm:hurtpr}.
We first interpret the first column.
The upward-sloping WTI price trend,
reflected by $E(X_\infty - X_0)$,
becomes more prominent as $\alpha$ increases,
boosting the negative effect of the WTI price trend on the demand for \texttt{Sport}.
In the meantime, the hurting factor $\Delta$ increases as shown in the first graph of the first column.
The second graph of the first column shows the same pattern, although for \texttt{Compact}, a stronger negative effect is reflected by a larger value of $E(X_0 - X_\infty)$.
Therefore, we can indeed use $\Delta$ to measure the magnitude of the asset price trend's negative effect on product demand.
The second column of the graphs shows that the  difference in expectations between $A_T^M$ (market size under risk-neutral measure) and $A_T$ (the actual market size) increases as  $\Delta$ increases.
This is because the strengthening asset price trend reduces $A_T$ but does not affect $A_T^M$.
The third column shows that $r^\circ$, defined in (\ref{PCirc}), increases with the negative effect for both car models, which aligns with the discussion following Theorem~\ref{thm:hurtpr}.

\begin{figure}[ht]
\FIGURE
{\includegraphics[width=0.6\textwidth]{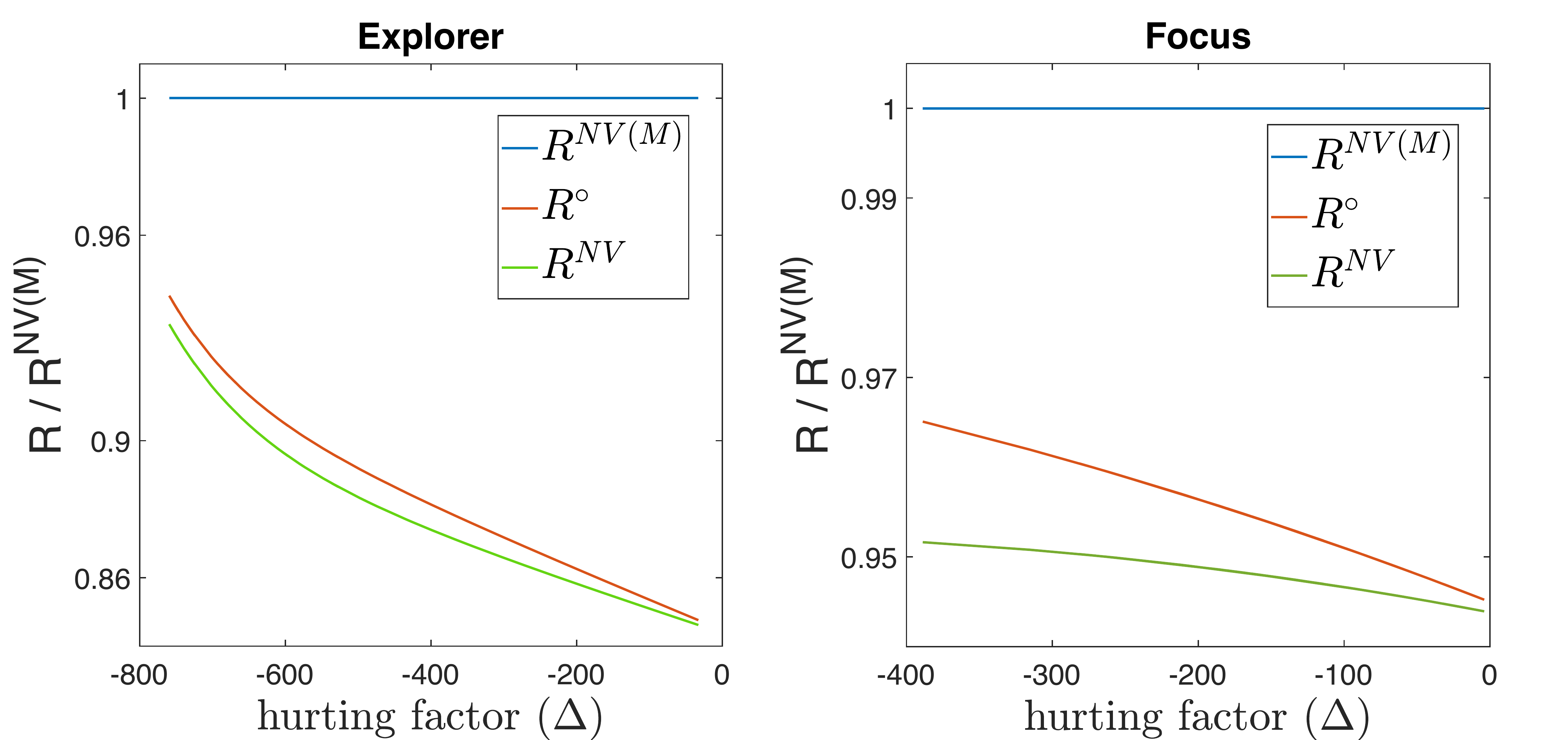}}
{Illustration of Theorem~\ref{thm:hurtpr}(iii). \label{fig:thm3part3}}
{}
\end{figure}

Figure~\ref{fig:thm3part3} illustrates Part (iii) of Theorem~\ref{thm:hurtpr}.
When the negative effect is not strong enough (i.e., $\Delta < 0$),
$R^\circ$
is close to $\nvr$ relative to its distance to $R^{\rm NV(M)}$,
confirming that in this case
$R_m^h$ can possibly exceed $\nvr$,
but only by a small margin
(recall that $R_m^h \leq R^\circ$).

\subsection{Risk Decomposition}
\label{app:risk-decomp}
\begin{figure}[ht]
\FIGURE
{\includegraphics[width=0.75\textwidth]{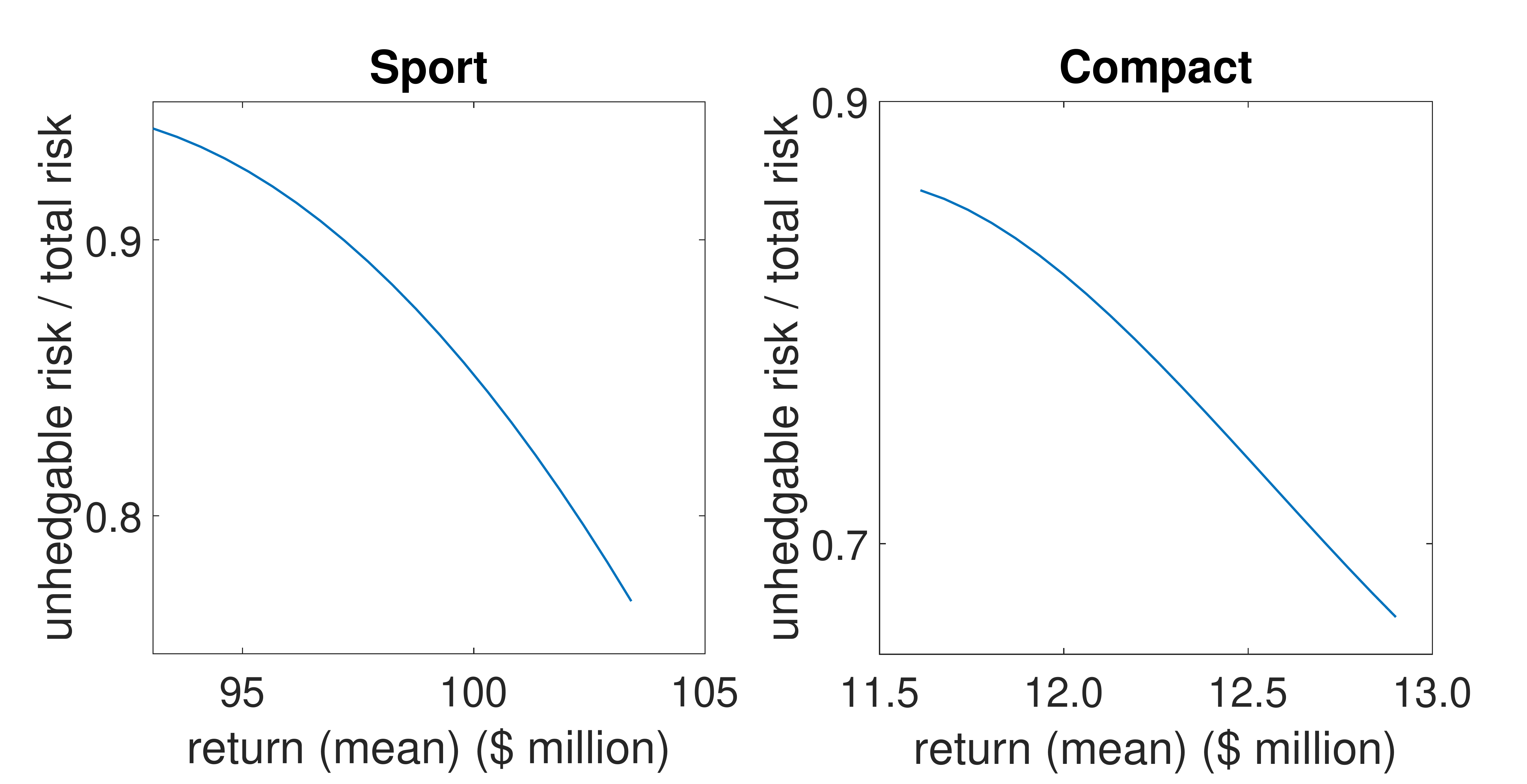} }
{Percentage Contribution of Squared Unhedgeable Risk to Total Variance.  \label{fig: unhedgeable_risk_ratio}}
{Initial WTI price is $ X_{0}=40$.}
\end{figure}

Recall, by Part (iii) of Theorem \ref{thm:hedgingsol}, the squared total risk (i.e., variance) is the sum of squared investment and unhedgeable risks.
Figure~\ref{fig: unhedgeable_risk_ratio} shows that the squared unhedgeable risk (see (\ref{unhedgeablerisk})) contributes to most of the squared total risk $B(m, P,R)$ for all instances---at least $77\%$ for \texttt{Sport} and $64\%$ for \texttt{Compact}---indicating that the main risk factor (after being hedged) is intrinsic demand volatility.
In other words, the investment risk (see (\ref{bmpr}) and (\ref{invrisk})) constitutes only a moderate part of the total risk.
This trait is desirable: a manufacturer, which is non-financial in nature, does not want to bear too much risk from financial investment.
{The risk decomposition for the instances of the other two starting oil prices
($X_0 = 70, 100$) exhibits analogous pattern as that in Figure~\ref{fig: unhedgeable_risk_ratio}.
}

\section{General Operational Payoff Functions}
\label{sec:generalization}

In this section, we extend the methodology developed so far to incorporate general operational payoff functions beyond the price-setting newsvendor.
We solve the corresponding hedging problem and identify crucial structural properties of the payoff function under which risk hedging reduces optimal operational levels when the asset price trend positively impacts the stochastic operational factors.
We apply the general result to two examples---(i) price-setting newsvendor with price-based substitutable products and (ii) price-advertising optimization.

Consider a non-financial firm operating over a time horizon $[0, T]$.
At time $0$, the firm needs to make an operational decision $\bq$, an $n$-dimensional vector.
Without loss of generality, we assume $\bq \ge \bzero$; that is, $q_i\geq 0$ for each $i=1,\ldots,n$.
At time $T$, the firm receives an operational payoff $H_T(\bq, \bA_T)$, where $\bA_T = (A_{jT})_{j=1}^T$ is a $J$-dimensional random vector, representing stochastic operational factors that may affect the payoff, such as demands, market sizes, costs, and exchange rates.
Suppose these factors partially depend on some financial asset price and are modeled as
\begin{equation}
\label{ATgeneral}
\bA_T = \bC_T + \tilde\bSigma\tilde\bB_T
\end{equation}
with the associated dynamics
\begin{equation}
\label{Atdynageneral}
d\bA_t = d\bC_t + \tilde\bSigma d\tilde\bB_t,
\end{equation}
where $\tilde\bB_t$ is a $K$-dimensional standard Brownian motion, independent of $B_t$ that drives the financial asset price $X_t$ in (\ref{Xt}).
In equation~(\ref{ATgeneral}),
$\tilde\bSigma\tilde\bB_T$ represents operational noise independent of the financial market (particularly $\{X_t, 0\leq t\leq T\}$), where $\tilde\bSigma = [\tilde\sigma_{jk}]_{j,k=1,1}^{J,K}$ is a $J$-by-$K$ matrix with $\tilde\bsigma_j = (\tilde\sigma_{jk})_{k=1}^K$ denoting the $j$-th row.
In equation~(\ref{Atdynageneral}),
$\bC_t = (C_{jt})_{j=1}^J$ is a $K$-dimensional stochastic process,
representing the operational components determined by $X_t$.
That is,  each component $C_{jt}$ is adapted to the filtration generated by  $B_t$ and we assume that $\bC_v - \bC_u$ is adapted to the  sigma-algebra generated by $\{X_t, u\leq t \leq v\}$.

A standard operations management problem is $\max_{\bq \ge \bzero} \ex[H_T(\bq, \bA_T) ]$.
However, maximizing the expected payoff without risk management may lead to highly risky decisions.
To develop a risk-management framework for the general setup,
we extend the mV optimization problem (\ref{mvprob0}) to
\begin{equation}
\label{probgenepayfunc}
\min_{\bq \ge \bzero,\, \vartheta \in \cala_X}\var[H_T(\mathbf{q}, \bA_T) + \chi_T(\vartheta)]
           \quad{ s.t.}\quad \chi_t(\vartheta):= \int_0^t \theta_s dX_s, \quad \ex[H_T(\mathbf{q}, \bA_T) + \chi_T(\vartheta)] = m.
\end{equation}

We first fix $\bq$ and consider the hedging problem (with $\bq$ dropped from the arguments of $H_T$):
\begin{equation}
\label{hedstraprobgenepayfunc}
B(m, \mathbf{q}) := \min_{\vartheta \in \cala_X}\var[H_T + \chi_T(\vartheta)]
           \quad {s.t.}\quad \chi_t(\vartheta):= \int_0^t \theta_s dX_s, \quad \ex[H_T+ \chi_T(\vartheta)] = m.
\end{equation}
The solution is characterized by Theorem \ref{thm:genpayhedsol} in Section~\ref{ECsec:general-payoff} of the e-companion,
which extends Theorem \ref{thm:hedgingsol}.
In Section~\ref{sec:genpayoptq},
we minimize the variance function $B(m, \bq)$ over $\bq$ to find the optimal operational decision in the presence of hedging.

\subsection{Optimal Operational Decision in the Presence of Hedging}
\label{sec:genpayoptq}

Theorem~\ref{thm:benefitpr} shows that for a price-setting newsvendor, hedging reduces both the optimal price and VPQ when asset price trend {\it positively} impacts demand.
We extend this result to general operational payoff functions under two technical assumptions, one concerning the payoff function's structure while the other the relationship between the unhedgeable risk and the operational decision.
\begin{assumption}
\label{assumption:genepayfunc}
The operational payoff function $H_T(\bq, \ba)$ satisfies the following conditions.
\begin{enumerate}[label=(\roman*)]
    \item For any $\ba$, $H_T(\bq, \ba)$ is concave in $q_i$ for  all $i=1,\ldots,n$.
    \item For any $\bq$, $H_T(\bq, \ba)$ is increasing in $a_j$ for all  $j=1,\ldots,J$.
    \item For any given $\ba$, $H_T(\bq, \ba)$ is supermodular in $\bq$.
    \item $H_T(\bq, \ba)$ has increasing differences in $(\bq, \ba)$; that is, $H_T(\bq_1, \ba_2) - H_T(\bq_1, \ba_1) \leq H_T(\bq_2, \ba_2) - H_T(\bq_2, \ba_1)$ for all $\bq_1 \leq \bq_2$ and $\ba_1 \leq \ba_2$.
\end{enumerate}
\end{assumption}

In Part~(i) of Assumption \ref{assumption:genepayfunc},
the concavity of the payoff function in each decision variable reflects the ``too-little-or-too-much'' trade-off.
For instance,
the newsvendor payoff function is concave in the production quantity,
reflecting the underage or overage.
In Part~(ii), the monotonicity of the payoff function in the stochastic operational factors is common for many practical applications.
In addition to the price-setting newsvendor example for which the payoff increases with the market size, a manufacturer's payoff may increase with the exchange rate or decrease with the (random) cost of raw materials.
(In the latter case, we can replace $a_i$ with $-a_i$ to recover the increasing property.)
In Part~(iii), the supermodularity property means that the operational variables are complementary to each other.
Examples include the production quantities of complementary goods.
In Part~(iv), the increasing-difference property indicates that the marginal gain from increasing the operational levels is higher for higher levels of the stochastic operational factors.
For example, marginal profit from increasing production is higher with a larger market size.

Under Assumption~\ref{assumption:genepayfunc}, we can define the following standard profit-maximizing decision:
\begin{equation}
    \label{optprobenefunc}
   \bq^* :=  \min \left\{\argmax_{\bq \ge \bzero} \ex[H_T(\bq, \bA_T) ]\right\},
\end{equation}
where the minimization takes effect entry-wise if the inner maximization problem has multiple optimal solutions.
The supermodularity property of $H_T(\bq,\ba)$ ensures that the optimal solutions of the inner maximization problem form a sublattice, so there exists a smallest optimal solution and $\bq^*$ is well defined \citep{topkis1978}.
We select the smallest optimal solution based on the economic intuition that a firm prefers a lower operational level to reach the same profit level.
For example, a manufacturer prefers to hold less inventory to reach the same level of production payoff.

The following assumption generalizes the relationship between the unhedgeable risk and the price/VPQ in the price-setting newsvendor model.
In this model, the unhedgeable risk is defined in equation~(\ref{unhedgeablerisk}) and increases in both $P$ and $R$.
This definition naturally extends to the general setting, in which the unhedgeable risk is a function of the operational decision $\bq$ (see Theorem \ref{thm:genpayhedsol} in  Section~\ref{ECsec:general-payoff} of the e-companion).
\begin{assumption}
\label{assumption:Psi}
The unhedgeable risk
increases in each component of $\bq$.
\end{assumption}

This assumption is based on the intuition that a higher operational level induces more exposure to the uncertainties associated with the operational noises, leading to higher unhedgeable risk.
\begin{theorem}
\label{thm:genpay}
Suppose Assumptions~\ref{assumption:genepayfunc}--\ref{assumption:Psi} hold and the asset price trend positively impacts the stochastic operational factors (i.e., $\bC_T^M \preceq \bC_T$, or equivalently,
$\pr(\bC_T^M \in U) \leq \pr(\bC_T \in U)$ for any upper set $U \subseteq \mathbb{R}^J$).
Let $m = \ex[H_T(\bq^*, \bA_T)]$ and $\cals_m^h = \argmin_{\bq \ge \bzero } B(m, \bq)$, where $\bq^*$ is defined in (\ref{optprobenefunc}).
Then, $\bq \leq \bq^*$ for all $\bq \in \cals_m^h$.
\end{theorem}
\begin{remark}
While Theorem~\ref{thm:benefitpr} extends to general operational payoff functions, it is not the case for Theorem~\ref{thm:hurtpr} for the following reason.
When the asset price trend has a positive effect (the setting of Theorems~\ref{thm:benefitpr} and \ref{thm:genpay}), hedging reduces the levels of the operational factors, thereby decreasing the optimal operational decision due to the increasing-difference property in Assumption~\ref{assumption:genepayfunc}.
Technically, a crucial step in proving Theorem~\ref{thm:genpay} is to show that if  $\bC_T^M \preceq \bC_T$, then the mV-optimal operational decision with hedging is lower than $\bq^{*M}$ and $\bq^{*M} \leq \bq^*$, where $\bq^{*M}$ is the counterpart of $\bq^*$ under the risk-neutral measure.
This is analogous to Proposition~\ref{pro:prx0-positive} (see the discussion therein).
However, when the asset price trend has a negative effect (the setting of Theorem~\ref{thm:hurtpr}),
although the mV-optimal operational decision with hedging is still lower than $\bq^{*M}$,
the relationship between $\bq^{*M}$ and $\bq^*$ is reversed (i.e., $\bq^{*M} \geq \bq^*$ if $\bC_T^M \succeq \bC_T$).
Thus, the relationship between $\bq^*$ and the mV-optimal operational decision with hedging is generally unclear.
This relationship becomes clear in  Theorem~\ref{thm:hurtpr} (i.e., the price-markdown result for a price-setting newsvendor when the asset price trend negatively impacts demand)
because we can exploit specific structures of the newsvendor payoff function.
The analysis, however, does not extend to general payoff functions.
\end{remark}

\subsection{Example: Price-Setting Newsvendor with Price-Based Substitutable Products}\label{sec:genexamples-news}

A firm manufactures $J$ products and sells them to the market over a time horizon $[0,T]$.
At time $t=0$, the firm must decide the production quantities $\bQ:= (Q_j)_{j=1}^J$ and prices (net costs) $\bP: = (P_j)_{j=1}^J$.
All the product demands are realized at time $t=T$.
The demand for product $j$ is $D_{jT}(\bP) = A_{jT}-\sum_{i=1}^J \gamma_{ji} P_i$, where $A_{jT}$ is the market size of this product and $\bGamma = [\gamma_{ij}]_{i,j=1}^{J,J}$ is a $J$-by-$J$ matrix.
For each $j$, $\gamma_{jj} > 0$ represents the sensitivity of the demand for product $j$ relative to its price $P_j$, and $\gamma_{ji}$ (with $j \neq i$) represents the effect of the demand for product $i$ on the demand for product $j$.
We also define $R_j = Q_j + \sum_{i=1}^J \gamma_{ji} P_i$, representing the VPQ for product $j$, and let $\bR = (R_j)_{j=1}^J$.
The payoff function is
\begin{equation}
    \label{multi-nv_profit_PR}
    H_{T}((\bP, \bR), \bA_T) = \sum_{j=1}^{J}\left[P_j \Big(R_{j}-\sum_{i=1}^J \gamma_{ji} P_i \Big)-\left(P_j + c_j\right)\left(R_{j}  -A_{j T} \right)^{+}\right],
\end{equation}
where $c_j$ is the unit production cost net unit salvage value.
Then, the mV risk-management problem for this multi-product price-setting newsvendor model is in the form of  (\ref{probgenepayfunc}) with $\bq = (\bP, \bR)$.
\begin{corollary}
\label{cor:multi}
Suppose $\gamma_{ij} < 0$
and $\tilde\bsigma_i\cdot \tilde\bsigma_j \ge 0$ for all $i \neq j$, where $\tilde\bsigma_j$ is the $j$-th row of $\tilde\bSigma$ in~(\ref{ATgeneral}).
Let $m = \ex[H_T((\bP^*, \bR^*), \bA_T)]$ and
$\cals_m^h  = \arg\min_{\bP \ge \bzero, \bR \ge \bzero} B(m, (\bP, \bR))$. Let $\bq^*=(\bP^*, \bR^*)$ be as defined in (\ref{optprobenefunc}).
If the asset price trend positively impacts the market sizes (i.e., $\bC_T^M \preceq \bC_T$), then
$\bP \leq \bP^*$ and $\bR \leq \bR^*$ for all $(\bP, \bR) \in \cals_m^h$.
\end{corollary}

Corollary \ref{cor:multi} indicates that if the asset price trend positively impacts each product's market size, risk hedging reduces all products' optimal price and VPQ.
The condition $\gamma_{ij} < 0$ is a standard assumption in the literature on multi-product problems  \citep{aydin2008joint,dong2009dynamic}.
It means that different products are {\it price-based substitutes}: a price increase in one product boosts the demand for another.
The condition $\tilde\bsigma_i\cdot \tilde\bsigma_j \ge 0$ means that the unhedgeable factors of different product demands are nonnegatively correlated.

\subsection{Example: Price-Advertising Optimization}
\label{sec:genexamples}

The example below is adapted from the classical price-advertising optimization problem in \cite{DS1954}; see also \cite{bagwell2007} for a survey of the optimal advertising literature.
Consider a make-to-order manufacturer which must decide, at time $t=0$, the unit selling price $P$ (net the unit production cost) and an {\it advertising level} $\gamma$.
The demand at time $t=T$ is $D_T$ with
\begin{equation}
\label{pgammademand}
D_T = d(A_T, P, \gamma),
\end{equation}
where $A_T$ is the market size following the model (\ref{ATgeneral}) and
$d(\cdot)$ is a twice-differentiable function,  increasing in $A_T$ and $\gamma$ and decreasing in $P$.
The manufacturer pays advertising cost $k(\gamma)$ that is increasing in $\gamma$ and does not depend on $D_T$.
The profit function is
$H_T((P, \gamma), A_T) = Pd(A_T, P, \gamma) - k(\gamma)$.
Then, the mV risk-management problem for this pricing-advertising model is in the form of (\ref{probgenepayfunc}) with $\bq = (P, \gamma)$.
\begin{corollary}
\label{cor:pgamma}
Suppose that $d(A_T, P, \gamma)$ defined  in (\ref{pgammademand})
is increasing in $A_T$, concave and decreasing in $P$, concave and increasing in $\gamma$, and supermodular in $(A_T, P, \gamma)$.
Suppose also that $k(\gamma)$ is convex and increasing in $\gamma$.
Let $m = \ex[H_T((P^*,\gamma^*), A_T)]$ and  $\cals_m^h = \arg\min_{P \ge 0, \gamma \ge 0} B(m, (P, \gamma))$. Let $\bq^* = (P^*, \gamma^*)$ be defined as in (\ref{optprobenefunc}).
If the asset price trend positively benefits the market size (i.e., $C_T \succeq C_T^M$), then
$P \leq P^*$ and $\gamma \leq \gamma^*$ for all $(P, \gamma) \in \cals_m^h$.
\end{corollary}

Corollary~\ref{cor:pgamma} states that if the asset price trend positively impact the market size, hedging decreases both the optimal pricing and advertising levels. %
The concavity of $d(A_T, P, \gamma)$ in $P$ means that the demand is more sensitive to price increment at higher pricing levels.
The concavity of $d(A_T, P, \gamma)$ in $\gamma$ suggests a diminishing marginal benefit of advertising on product demand.
The convexity of $k(\gamma)$ means that
the marginal advertising cost increases as advertising level increases.

Now we explain the supermodularity of $d(A_T, P, \gamma)$ in
all three arguments.
This property is equivalent to the supermodularity of the function in each pair of the arguments, and we discuss them one by one.
First, the supermodularity in the pricing and advertising levels $(P, \gamma)$ indicates that the demand is more sensitive to the advertising level with a higher selling price.
The same condition is also assumed in \cite{topkisbook}; see its Example 3.3.10, where the ``quality'' there can stand for the advertising level.
Second, the supermodularity in the market size and advertising level $(A_T, \gamma)$ means that the marginal benefit on product demand from a higher advertising level increases with the market size.
This is intuitive: a larger population of potential customers implies that advertising achieves a broader audience coverage.

Last, the supermodularity in the market size and pricing level $(A_T, P)$ means that the detrimental effect on product demand from a price markup decreases with the market size.
It may be attributed to factors---either asset-related or unrelated to the financial market---that boost the market size and decrease customers' sensitivity to price increments.
For example, as the economy strengthens, the broad market stock index rises, and the financial component of the market size increases.
Meanwhile, income increases amid better economic conditions, and people are less sensitive to price markups.
For another example, when customers' preferences and tastes shift in favor of the product in concern, market size increases due to the increment of its non-financial component.
Meanwhile, favoring this product, customers tend to be less sensitive to its price increment.

\clearpage
\vspace{18pt}
\noindent
\begin{center}
\large \textbf{Part II: Proofs}
\end{center}

\section{Proofs for Section~\ref{sec:NV}}
\subsection{Proof of Proposition~\ref{pro:nvsol}} \label{appendix:nvsol}
   The expected profit, as a function of $(P, R)$, is
	$$\ex[H_{T}(P, R)]=(P-c) (R-bP) - (P-s)\ex[\left(R-A_{T}\right)^{+}].$$
	Given $P$, differentiate $\ex[H_{T}(P, R)]$ with respect to $R$:
	\begin{equation}
	\frac{\partial \ex[H_{T}(P, R)]}{\partial R} = (P-c) -(P-s)F( R).
	\end{equation}
	Then, take second-order derivative to $R$:
	\begin{equation}
	\frac{\partial^2 \ex[H_{T}(P, R)]}{\partial R^2} =  -(P-s)f( R) <0
	\end{equation}
	Thus, for a given $P$, the expected profit is a concave function in  $R$,
	and the optimal solution, denoted $R^{\rm NV}(P)$,
	is solved from setting $\partial\ex[H_T(P,R)] / \partial R$ to zero.
	This leads to
	$$R^{\rm NV}(P) =F^{-1}(\frac{P-c}{P-s}).$$
	Clearly, $R^{\rm NV}(P)$ increases in $P$.
	Define:
	$$m(P):=\ex[H_T(P, R^{\rm NV}(P) )]  = (P-c)(R^{\rm NV}(P) -bP)-(P-s)\ex[(R^{\rm NV}(P) -A_T)^+].$$
	As $P \rightarrow c$, $R^{\rm NV}(c) \rightarrow -\infty $,  $(P-s)\ex[(R^{\rm NV}(P) -A_T)^+] \rightarrow 0$,  then,
\begin{eqnarray*}
\lim_{P \rightarrow c}m(P) &=& \lim_{P \rightarrow c} (P-c)(R^{\rm NV}(P) -bP)
                           = \lim_{P \rightarrow c} \frac{R^{\rm NV}(P) -bP}{\frac{1}{P-c}}
                           = \lim_{P \rightarrow c} \frac{\frac{(c-s)f(R^{\rm NV}(P))}{(1-F(R^{\rm NV}(P))^2)}-b}{-\frac{1}{(P-c)^2}} \\
                           &=&-\lim_{P \rightarrow c} \frac{1}{f(R)}(c-s)(\frac{P-s}{P-c})^2 + \lim_{P \rightarrow c} b(P-c)^2
                           =-(c-s)\lim_{R \rightarrow -\infty} \frac{F^2(R)}{f(R)} \\
                           &=&-(c-s)\lim_{R \rightarrow -\infty} \frac{F^2(R)}{r(R)}(1-F(R)).
\end{eqnarray*}
Under Assumption~\ref{assumption:r}, $F^2(a)/r(a) \to 0$ as $a \to -\infty$, so the limit above is $0$.

\smallskip
Next, differentiate $m(P)$ with respect to $P$:
	\begin{align}
	m'(P) = R^{\rm NV}(P)-2bP+bc -\ex[(R^{\rm NV}(P) -A_T)^+]
	\end{align}
	As $P \to c$, $ R^{\rm NV}(c) \to -\infty$ and $ m'(c) \to -\infty$.

	Differentiate $m'(P)$ with respect to $P$:
	\begin{align}
	m''(P)= -2b+ (1-F(R^{\rm NV}(P)))\frac{1}{\frac{dP}{d R^{\rm NV}(P)}}= -2b+ \frac{(1-F(R^{\rm NV}(P)))^3}{(c-s)f(R^{\rm NV}(P))} = -2b+ \frac{(1-F(R^{\rm NV}(P)))^2}{(c-s)r(R^{\rm NV}(P))}
	\end{align}
	where we use the definition of the hazard rate for $A_T$, i.e., $r(a) = \frac{f(a)}{1-F(a)}$.
	Further differentiating $m''(P)$ with respect to $P$:
	$$\frac{dm''(P)}{d P} = -\frac{(2r(R^{\rm NV}(P))^2+r^{'}(R^{\rm NV}(P)))(1-F(R^{\rm NV}(P)))^2}{(c-s)r^2(R^{\rm NV}(P))} \frac{d R^{\rm NV}(P) }{d P}.$$
  	By Assumption~\ref{assumption:r}, we have $\frac{dm''(P)}{d P}<0$, so $m''(P)$ is decreasing in $P$. Thus, as $P$ increases from $c$ to $\infty$, $m''(P)$  decreases from $\infty$ to $-2b$.
  	  (As $P \to c$, $R^{\rm NV}(P) \to -\infty$, $r(R^{\rm NV}(P)) \to 0$, leading to $m''(P) \to \infty$;
  	  As $P\to \infty$, $\frac{(1-F(R^{\rm NV}(P)))^3}{f(R^{\rm NV}(P))} \to 0$ by Assumption~\ref{assumption:r}).
  	  Combining the above, $m'(P)$ increases first and then decreases with $ m'(P) \to -\infty$ as $P\to c$ and $m'(P)\to -\infty$ as $P \to \infty$.

    Combining the analysis above and Assumption~\ref{assumption:nvm}, there must exist some $P^0>c$ such that $m'(P^0)>0$, otherwise, the optimal expected profit is negative.
    Therefore, $m'(P)=0$ has two zeros for $P>c$.
    The smaller one is the minimizer, the larger one is the  maximizer.
    As $P$ increases over $c$, $m(P)$ first decreases from $0$ to a negative value, which is the minimum level of $m(P)$.
    Then, it increases to a positive value, which is the maximum level of $m(P)$ and the associated optimal price is denoted as $P^{\rm NV}$, which solves the following optimality equation:
	\begin{equation}
	m'(P^{\rm NV}) = R^{\rm NV}(P) -2bP^{\rm NV}+bc - \ex[(R^{\rm NV}(P) -A_T)^+] = 0.
	\end{equation}
	The equation above is equivalent to:
	\begin{equation}
	2bP^{\rm NV}-bc  = \ex(R^{\rm NV}(P^{\rm NV}) \wedge A_T),
	\end{equation}
    and $m(P)$ decreases when $P>P^{\rm NV}$.  Let $\underline{P}$ be the smaller zero of $m(P)=0$ and then $\nvp >\underline{P}>c$.
    Let $R^{\rm NV} = R^{\rm NV}(P^{\rm NV})$, which satisfies the other optimality equation:
    $$ \nvp - c - (\nvp - s) F(\nvr) = 0.$$
    Combining the optimality equations of $(P^{\rm NV}, R^{\rm NV})$ leads to (\ref{nvopteqns}),
    and this completes the proof.
$\hfill\Box$

\subsection{Proof of Proposition~\ref{pro:nvfrontier}}
\label{appendix:pronvfrontier}
We prove the three results stated in this proposition one by one.

For Part (i),  we have already proved that $m(P)$ is increasing in $P\in [\underline{P}, P^{\rm NV}]$ in  Section~\ref{appendix:nvsol}.
The variance of production payoff is:
$$\var[H_T(P,R)] = (P-s)^2\{\ex\big[[(R-A_{T})^{+}]^2\big]-E^2[(R-A_{T})^{+}]\}.$$
Differentiate $\var[H_T(P,R)]$ with respect to $P$:
$$\frac{\var[H_T(P,R)]}{\partial P} =2(P-s)\big[\ex[(R-A_{T})^{+}]^2-E^2[(R-A_{T})^{+}]\big]  >0 .$$
Differentiate $\var[H_T(P,R)]$ with respect to $R$:
$$\frac{\partial \var[H_T(P,R)]}{\partial R} =2(P-s)^2
\{  \ex[\left(R-A_{T}\right)^{+}](1-F(R))>0.
$$
Next, differentiating $v(P) = \var[H_T(P, R^{\rm NV}(P))]$ with respect to $P$, we have
\begin{align}
\frac{d v(P)}{d P} = \frac{\partial \var[H_T(P,\nvr(P))] }{\partial P} +\frac{\partial \var[H_T(P,\nvr(P))]] }{\partial R}\frac{d R^{\rm NV}(P)}{d P}>0
\end{align}
Then ${\rm Var}[H_T(P, R^{\rm NV}(P))]$ is increasing in $P$ for $P \ge c$.

\medskip
For Part (ii), fixing $R$, define
$$m(R)=\ex[H_T(P^{\rm NV}(R), R)] = (P^{\rm NV}(R)-c)(R-bP^{\rm NV}(R))-(P^{\rm NV}(R)-s)\ex[(R-A_T)^+].$$
Recall, $P^{\rm NV}(R) = (\ex[R \wedge A_T] +bc)/(2b)$.
It is straightforward to verify that $P^{\rm NV}(bc)<c$.
Further, from Assumption~\ref{assumption:Abc}, we have:
$$\ex[(bc - A_T)^+] \leq 2b(c-s).$$
Then,
$$P^{\rm NV}(bc) = \frac{1}{2b}(\ex[bc \wedge A_T] +bc)= \frac{1}{2b}( bc - \ex[(bc - A_T)^+] +bc)> s,$$
so
$$m(bc) = b(P^{\rm NV}(bc)-c)(c-P^{\rm NV}(bc))- (P^{\rm NV}(bc) -s)\ex[(bc-A_T)^+]<0.$$
Next, differentiate $m(R)$ with respect to $R$ and taking into account
$\partial \ex[H_T(P^{\rm NV}(R), R)] / \partial P = 0$:
$$\frac{dm(R)}{d R} = \frac{d \ex[H_T(P^{\rm NV}(R), R)]}{d R} = \frac{\partial \ex[H_T(P^{\rm NV}(R), R)]}{\partial R}
=(P^{\rm NV}(R)-c)-(P^{\rm NV}(R)-s)F(R).$$
Let
$$J(R) =: \frac{dm(R)}{d R}=(P^{\rm NV}(R)-c)-(P^{\rm NV}(R)-s)F(R) ,$$
and differentiate $J(R)$ with respect to $R$:
$$\frac{dJ(R)}{d R} = f(R)\Big[\frac{1-F(R)}{2br(R)}-(P^{\rm NV}(R)-s)\Big].$$
Then, differentiate $\frac{dJ(R)}{d R}$ with respect to $R$:
$$\frac{d^2J(R)}{d R^2} =\frac{df(R)}{d R} \frac{\frac{dJ(R)}{d R} }{f(R)}-\frac{f(R)(1-F(R))}{2b r^2(R)}(2r^2(R)+r'(R)).$$
We claim that $J(R)$ is either monotone or unimodal in $R$.
First note for any $R$, $\frac{f(R)(1-F(R))}{2b r^2(R)}(2r^2(R)+r'(R))>0$ by Assumption~\ref{assumption:r}.
Then, there are two cases.
The first one is that $\frac{dJ(R)}{d R}>0$ or $\frac{dJ(R)}{d R}<0$ for all $R$,
then $J(R)$ is monotone in $R$.
The other case is that there exists $R$ such that $\frac{dJ(R)}{d R}=0$.
Then, at this $R$, $\frac{d^2J(R)}{d R^2}<0 $, which indicates that for any stationary point $R$,
it must be a local maximizer.
Thus, $J(R)$ is unimodal for this case.
Note, $J(bc) <0$ and $J(R) \rightarrow -(c-s) < 0$ as $R \rightarrow \infty$.
Then, by Assumption~\ref{assumption:nvm}, there must exist some $R>bc$ such that $J(R)>0$.
Otherwise $J(R)<0$ for all $R\geq bc$ which implies that $m(R)$ is decreasing in $R$ and $m(R)\leq m(bc)<0$,
and the optimal VPQ is $R=bc$, which contradicts this assumption.
By the analysis above, we conclude that $J(R)$ is unimodal in $R\geq bc$.
Furthermore, there are two roots of $J(R)=0$ and
let $R_{1}^0$ (resp.,\ $R_2^0$) be the smaller (resp., larger) one.
Clearly, $m(R)$ is decreasing in $R<R_1^0$, increasing in $R_1^0<R<R_2^0$ and decreasing in $R>R_2^0$.
Since $m(bc)<0$, there are two roots for $m(R) = 0$ for $R\geq bc$ and let $\underline{R}$ denote the smaller one.
By Proposition~\ref{pro:nvsol}, $R^{NV} = R_2^0$.
Thus, we conclude that $m(R)$ is increasing in $\underline{R}<R<R^{NV}$ (and decreasing in $R>R^{NV}$).

Analogous to Proposition~\ref{pro:nvfrontier} (i), it is straightforward to prove that $\var[H_T(P^{\rm NV}(R), R)]$ increase in $R$.

\medskip
For Part (iii), let $\bar{m} =\ex[H_T(\nvp, \nvr)]$.
For any $m\in(0, \bar{m})$, we claim that both $\partial \ex[H_{T}(P^{\rm NV}_m, R^{\rm NV}_m))] / \partial P$ and $\partial \ex[H_{T}(P^{\rm NV}_m, R^{\rm NV}_m]/\partial R$ are positive.
Otherwise, suppose $\partial \ex[H_{T}(P^{\rm NV}_m, R^{\rm NV}_m))] / \partial P \leq 0 $.
As $ \ex[H_{T}(P, R^{\rm NV}_m)]$ is concave in $P$ and  $\ex[H_{T}(c, R^{\rm NV}_m)]\leq 0<\ex[H_{T}(P^{\rm NV}_m, R^{\rm NV}_m)] =m$,  we know that $ \ex[H_{T}(P, R^{\rm NV}_m)]$ is increasing in $P\in(c, P^{\rm NV}(R^{\rm NV}_m))$ and decreasing in $P>P^{\rm NV}(R^{\rm NV}_m)$.
Then $P^{\rm NV}_m \ge P^{\rm NV}(R^{\rm NV}_m)$ as we have assumed $\partial \ex[H_{T}(P^{\rm NV}_m, R^{\rm NV}_m))] / \partial P \leq 0 $.
So there exists $P^{'}_m$ such that  $P^{'}_m \leq P^{\rm NV}(R^{\rm NV}_m) \leq P^{\rm NV}_m$ and we have:
	$$ \ex[H_{T}(P^{\rm NV}_m, R^{\rm NV}_m)] = \ex[H_{T}(P^{'}_m, R^{\rm NV}_m)] = m,\
	\var[H_T((P^{'}_m, R^{\rm NV}_m))] \leq \var[H_T(\nvp_m, \nvr_m)].$$
	In other words, keeping the target return $m$, $(P^{'}_m, R^{\rm NV}_m)$ has a smaller variance than $(P^{\rm NV}_m, R^{\rm NV}_m)$ does, and thus $(P^{\rm NV}_m, R^{\rm NV}_m)$ cannot be the solution to the problem in  (\ref{nvmv}).
	By contradiction, $\partial \ex[H_{T}(P^{\rm NV}_m, R^{\rm NV}_m))] / \partial P$ must be positive.
	Similarly, $\partial \ex[H_{T}((P^{\rm NV}_m, R^{\rm NV}_m))] / \partial R > 0$ always holds.

	Next, introducing the Lagrange multiplier $\lambda$, the Lagrangian function of the problem in (\ref{nvmv}) is:
	\begin{align}
	\mathcal{L} = \mathrm{Var}[H_T(P, R)]-\lambda( \ex[H_{T}(P, R)]-m).
	\end{align}
	Then,  $(P^{\rm NV}_m, R^{\rm NV}_m  , \lambda^*_m)$ satisfies the Karush–Kuhn–Tucker (KKT) equations:
	\begin{align} \label{kktcondition1}
	\frac{\partial \mathrm{Var}[H_{T}(P, R)]}{\partial P} - \lambda \frac{\partial \ex[H_{T}(P, R)]}{\partial P} =0;\\
	\label{kktcondition2}
	\frac{\partial \mathrm{Var}[H_{T}(P, R)]}{\partial R} - \lambda \frac{\partial \ex [H_{T}(P, R)]}{\partial R} =0;\\
	\ex [H_T(P,R)] -m  =0.
	\end{align}
	we have  $\lambda^*_m>0$  as we have proved that $\partial \ex[H_{T}(P^{\rm NV}_m, R^{\rm NV}_m)] / {\partial P} >  0 $ and   $\partial \ex[H_{T}(P^{\rm NV}_m, R^{\rm NV}_m)] / {\partial R} >0 $ hold.
	By Envelop Theorem, we have:
	$$\frac{d}{dm}\mathrm{Var}[H_T(P^{\rm NV}_m,R^{\rm NV}_m)] =\lambda^*>0$$
	this proves that $\var[H_T(\nvp_m, \nvr_m)]$ is increasing in $m$ for  $m\in(0, \ex[H_T(\nvp, \nvr)])$.

	When $(P, R)\rightarrow (P^{\rm NV}, R^{\rm NV})$ , $m \rightarrow \ex[H_T(\nvp, \nvr)]]$,  $\partial \ex[H_{T}(P, R)] / {\partial P}\rightarrow 0$ and $\partial \ex[H_{T}(P, R)]/{\partial R}\rightarrow 0$. Therefore,  the KKT condition (\ref{kktcondition1}) or condition (\ref{kktcondition2}) implies that $\lambda^* \rightarrow \infty$.
 Next, note that by Assumption~\ref{assumption:nvm}, $\pr(A_T \leq \nvr) = (\nvp-c)/(\nvp-s) \in (0,1)$, hence $\var(H_T(\nvp,\nvr)) = (\nvp-s)^2\var[(\nvr-A_T)^+] > 0$.
 Also, $\var[(H_T(\nvp, \nvr))] = (\nvp-s)^2\var[(\nvr-A_T)]^+ \leq (\nvp-s)^2\var(A_T) < \infty$.
 So, as $m \to \infty$, $\var[H_T(P,R)] \to \var[H_T(\nvp, \nvr)] \in (0, \infty).$
 Hence, $d\sqrt{\var[H_T(P,R)]} / dm = \g^*/(2\sqrt{\var[H_T(P,R)]}) \to \infty$.
 This proves that the incremental risk approaches infinity when the expected payoff approaches the newsvendor's maximum profit in the base model.
	$\hfill\Box$

\section{Proofs for  Section~\ref{sec:hedging}}
\subsection{Martingale Representation}
\label{appendix:mrt}
Here we show that any martingale under $\pr^M$ admits a martingale representation based on $dX_t$ and $d\tilde B_t$.
The result is summarized in the following proposition.
\begin{proposition}
\label{pro:mrt}
Let $Z_t$ and $\pr^M$ be defined in (\ref{genZt}) and suppose $Z_t$ satisfies Assumption \ref{assumption:Z}.
\begin{itemize}
    \item[(i)] For any martingale $\tilde M_t$ under $\pr^M$, it has the following representation:
\begin{equation}
\label{MRT}
\tilde M_t = \tilde M_0 + \int_0^t a_s dX_s + \int_0^t b_s d\tilde B_s
\end{equation}
for some adapted processes $\{a_t\}$ and $\{b_t\}$.
\item[(ii)] For any martingale $\{\tilde M_t\}$ with respect to $\calg_t$ under $\pr^M$
(recall, $\calg_t$ is generated by $\{B_t\}$ and $\calg_t \subseteq \calf_t$),
\begin{equation}
\label{MRT1}
\tilde M_t = \tilde M_0 + \int_0^t \gamma_s dX_s
\end{equation}
for some process $\{\gamma_t\}$ that is adapted to $\calg_t$.
\end{itemize}
\end{proposition}
{\startb Proof.}
To start with, note that $\{Z_t\}$ is also a martingale with respect to $\calg_t$ under $\pr$.
To see this, first note that $\{Z_t\}$ is clearly adapted to $\calg_t$,
and
$$\ex[Z_T | \calg_t] = \ex[\ex[Z_T|\calf_t] | \calg_t] = \ex[Z_t | \calg_t] = Z_t,$$
where the first equality is the iterated conditioning and the second one follows from Assumption \ref{assumption:Z}; the last one is due to $Z_t \in \calg_t$.
Applying It\^{o}'s Lemma to $1/Z_t = \exp\{\int_0^t \eta_s dB_s + (1/2)\int_0^t \eta_s^2ds\}$:
\begin{eqnarray*}
d\frac{1}{Z_t} = \frac{1}{Z_t}[\eta_t dB_t + \frac{1}{2}\eta_t^2 dt] + \frac{1}{2}\frac{1}{Z_t}\eta_t^2dt
               = \frac{1}{Z_t}[\eta_t^2 dt + \eta_t dB_t].
\end{eqnarray*}

Next, by Girsanov Theorem \citep{phambook}, $B_t^M := B_t + \int_0^t \eta_s ds$ is a Brownian motion under $\pr^M$.
Note that since the MPR process associated with $\tilde B_t$ is 0,
$\{\tilde B_t\}$ is also a Brownian motion under $\pr^M$,
i.e., the change of measure does not affect $\{\tilde B_t\}$.
Therefore,
\begin{equation}
\label{XtM}
dX_t = \sigma_tX_t(\eta_t dt + dB_t) = \sigma_t X_t dB_t^M.
\end{equation}
Clearly, $\{X_t\}$ is adapted to $\calg_t$.

\medskip
Now we prove Part (i).
Define $M_t = Z_t \tilde M_t$, which is clearly an adapted process.
In particular, $\ex[|M_t|] = \ex[Z_t|\tilde M_t|] = \ex^M[|\tilde M_t|] < \infty$ hence $\{M_t\}$ is integrable under $\pr$.
Then, for any $0 \leq t \leq u \leq T$,
\begin{eqnarray*}
\ex[M_u \,|\, \calf_t] = \ex[Z_u \tilde M_u \,|\, \calf_t]
                       = \ex\Big[\frac{Z_u}{Z_t}Z_t\tilde M_u \,\Big{|}\, \calf_t\Big]
                       = Z_t\ex^M[\tilde M_u \;|\; \calf_t]
                       = Z_t \tilde M_t
                       = M_t.
\end{eqnarray*}
The third equality follows from $Z_t \in \calf_t$ and Bayes formula for change of measure
\citep{phambook}.
Thus, $M_t$ is a martingale under $\pr$
and admits the following martingale representation \citep{phambook}:
$$dM_t = \Gamma_t dB_t + \Lambda_t d\tilde B_t.$$
$\{\Gamma_t\}$ and $\{\Lambda_t\}$ are adapted processes.
Then, apply It\^{o}'s Lemma on $\tilde M_t = M_t \frac{1}{Z_t}$:
\begin{eqnarray*}
d\tilde M_t = d \Big(M_t \frac{1}{Z_t}\Big) &=& M_t d\frac{1}{Z_t} + \frac{1}{Z_t}dM_t + dM_td\frac{1}{Z_t} \\
                                  &=& M_t\frac{1}{Z_t}(\eta_t^2dt +\eta_t dB_t)
                                  +\frac{1}{Z_t}(\Gamma_t dB_t + \Lambda_t d\tilde B_t) + \Gamma_t\frac{\eta_t}{Z_t}dt \\
                                  &=& \tilde M_t \eta_t(\eta_tdt + dB_t) + \frac{\Gamma_t}{Z_t}(dB_t + \eta_t dt)
                                     + \frac{\Lambda_t}{Z_t}d\tilde B_t \\
                                  &=& \frac{\Big[\tilde M_t \eta_t + \frac{\Gamma_t}{Z_t}\Big]}{\sigma_t X_t}dX_t + \frac{\Lambda_t}{Z_t}d\tilde B_t.
\end{eqnarray*}
This proves the first part of the proposition.

\medskip
Now we proceed to Part (ii),
under which $\{\tilde M_t\}$ is an $\pr^M$-martingale with respect to $\calg_t$.
We have:
\begin{eqnarray*}
\tilde M_t = \ex^M[\tilde M_T | \calg_t]
=\ex\Big[\frac{Z_T}{Z_t} \tilde M_T \Big{|} \calg_t\Big]
=\frac{1}{Z_t} \ex(Z_T\tilde M_T | \calg_t).
\end{eqnarray*}
To show the second equality, we verify that the definitions of conditional expectation are satisfied.
First, $\ex\Big[\frac{Z_T}{Z_t} \tilde M_T \Big{|} \calg_t\Big]$ is clearly adapted to $\calg_t$.
And, for any set $B \in \calg_t$,
$$\ex^M\Big[ \ex\Big[\frac{Z_T}{Z_t} \tilde M_T \Big{|} \calg_t\Big] \ind_B\Big]
= \ex\Big[Z_t \ex\Big[\frac{Z_T}{Z_t} \ind_B \tilde M_T \Big{|} \calg_t\Big]\Big]
=  \ex\Big[\frac{Z_t}{Z_t} \ex\Big[Z_T \ind_B \tilde M_T \Big{|} \calg_t\Big]\Big]
= \ex\Big[Z_T \ind_B \tilde M_T\Big]
= \ex^M(\ind_B \tilde M_T). $$
The first equality is change of measure and the second one follows from $Z_t \in \calg_t$;
the third equality is the iterated conditioning and the last one is again the change of measure.

Define $M_t := \ex(Z_T\tilde M_T | \calg_t)$,
which is clearly a martingale with respect to $\calg_t$ under $\pr$
hence admits the martingale representation:
$$dM_t = \Gamma_t dB_t,$$
where $\{\Gamma_t\}$ is a process adapted to $\calg_t$.
Now, apply It\^{o}'s Lemma to $\tilde M_t$:
\begin{eqnarray*}
d\tilde M_t = d\Big(\frac{1}{Z_t}M_t\Big)
            &=& \frac{1}{Z_t}dM_t + M_td\frac{1}{Z_t} + dM_td\frac{1}{Z_t} \\
            &=& \frac{1}{Z_t}\Gamma_t dB_t + M_t\frac{1}{Z_t}[\eta_t^2 dt + \eta_t dB_t] + \frac{\Gamma_t}{Z_t}\eta_t dt \\
            &=& \frac{M_t\eta_t}{Z_t}[\eta_t dt + dB_t] +\frac{\Gamma_t}{Z_t}[\eta_t dt + dB_t] \\
            &=& \Big[\frac{M_t\eta_t+\Gamma_t}{Z_t}\Big]dB_t^M \\
            &=& \Big[\frac{M_t\eta_t+\Gamma_t}{Z_t\sigma_tX_t}dX_t^M\Big]
\end{eqnarray*}
Since $\{M_t\}$, $\{\Gamma_t\}$, $\{\eta_t\}$, $\{Z_t\}$, $\{\sigma_t\}$ and $\{X_t\}$ are all adapted to $\calg_t$,
this completes the proof.
$\hfill\Box$

\begin{remark}
\label{rem:mrthd}
If $\calf_t$ is generated by $\{B_t\}$ and $\{\tilde \bB_t\}$,
where the latter is a multi-dimensional Brownian motion that is independent from $\{B_t\}$,
(\ref{MRT}) in Part (i) of Proposition~\ref{pro:mrt} still holds and the only modification is to
replace $b_sd\tilde B_s$ with $\mathbf{b}_s\cdot d\tilde \bB_s$ for some adapted (vector) process $\{\mathbf{b}_t\}$.
The proof is completely analogous.
\end{remark}

\begin{corollary}
\label{cor:calfg}
For any nonnegative and $\pr^M$-integrable random variable $M_T \ge 0$ that is measurable with respect to $\calg_T$,
\begin{equation}
\label{calfg}
\ex^M(M_T \,|\, \calf_t) = \ex^M(M_T \,|\, \calg_t) = M_0 + \int_0^t \gamma_s dX_s,
\quad \forall 0\leq t \leq T,
\end{equation}
where $\{\gamma_t\}$ is a process that is adapted to $\calg_t$.
In particular, the above holds for $M_T = Z_T$ defined in (\ref{genZt}).
\end{corollary}
{\startb Proof.}
Define $Y_t := \ex^M(M_T \,|\, \calg_t)$, which is clearly a nonnegative martingale under $\pr^M$ with respect to $\calg_t$.
Since $M_T \in \calg_T$, we have $Y_T = M_T$.
By Proposition \ref{pro:mrt} Part (ii),
$Y_t$ admits the martingale representation
(recall, $\{B_t^M = B_t + \int_0^t \eta_s ds\}$ is a Brownian motion under $\pr^M$):
$$dY_t = \gamma_tdX_t = \gamma_t \sigma_t X_t dB_t^M$$
for some process $\{\gamma_t\}$ adapted to $\calg_t$;
recall, $\{\sigma_t\}$ and $\{X_t\}$ are also adapted to $\calg_t$.
So, $Y_t$ is a local martingale under $\pr^M$ and, being nonnegative, a supermartingale under $\pr^M$ \citep{yorbook}.
Next, for any $0 \leq t \leq T$,
$$\ex^M(Y_t) = \ex^M[\ex^M(M_T \,|\, \calg_t)] = \ex^M(M_T),$$
which is a constant.
Therefore, being a supermartingale with constant expectation, $\{Y_t\}$ is a martingale under $\pr^M$ \citep{yorbook}.
Then, we have
\begin{eqnarray*}
Y_t = \ex^M(Y_T \,|\, \calf_t) = \ex^M(M_T  \,|\, \calf_t ).
\end{eqnarray*}
This establishes (\ref{calfg}).
Since $Z_T > 0$ and $Z_T \in \calg_T$, (\ref{calfg}) holds for $M_T = Z_T$.
This completes the proof.
$\hfill\Box$

\subsection{Proof of Lemma~\ref{lem:Vt}}
\label{appendix:emVt}
{Recall, $X_t$ follows the dynamics in (\ref{XtM}),
which we copy below for easy reference,
and hence is a (local) martingale under $\pr^M$:
\begin{equation*}
dX_t = \sigma_tX_t dB_t^M.
\end{equation*}
}
According to the definition of $A_{{t}}$,
{for all $u \ge t \ge 0$,}
\begin{eqnarray}
\label{Audecomp}
A_{{u}} = C_{{u}} + \tilde{\sigma} \tilde{B}_{{u}}
=C_t + \tilde{\sigma} \tilde{B}_t + C_{{u}}
-C_t +  \tilde{\sigma} (\tilde{B}_{{u}} -\tilde{B}_t)
=  A_t +  C_{{u}} -C_t +  \tilde{\sigma} (\tilde{B}_{{u}} -\tilde{B}_t).
\end{eqnarray}
{By the assumption that $C_u - C_t \in \sigma(\{X_s, \, t\leq s \leq u \})$ and $(X_t, \sigma_t)$ is Markovian under $\pr^M$,
the distribution of $C_u - C_t$ (conditional on $\calf_t$) is in turn determined by $(X_t, \sigma_t)$.
Then,
as already argued,
$(A_t, X_t, \sigma_t)$ is Markovian under $\pr^M$. }

We reiterate the projected production payoff process in (\ref{Vt}):
\begin{eqnarray*}
V_t(P,R) &=& \ex^M[H_T(P,R) \,|\, \calf_t]  \\
&=&  (R-bP)(P-c) - (P-s)\ex^M[(R - A_T)^+\,|\, \calf_t] \\
&=& (R-bP)(P-c) - (P-s)\ex^M[(R - A_T)^+\,|\, X_t , A_t ,  \sigma_t] \\
&=& (R-bP)(P-c) - (P-s)f(t,{X_t,A_t, \sigma_t}),
\end{eqnarray*}
where the third equality follows from {the Markovian property of $(A_t, X_t, \sigma_t)$}
and the {function in the last line is defined as follows:}
$$f(t, x,a, \sigma) := \ex^M[(R - A_T)^+ \,|\, X_t = x, A_t = a, \sigma_t =\sigma].$$
{Clearly}, $V_t$ is a martingale under $\pr^M$.
By Proposition~\ref{pro:mrt} and accounting for (\ref{XtM}),
$V_t(P,R)$  has the following representation,
with $V_0(P,R) = \ex^M[H_T(P,R)]$:
\begin{equation*}
V_t(P,R) = V_0(P,R) + \int_0^t {\psi_s(P,R)} d{B_t^M} + \int_0^t \delta_s(P,R)d\tilde B_s.
\end{equation*}
$\psi_t$ and $\delta_t$ are processes adapted to $\calf_t$
{and they lead to (\ref{GKW})} by applying It\^{o}'s lemma to $V_t(P, R)$:
\begin{eqnarray*}
dV_t(P, R) & = & -(P-s)df(t, x,a, \sigma)\\
&=& -(P-s) \{ f_tdt + f_xdX_t + f_adA_t + f_{\sigma}d\sigma_t + \frac{1}{2}f_{xx} x^2\sigma_t^2dt
+ \frac{1}{2} f_{aa}\tilde{\sigma}^2dt
+ \frac{1}{2} f_{\sigma \sigma}{(d\sigma_t)^2} \\
& &{+f_{xa}(dX_tdA_t) + f_{a\sigma}(dA_td\sigma_t)}
+ f_{x\sigma}(dX_td\sigma_t)\}\\
&=&{-(P-s)\{[...]dt + \psi_t dB_t^M + f_a\tilde\sigma d\tilde B_t\}} \\
&=&{ -(P-s)\Big\{\frac{\psi_t}{\sigma_t X_t}dX_t + f_a\tilde\sigma d\tilde B_t\Big\}.}
\end{eqnarray*}
{In the last line, the $dt$-term vanishes due to martingale property of $V_t$.
Furthermore, $dB_t^M$ is related to $dX_t$ based on the dynamics of $X_t$ in $\pr^M$; see (\ref{XtM}).}
Comparing {the last line} with the martingale representation of $V_t(P, R)$ {above} and matching the coefficient of the term $d \tilde{B}_t$ leads to:
\begin{equation*}
\delta_t(P,R) = -\tilde\sigma(P-s)f_a(t, X_t, A_t, \sigma_t).
\end{equation*}
{(Matching the $dX_t$-term leads to $\xi_t = \psi_t/(\sigma_t X_t)$).}

{Applying (\ref{Audecomp})},
\begin{eqnarray*}
f(t, x,a,\sigma) &=& \ex^M[(R - A_T)^+ \,|\, X_t = x, A_t = a, \sigma_t =\sigma] \\
&=&\ex^M\Big[(R - (a + C_T -C_t +\tilde{\sigma} (\tilde{B}_T -\tilde{B}_t)  ))^+ \,|\, X_t = x, A_t = a, \sigma_t =\sigma\Big] \\
&=&\ex^M\Big[(R - (a + C_T -C_t  +\tilde{\sigma}\sqrt{T-t}Z))^+ \,|\, X_t = x, A_t = a, \sigma_t =\sigma\Big]\\
&=&{\ex^M\Big[(R - (a + C_T -C_t  +\tilde{\sigma}\sqrt{T-t}Z))^+ \,|\, X_t = x,  \sigma_t =\sigma\Big]}
\end{eqnarray*}
where $Z= (\tilde{B}_T -\tilde{B}_t)/{\sqrt{T-t}}$ follows standard normal distribution independent
{of all other random variables.}
{For the last equality,
note that as argued above,
the distribution of $C_T - C_t$ (given $\calf_t$) depends on realization of $(X_t, \sigma_t)$ only.
In other words, the conditional distribution
$[C_T - C_t \,|\, X_t = x, A_t = a, \sigma_t = \sigma]$ is the same as $[C_T - C_t \,|\, X_t = x, \sigma_t = \sigma]$.}
{Therefore, we can}
take first derivative of $f(t, x,a,\sigma)$ with to $a$ as follows:
\begin{eqnarray*}
f_a(t, x,a,\sigma) &=&
- \ex^M\Big[\ind\{a + C_T -C_t  +\tilde{\sigma}\sqrt{T-t}Z\leq R\} \,|\, X_t = x,  \sigma_t =\sigma\Big] \\
&=&-\pr^M(A_T \leq R \,|\,X_t = x, A_t = a, \sigma_t =\sigma)\\
&=&-\pr^M(A_T \leq R \,|\, \calf_t ).
\end{eqnarray*}
Therefore, we have:
$$\delta_t(P,R) = \tilde\sigma(P-s)\pr^M(A_T \leq R \,|\, \calf_t ).$$
Clearly, $\delta_t(P,R)$ increases in both $P$ and $R$.
$\hfill\Box$

\subsection{Proof of Theorem~\ref{thm:hedgingsol}}
\label{appendix:thmhedgingsol}
In this part, we apply the quadratic hedging technique in \cite{gourieroux1998mean} to solve the hedging problem.
The setup in \cite{gourieroux1998mean} is semimartingale-based and abstract, hence the solution is not as explicit as ours since our setup is based on Brownian motions.
In Section~\ref{appendix:mvgeneraltech}, we lay out technical preparations that are needed in the proof of Theorem~\ref{thm:hedgingsol} in Section~\ref{appendix:thmhedgingsol}.
\subsubsection{Technical Preparation}
\label{appendix:mvgeneraltech}
Recall, the risk-neutral measure $\pr^M$ is defined via the associated Radon-Nikodym (R-N) derivative:
$$
Z_T = \frac{d\pr^M}{d\pr}:= e^{-\int_0^T \eta_t dB_t - \frac{1}{2}\int_0^T \eta_t^2 dt},
$$
and $Z_t^M:= \ex^M(Z_T \,|\, \calf_t) = \ex(Z_T^2 \,|\, \calf_t)/{Z_t}$ in (\ref{genZtM}), and thus $Z_0^M = \ex(Z_T^2)$.
Note that $Z_t^M > 0$ since $Z_T > 0$.
We introduce a process $N_t$ and another probability measure $\pr^R$.
With Assumption~\ref{assumption:Z}, $\pr^R$ below is well-defined:
 \begin{equation}
\label{genpr}
\frac{d \pr^R}{d \pr} = \frac{\big(Z_T^M\big)^2}{Z_0^M}, \qquad \mbox{thus}\quad \frac{d \pr^R}{d \pr^M} = \frac{Z_T^M}{Z_0^M}.
\end{equation}
Note that $Z_t^M$ is a positive $\pr^M$-martingale,
and by Corollary~\ref{cor:calfg},
it admits the following martingale representation:
\begin{equation}
\label{genpsi}
dZ_t^M = \zeta_t dX_t^M = \zeta_t \sigma_tX_t dB_t^M,
\end{equation}
where $\zeta_t$ is an adapted process to $\calg_t$
(recall, $\calg_t$ is the filtration generated by $B_t$, hence independent {of} $\tilde B_t$).
$B_t^M = \int_0^t\eta_s dt +  B_t$, which is a Brownian motion under $\pr^M$.
The second equality in (\ref{genpsi}) follows from (\ref{XtM}).
Define:
\begin{equation}
\label{genzetapsi}
\psi_t := \sigma_t X_t \zeta_t.
\end{equation}
 Now, from (\ref{genpr}), $Z_t^M/ Z_0^M$ is the density process associated with $d\pr^R/{d \pr^M}$.
 Then, applying Girsanov's Theorem and accounting for (\ref{genpsi}),
 the MPR process associated with $d\pr^R / {d\pr^M}$ is
\begin{equation}
\label{etaM}
\eta_t^M := -\frac{\psi_t}{Z_t^M}.
\end{equation}
In addition, $B_t^R$ defined below is a Brownian motion under $\pr^R$:
\begin{eqnarray}
\label{genBtR}
dB_t^R := dB_t^M + \eta_t^M dt = dB_t^M - \frac{\psi_t}{Z_t^M}dt.
\end{eqnarray}
Next, we introduce $N_t = (N_t^0, N_t^1)$ as follows:
\begin{equation}
N_t^0 := \frac{1}{Z_t^M}, \qquad N_t^1:= \frac{X_t}{Z_t^M}. \label{genNt}
\end{equation}
Recall, $N_t^0$ and $N_t^1$ are interpreted as the original assets, dollar (i.e. $1$) and financial asset ($X_t$), {\it denominated} in $Z_t^M$.
By change of measure and Jensen's inequality,
$$Z_t^M = \frac{1}{Z_t}\ex(Z_T^2 \,|\, \calf_t) \ge \frac{1}{Z_t}\ex^2(Z_T \,|\, \calf_t) = Z_t > 0, $$
with $Z_t$ defined in (\ref{genZt}). By Assumption (\ref{assumption:X}), $X_t > 0$. Therefore, $N_t^0$ and $N_t^1$ are well-defined, and both are strictly positive.

Applying It\^{o}'s lemma and accounting for (\ref{genBtR}):
\begin{eqnarray}
\label{gendNt}
dN_t^0 &=&   -(N_t^0)^2\psi_t[dB_t^M - \psi_t N_t^0dt] = -(N_t^0)^2\psi_t dB_t^R \nonumber,\\
dN_t^1 &=&    -N_t^0[N_t^1\psi_t - \sigma_tX_t][[dB_t^M - \psi_t N_t^0dt]] =-N_t^0[N_t^1\psi_t - \sigma_tX_t]dB_t^R.
\end{eqnarray}
\noindent
Clearly, both $N_t^0$ and $N_t^1$ are {\it local martingales} under $\pr^R$;
being nonnegative, they are also {\it supermartingales}.
It is easy to verify that they are indeed $\pr^R$-martingales by having constant means
(and being supermartingales):
\begin{eqnarray}
\ex^R(N_t^0) = \frac{1}{Z_0^M}\ex^M\Big[Z_T^M\frac{1}{Z_t^M}\Big] \nonumber
			= \frac{1}{Z_0^M}\ex^M\Big[\frac{1}{Z_t^M} \ex^M(Z_T^M | \calf_t)\Big] \nonumber
			= \frac{1}{Z_0^M};
\end{eqnarray}
the first equality is change of measure using (\ref{genpr}),
the second one uses iterated conditioning on $\calf_t$ and the fact that $Z_t^M$ is adapted to $\calf_t$.
The last equality uses the definition of $Z_t^M$ in (\ref{genZtM}), noting $Z_T = Z_T^M$.
Analogous verification can be applied to $N_t^1$:
$$\ex^R(N_t^1) = \ex^M\Big[\frac{Z_t^M}{Z_0^M}\frac{X_t}{Z_t^M}\Big] = \frac{X_0}{Z_0^M} = N_0^1;$$
\noindent the first equality uses the fact that $Z_t^M / Z_0^M$ is the density process for $d\pr^R / d\pr^M$ and $N_t^1$ is adapted to $\calf_t$, and the second equality is by martingale property of $X_t$. We summarize the analysis above into the following lemma.
\begin{lemma}
\label{lem:genNtmg}
{
$N_t^0$ and $N_t^1$ in (\ref{genNt}) are both martingales under $\pr^R$.
}
\end{lemma}

\medskip
Now, we are ready to define $\calm_X$, $\calm_N$, $\cala_X$ and $\cala_N$, all of which are technically crucial in defining {\it admissible class} of hedging strategies.
($\sim$ stands for the equivalence between probability measures.)
\begin{equation}
\label{genlocXM}
\calm_X := \Big\{\pr^{\bar M} \sim \pr: \frac{d\pr^{\bar M}}{d\pr} \in L_2(\pr), \, X_t \mbox{ is a $\pr^{\bar M}$-martingale} \Big\} .
\end{equation}
\noindent $\calm_X$ contains the equivalent martingale measures that have square-integrable R-N derivatives. By Assumption~\ref{assumption:X}, $\pr^M \in \calm_X$, hence $\calm_X \neq \emptyset $.
Similarly for $N_t$, we define %
\begin{equation}
\label{genlocNM}
\calm_N := \Big\{\pr^{\bar R} \sim P: \, \frac{1}{Z_T^M}\frac{d\pr^{\bar R}}{d\pr} \in L_2(\pr),\ N_t^0 \mbox{ and } N_t^1 \mbox{ are } \pr^{\bar R}\mbox{-martingales}  \Big\}.
\end{equation}
\noindent It is straightforward to verify that for $\pr^R$ defined in (\ref{genpr}), $\frac{1}{Z_T^M}\frac{d\pr^R}{d\pr} \in L_2(\pr)$, and together with Lemma~\ref{lem:genNtmg} this implies $\pr^R \in \calm_N$, hence $\calm_N \neq \emptyset$. %

Based on $\calm_X$ and $\calm_N$, we define admissible classes of trading strategies.
We start with $\cala_X$, the admissible class of hedging strategies in (\ref{mvprob}).
An adapted process $\vartheta = \{\theta_t, t\in[t,T]\}$ is admissible by belonging to the following set:
\begin{eqnarray}
\label{genadX}
\cala_X &:=& \{\vartheta :\, \vartheta \mbox{ is $X_t$-integrable}; \,  \chi_T(\vartheta) \in L_2(\pr);
\forall \pr^{\bar M} \in \calm_X, \, \{\chi_t(\vartheta), t\in [0,T]\} \mbox{ is  a } \pr^{\bar M}\mbox{-martingale}\}. \nonumber\\
\end{eqnarray}
\noindent
(Recall, $\vartheta = \{\theta_t, t\in[0,T]\}$ and $\chi_t(\vartheta) = \int_0^t \theta_s dX_s$.)
Next, we define the set of all terminal wealth attainable by admissible trading strategies:
\begin{equation}
\label{gentermwealth}
\chi_T(\cala_X) :=\{\chi_T(\vartheta) \,|\, \vartheta \in \cala_X \}.
\end{equation}
We remark that $\chi_T(\cala_X)$ is closed in $L^2(\pr)$;
see Lemma~2.6 and Theorem~2.8 of \cite{vcerny2008mean};
and for a brief review on this, see Theorem~A.1 of \cite{wang2013mean}.
This property of $\chi_T(\cala_X)$ allows us to establish the following technical result,
with proof collected in Section~\ref{appendix:genztmmg}.
\begin{lemma}
\label{lem:genztmmg}
{
Let $Z_t^M$ be defined in (\ref{genZtM}), with dynamics specified in (\ref{genZtM}) which is reiterated below:
$$dZ_t^M = \zeta_t dX_t.$$
Under Assumptions \ref{assumption:X} and \ref{assumption:Z}, $\zeta_t \in \cala_X$; in other words, $\zeta_t$ is an admissible hedging strategy with respect to $X_t$.
Hence, by definition of the set of admissible strategies $\cala_X$ in (\ref{genadX}), $Z_t^M$ is a $\pr^{\bar M}$-martingale for each $\pr^{\bar M}$ in $\calm_X$.
}
\end{lemma}
It will become clear later that Lemma~\ref{lem:genztmmg} is crucial in establishing connection between $\calm_X$ and $\calm_N$, which plays a key role in solving the quadratic hedging problem.

Next, recall that $N_t$ in (\ref{genNt}) can be viewed as asset prices  {\it denominated}  in $Z_t^M$,
hence we can also define admissible trading strategies with respect to $N_t$.
An adapted process $\varphi = \{\phi_t = (\phi_t^0,\phi_t^1), t\in[0,T]\}$ is admissible if it belongs to the following set:
\begin{equation}\label{genadN}
    \begin{aligned}
\cala_N := \Bigl\{\varphi :\,  & \varphi \mbox{ is } N_t\mbox{-integrable and }  \\
& Z_T^M\pi_T(\varphi) \in L_2(\pr), \;
 \forall \pr^{\bar R} \in {\cal M}_N, \, \{\pi_t(\varphi), t \in [0,T] \}\mbox{ is a }  \pr^{\bar R}\mbox{-martingale} \Bigr \},
\end{aligned}
\end{equation}
where the notation parallels those for $\cala_X$: $\varphi = \{\phi_t, t\in[0,T]\}$, with $\phi_t=
(\phi_t^0,\phi_t^1)$ being a two-dimensional adapted process.
And
\begin{equation}
\label{genpit}
\pi_t(\varphi) := \int_0^t \phi_s \cdot dN_s = \int_0^t\phi_s^0dN_s^0 + \int_0^t\phi_s^1dN_s^1.
\end{equation}
Similar to (\ref{gentermwealth}), we define $\pi_T({\cal A}_N) := \{\pi_T(\varphi)| \varphi \in {\cal A}_N\}$ to be the attainable terminal wealth by admissible strategies in ${\cal A}_N$.

Now, we establish bijection between $\calm_X$ in (\ref{genlocXM}) and $\calm_N$ in (\ref{genlocNM}),
which will be used later in proving the key lemma of this section. The following lemma is a special case of Proposition~3.1 in \cite{gourieroux1998mean}, and our proof here will make explicit uses of Bayes formula based on Doob's martingale. %

\begin{lemma}
\label{lem:keystone}
{
Recall $\pr^M$ and $\pr^R$ are defined in (\ref{genZt}) and (\ref{genpr}), respectively.

(i) $\forall \pr^{\bar M} \in \calm_X$, the probability measure defined below is in $\calm_N$.
$$\frac{d\pr^{\bar R}}{d\pr} := \frac{d\pr^{\bar M}}{d\pr} \cdot \frac{Z_T^M}{Z_0^M} = \frac{d\pr^{\bar M}}{d\pr}\cdot \frac{d\pr^R}{d\pr^M}.$$

(ii) $\forall \pr^{\bar R} \in \calm_N$,the probability measure defined below is in $\calm_X$.
$$\frac{d\pr^{\bar M}}{d\pr} := \frac{d\pr^{\bar R}}{d\pr}\cdot \frac{Z_0^M}{Z_T^M} = \frac{d\pr^{\bar R}}{dP}\cdot \frac{1}{\frac{d\pr^R}{d\pr^M}}$$
}
\end{lemma}
\noindent The proof is to check the conditions specified in (\ref{genZtM}) and (\ref{genlocXM}) for each case respectively, and we collect the details in Section~\ref{appendix:genztmmg}.

To this point,
we are ready to present the key lemma of this section, which spells out an one-to-one relationship between the two admissible classes $\cala_X$ in (\ref{genadX}) and $\cala_N$ in (\ref{genadN}).
\begin{lemma}
\label{lem:genmvkeytech}
{
(i) For any given $X_t$-admissible trading strategy $\vartheta = \{\theta_t, \, t\in[0,T]\} \in {\cal A}_X$, there exists an $N_t$-admissible strategy $\varphi = \{\phi_t = (\phi_t^0, \phi_t^1), \, t\in[0,T]\} \in {\cal A}_N$,
such that $\forall t \in [0,T]$,
$$\frac{\chi_t(\vartheta)}{Z_t^M} = \pi_t(\varphi), \quad {\rm and} \qquad  \phi_t = (\chi_t(\vartheta)-\theta_t X_t , \, \theta_t) .$$

\noindent
(ii) Conversely,
given any $N_t$-admissible strategy $\varphi = \{\phi_t = (\phi_t^0, \phi_t^1), \, t\in[0,T] \} \in {\cal A}_N$, there exists an $X_t$-admissible trading strategy $\vartheta = \{\theta_t, \, t\in[0,T]\} \in {\cal A}_X$, such that  $\forall t \in [0,T]$,
$$\frac{\chi_t(\vartheta)}{Z_t^M} = \pi_t(\varphi), \quad  {\rm and}\qquad \theta_t = \zeta_t(\pi_t(\varphi)-\phi_t \cdot N_t) + \phi_t^1
\quad\mbox{ with } \; \zeta_t \; \mbox{ defined in (\ref{genZtM})}.$$

\noindent
(iii) Combining (i) and (ii), we have:
$$\pi_T({\cal A}_N) = \frac{\chi_T({\cal A}_X)}{Z_T^M} := \Big\{\frac{\chi_T(\vartheta)}{Z_T^M}: \, \vartheta \in {\cal A}_X \Big\} ;$$
\noindent recall, $\chi_T({\cal A}_X)$ is the set of attainable wealth defined in (\ref{gentermwealth}) and $\pi_T({\cal A}_N)$ is similarly defined.

}
\end{lemma}
{\startb Proof.}
We first show Part (i) of this lemma.
Given $\vartheta = \{\theta_t, t\in[0,T]\}$, write $\chi_t = \chi_t(\vartheta) = \int_0^t \theta_s dX_s $ for lighter notation. Apply It\^{o}'s lemma on $\frac{\chi_t}{Z_t^M}$, accounting for the dynamics of $Z_t^M$, $N_t^0$ and $N_t^1$ in, respectively, (\ref{genpsi}) and (\ref{gendNt}):
\begin{eqnarray}
d\chi_t N_t^0 &=& \chi_t dN_t^0 + N_t^0d\chi_t + d\chi_t dN_t^0 \nonumber\\
&=& \chi_t dN_t^0 + \theta_t N_t^0\sigma_tX_tdB_t^M - \theta_t\sigma_tX_tN_t^0(N_t^0\psi_t)dt \nonumber\\
&=&(\chi_t - \theta_t X_t)dN_t^0 + \Big[\theta_t X_t dN_t^0 + \theta_t N_t^0\sigma_tX_tdB_t^M - \theta_t\sigma_tX_tN_t^0(N_t^0\psi_t)dt \Big] \nonumber\\
&=& (\chi_t - \theta_t X_t)dN_t^0 + \Big[(\theta_t X_t)[-N_t^0(N_t^0\psi_t)dB_t^R] + \theta_t\sigma_tX_tN_t^0[dB_t^M - (N_t^0\psi_t)dt] \Big] \nonumber\\
&=& (\chi_t - \theta_t X_t)dN_t^0 + \theta_t\Big[-(N_t^0\psi_t)N_t^1 + N_t^0\sigma_tX_t\Big]dB_t^R \nonumber \\
&=& (\chi_t - \theta_t X_t)dN_t^0 + \theta_t dN_t^1 \nonumber\\
&=& \phi_t \cdot dN_t,
\end{eqnarray}
and this leads to the equality stated in (i).

The rest is to show $\varphi \in \cala_N$.
First,
$Z_T^M\int_0^T \phi_t \cdot dN_t = \chi_T(\vartheta) \in L^2(P)$, since $\vartheta \in {\cal A}_X$. Next, $\forall \pr^{\bar R} \in {\cal M}_N$, we have
$ \ex^{\bar R}\big[|\frac{\chi_T(\vartheta)}{Z_T^M}\big{|}\big] = \ex\Big [|\chi_T(\vartheta)|(\frac{d\pr^{\bar R}}{d\pr}/{Z_T^M})\Big] < \infty$, by Cauchy-Schwarz inequality, the fact $\frac{d\pr^{\bar R}}{d\pr}/{Z_T^M} \in L^2(\pr)$ (since $\pr^{\bar R} \in \calm_N$) and $\chi_T(\vartheta) \in L^2(\pr)$ (by $\vartheta \in \cala_X$). Then $\pi_T(\varphi) \in L^1(\pr^{\bar R})$ and we can take conditional expectation:
\begin{equation}
\ex^{\bar R}\Big[\frac{\chi_T(\vartheta)}{Z_T^M} \Big{|} \calf_t \Big] = \ex\Big[\chi_T(\vartheta)\frac{\frac{d\pr^{\bar R}}{d\pr}}{(Z_T^M/Z_0^M)} \Big{|} \calf_t \Big] \frac{1}{\ex[\frac{d\pr^{\bar R}}{d\pr}|\calf_t]Z_0^M}. \nonumber
\end{equation}

\noindent Applying Bayes formula:
\begin{eqnarray}
\ex\Big[\frac{\frac{d\pr^{\bar R}}{d\pr}}{(Z_T^M/Z_0^M)} \Big{|} \calf_t\Big] &=& Z_0^M\bigg[\ex\Big[\frac{d\pr^{\bar R}}{d\pr}N_T^0 \Big{|} \calf_t\Big]\frac{1}{\ex[\frac{d\pr^{\bar R}}{d\pr}|\calf_t]}\bigg] \ex\Big[\frac{d\pr^{\bar R}}{d\pr}\Big{|}\calf_t\Big] \nonumber\\
&=& Z_0^M\ex^{\bar R}[N_T^0|\calf_t]\ex\Big[\frac{d\pr^{\bar R}}{d\pr} \Big{|} \calf_t \Big] \nonumber\\
&=& Z_0^MN_t^0\ex\Big[\frac{d\pr^{\bar R}}{d\pr} \Big{|} \calf_t \Big]; \nonumber
\end{eqnarray}
\noindent the above accounts for the fact that $N_t^0$ is a $\pr^{\bar R}$-martingale.

\noindent Combining the above, we derive:
\begin{eqnarray*}
\ex^{\bar R}\Big[\frac{\chi_T(\vartheta)}{Z_T^M} \Big{|} \calf_t \Big] &=& \Bigg[\ex\Big[\chi_T(\vartheta)\frac{\frac{d\pr^{\bar R}}{d\pr}}{(Z_T^M/Z_0^M)} \Big{|} \calf_t \Big] \frac{1}{\ex\Big[\frac{\frac{d\pr^{\hat R}}{d\pr}}{Z_T^M/Z_0^M} \Big{|} \calf_t \Big]} \Bigg]N_t^0\\
&=& \ex^{\bar M}[\chi_T(\vartheta) | \calf_t]N_t^0 \\
&=&\frac{\chi_t(\vartheta)}{Z_t^M};
\end{eqnarray*}
where $\pr^{\bar M}$ is defined by:
$$\frac{d\pr^{\bar M}}{d\pr} := \frac{d\pr^{\bar R}}{d\pr}\frac{Z_0^M}{Z_T^M},$$
\noindent and by Lemma~\ref{lem:keystone}, $\frac{d\pr^{\bar M}}{d\pr} \in {\cal M}_X$. Then, since $\vartheta \in \cala_X$, $\chi_t(\vartheta)$ must be a martingale under $\pr^{\bar M}$, which gives the last equality in the derivation above. Then, the above implies that $\pi_t(\varphi) = \chi_t(\vartheta)/Z_t^M$ is a $\pr^{\bar R}$-martingale $\forall \pr^{\bar R} \in {\cal M}_N$, hence $\varphi \in {\cal A}_N$ and (i) is proved.

\medskip
Now we prove Part (ii).
Given $\varphi = \{\phi_t = (\phi_t^0, \phi_t^1), t \in[0,T] \} \in \cala_N$, we apply It\^{o}'s lemma on $Z_t^M\int_0^t\phi_s\cdot dN_s$:
\begin{eqnarray}
dZ_t^M\int_0^t\phi_s\cdot dN_s
&=& Z_t^M(\phi_t^0dN_t^0 + \phi_t^1dN_t^1) + \Big(\int_0^t\phi_s\cdot dN_s \Big)\psi_tdB_t^M + \phi_t^0 dZ_t^MdN_t^0 + \phi_t^1 dZ_t^MdN_t^1 \nonumber\\
&=&  Z_t^M\phi_t^0dN_t^0 + Z_t^M\phi_t^1dN_t^1 + \Big(\int_0^t\phi_s\cdot dN_s \Big)\psi_tdB_t^M - \phi_t^0(\psi_t N_t^0)^2dt \nonumber\\
&-& \phi_t^1(\psi_t N_t^0)[\psi_t N_t^1 - \sigma_tX_t]dt .
\end{eqnarray}
Now, use (\ref{gendNt}) to express $dN_t^0$ and $dN_t^1$ (and choose the representation involving $dB_t^M$), then it is straightforward to verify that $dt$-term vanishes and the equation above reduces to:
\begin{eqnarray}
dZ_t^M\int_0^t\phi_s\cdot dN_S
&=& \Big[-\phi_t^0 N_t^0 \psi_t - \phi_t^1(N_t^0X_t\psi_t - \sigma_t) + \psi_t\int_0^t\phi_s\cdot dN_s \Big]dB_t^M \nonumber\\
&=& \Big[-\phi_t^0 N_t^0 \zeta_t - \phi_t^1(N_t^0X_t\zeta_t - 1) + \zeta_t\int_0^t\phi_s \cdot dN_s \Big]\sigma_tdB_t^M \nonumber\\
&=&\Big[\zeta_t \Big(\int_0^t\phi_s \cdot dN_s - \phi_t^0 N_t^0 - \phi_t^1 N_t^1 \Big) + \phi_t^1\Big]dX_t,
\end{eqnarray}
where the second equality uses $N_t^1 = X_tN_t^0$, as well as the relation between $\zeta_t$ and $\psi_t$ in (\ref{genzetapsi}); the third equality uses the $\pr^M$-dynamics of $X_t$: $dX_t = \sigma_tX_tdB_t^M$. The integrand with respect to $dX_t$ in the last line above gives the expression for $\theta_t$ specified in (ii).

What remains is to show $\vartheta $ stated in (ii) is in $\cala_X$.
First, note $\chi_T(\vartheta) = Z_T^M\pi_T(\varphi)) \in L^2(P)$ follows from $\varphi \in \cala_N$.
Next, $\forall \pr^{\bar M} \in {\cal M}_X$, define $\pr^{\bar R}$ by
$$\frac{d\pr^{\bar R}}{d\pr} := \frac{d\pr^{\bar M}}{d\pr}\frac{Z_T^M}{Z_0^M}. $$
\noindent By Lemma~\ref{lem:keystone}, $\pr^{\bar R} \in {\cal M}_N$. Note $\ex^{\bar M}[|\chi_T(\vartheta)|] = \ex\Big[\frac{d\pr^{\bar M}}{d\pr} \Big{|} \pi_T(\varphi)Z_T^M\Big{|}\Big] < \infty$ is easily verified by using Cauchy-Schwarz inequality, $\frac{d\pr^{\bar M}}{d\pr} \in L^2(\pr)$ and $\pi_T(\varphi) Z_T^M \in L^2(\pr)$ (since $\varphi \in \cala_N$).
Hence we compute the following conditional expectation under $\pr^{\bar M}$ and apply Bayes formula to change the measure:
\begin{eqnarray*}
\ex^{\bar M}[Z_T^M\pi_T(\varphi) | \calf_t] &=& \ex\Big[(Z_T^M\pi_T(\varphi) \frac{d\pr^{\bar M}}{d\pr}) \Big{|} \calf_t\Big]\frac{1}{\ex[\frac{d\pr^{\bar M}}{d\pr}|\calf_t]}
\end{eqnarray*}
By Bayes formula,
\begin{eqnarray*}
\ex\Big[\frac{d\pr^{\bar R}}{d\pr} \Big{|} \calf_t \Big]
= \ex\Big[ \frac{d\pr^{\bar M}}{d\pr}\frac{Z_T^M}{Z_0^M} \Big{|} \calf_t \Big]
= \ex^{\bar M}\Big[\frac{Z_T^M}{Z_0^M} \Big{|} \calf_t \Big]\ex\Big[\frac{d\pr^{\bar M}}{d\pr} \Big{|}\calf_t\Big]
= \frac{Z_t^M}{Z_0^M}\ex\Big[\frac{d\pr^{\bar M}}{d\pr} \Big{|} \calf_t \Big]; \nonumber
\end{eqnarray*}
\noindent where the first quality uses definition of $\frac{d\pr^{\bar R}}{d\pr}$ above;
the second equality switches measure between $\pr^{\bar M}$ and $\pr$;
and the last equality is by the fact that $Z_t^M$ is a martingale under $\pr^{\bar M}$ as implied by
Lemma~\ref{lem:genztmmg}.

\noindent Combining all above,
\begin{eqnarray}
\ex^{\bar M}[Z_T^M\pi_T(\varphi) | \calf_t] &=& \ex\Big[Z_T^M\pi_T(\varphi) \frac{d\pr^{\bar M}}{d\pr} \Big{|} \calf_t \Big] \frac{1}{\ex\Big[ \frac{d\pr^{\bar M}}{d\pr} \Big{|} \calf_t \Big]} \nonumber\\
&=& \ex\Big[Z_T^M\pi_T(\varphi) \frac{d\pr^{\bar M}}{d\pr} \Big{|} \calf_t \Big]\frac{1}{\ex[\frac{d\pr^{\bar R}}{d\pr}|\calf_t]}\frac{Z_t^M}{Z_0^M} \nonumber\\
&=& \ex\Big[\frac{d\pr^{\bar R}}{d\pr}\pi_T(\varphi) \Big{|} \calf_t \Big]\frac{1}{\ex[\frac{d\pr^{\bar M}}{d\pr}|\calf_t]}Z_t^M \nonumber\\
&=& \ex^{\bar R}[\pi_T(\varphi)|\calg_t]Z_t^M \nonumber\\
&=&\pi_t(\varphi)Z_t^M  \nonumber%
\end{eqnarray}
\noindent where the last equality is due to $\pi_t(\varphi)$ is a $\pr^{\bar R}$-martingale implied by $\varphi \in \cala_N$ (recall, $\pr^{\bar R} \in {\cal M}_N$ by Lemma~\ref{lem:keystone}). This concludes that $\chi_t(\vartheta) = \pi_t(\varphi)Z_t^M$ is a $\pr^{\bar M}$-martingale $\forall \pr^{\bar M} \in {\cal M}_X$; hence $\vartheta$ defined in (ii) is in $\cala_X$. This completes the proof of (ii).
$\hfill\Box$

\subsubsection{Proof of Lemma~\ref{lem:genztmmg} and and \ref{lem:keystone}.}
\label{appendix:genztmmg}
We first prove Lemma~\ref{lem:genztmmg}
and here is an outline of the proof.
Recall, $\chi_T(\cala_X)$, the set of wealth attainable by admissible strategies with $X_t$ is introduced in (\ref{gentermwealth}),
which is a nonempty set closed in $L^2(\pr)$. Our approach is to first show $Z_T^M - Z_0^M \in \chi_T(\cala_X)$.
Once this is established, then $\exists \{\theta_t, t\in[0,T]\} \in \cala_X $ such that $Z_T^M - Z_0^M = \int_0^T \theta_t dX_t$, i.e., $\theta_t$ is an admissible strategy that attains $Z_T^M - Z_0^M$. Then, we have
\begin{eqnarray*}
\ex^M(Z_T^M - Z_0^M \,|\, \calf_t) &=& Z_t^M - Z_0^M
= \int_0^t \zeta_s dX_s
= \ex^M \Big(\int_0^T \theta_t dX_t \,\Big{|}\, \calf_t \Big)
= \int_0^t \theta_s dX_s;
\end{eqnarray*}
\noindent the first equality follows the definition of $Z_t^M$ in (\ref{genZtM}) (note $Z_T = Z_T^M$), the second equality uses the dynamics of $Z_t^M$ in (\ref{genZtM}); the third equality uses definition of $\theta_t$: an admissible strategy attaining $Z_T^M - Z_0^M$; the last equality uses the admissibility of $\theta_t$: the induced wealth process is an martingale under any measure from $\calm_X$ (defined in (\ref{genlocXM})), and in particular, recall that $\pr^M \in \calm_X$. In this way, we establish that $\int_0^t \zeta_s dX_s = \int_0^t \theta_s dX_s$; note both integrals are continuous martingales under $\pr^M$, hence $\zeta_t = \theta_t$. %
Thus, we can establish $\zeta_t \in \cala_X$, which in turn implies $Z_t^M = Z^0_M + \int_0^t \zeta_s dX_s$ is an martingale under any measure from $\calm_X$, and the lemma is proved.

Now we proceed with showing $Z_T^M-Z_0^M \in \chi_T(\cala_X)$. Recall, $\chi_T(\cala_X)$ is closed in $L^2(\pr)$, hence it is sufficient to find a sequence of elements in this set, with limit (in $L^2(\pr)$) as $Z_T^M - Z_0^M$, then the desired result will follow from the closedness. To do this, we follow the approach similar to that in the proof of Theorem~3.5 in \cite{wang2013mean}.

Define the sequence of stopping times with respect to $\calg_t$:
\begin{equation}
\label{gentau}
\tau_k := \inf \{t \ge 0: |\eta_t| \ge k \} \wedge T; \; k \in \mathbb{N};
\end{equation}
\noindent recall $\eta_t$ is the MPR process defined in (\ref{genZtM}). Since $\eta_t$ is continuous, we have $\tau_k \uparrow T$ as $k \to \infty$.
Clearly,
\begin{equation}
\label{genztau}
Z_{\tau_k} \to Z_T \quad a.s.
\end{equation}
\noindent Recall $Z_t$ is assumed to be a continuous square-integrable martingale under $\pr$ (by Assumption~\ref{assumption:Z}), hence Doob's $L^p$ inequality implies:
\begin{equation}
\label{gendoob}
\ex\bigg[\sup_{t\in [0, T]}\, Z_t^2 \bigg] \leq 2 \ex(Z_T^2) < \infty ;
\end{equation}
\noindent Clearly, $\sup_{k \in \mathbb{N}}\, Z_{\tau_k}^2 \leq \sup_{t\in [0, T]}\, Z_t^2$ (note $Z_t$ is a positive process), hence (\ref{gendoob}) implies
$$\ex\bigg[\sup_{k \in \mathbb{N}}\, Z_{\tau_k}^2 \bigg] < \infty .$$
\noindent The above invokes dominated convergence in (\ref{genztau}), and we establish
\begin{equation}
\label{genztaul2}
Z_{\tau_k} \to Z_T \quad\mbox{in}\quad L^2(\pr).
\end{equation}
Clearly, $Z_{\tau_k} \in L^2(\pr)$, hence also in $L^1(\pr^M)$.
So, for each $k \in \mathbb{N}$, define
\begin{eqnarray}
\label{genmk}
M_t^{(k)} := \ex^M(Z_{\tau_k} | \calf_t ) = \frac{1}{Z_t} \ex(Z_T Z_{\tau_k} \,|\, \calf_t)
		= M_0^{(k)} + \int_0^t \theta^{(k)}_s dX_s.
\end{eqnarray}
\noindent the first equality is the change of measure formula, and the second equality is martingale representation
with $\theta^{(k)}_t$ being a process adapted to $\calg_t$
(see Corollary~\ref{cor:calfg}, accounting for the fact that $Z_{\tau_k} \ge 0$ and $Z_{\tau_k} \in \calg_T$).
In particular, note $M_T^{(k)} = Z_{\tau_k}$, and also $M_0^{(k)} = \ex(Z_T Z_{\tau_k}) \to \ex(Z_T^2) = Z_0^M$ by (\ref{genztaul2}); so we have
\begin{equation}
\label{genmkconv}
M_T^{(k)} - M_0^{(k)} \;\to\; Z_T - Z_0^M = Z_T^M - Z_0^M \quad\mbox{in}\quad L^2(\pr); \nonumber\\
\end{equation}
Now we have found a sequence of elements, $M_T^{(k)} - M_0^{(k)}$, converging to $Z_T^M - Z_0^M$ in $L^2(\pr)$. As outlined, the next step is to show $M_T^{(k)} - M_0^{(k)} \in \chi_T(\cala_X)$ for each $k$.
To this end, fix $k$ and $\pr^{\bar M} \in \calm_X$, we will show $M_t^{(k)} - M_0^{(k)} = \int_0^t \theta^{(k)}_s dX_s $ defined in (\ref{genmk}) is a martingale in $\pr^{\bar M}$, as follows. Clearly, $\int_0^t \theta^{(k)}_s dX_s $ is a $\pr^{\bar M}$-local-martingale, since $X_t$ is a martingale under this probability measure. To proceed, the crux is to examine the following. Recall $Z_t$ has the exponential form as defined in (\ref{genZt}), hence
\begin{eqnarray}
Z_{t \wedge \tau_k} &=& \exp\Big\{-\int_0^{t \wedge \tau_k} \eta_s dB_s - \frac{1}{2}\int_0^{t \wedge \tau_k} \eta_s^2 ds\Big\} \nonumber\\
&=& \exp\Big\{-\int_0^t \eta_s \ind\{ s \leq \tau_k \} dB_s - \frac{1}{2}\int_0^t \eta_s^2 \ind\{ s \leq \tau_k \} ds \Big\} \nonumber\\
&=& \exp\Big\{-\int_0^t \hat\eta_s (dB_s^M - \eta_s ds) - \frac{1}{2}\int_0^t \hat\eta_s^2  ds \Big\} \nonumber\\
&=& \exp\Big\{-\int_0^t \hat\eta_s dB_s^M  - \frac{1}{2}\int_0^t \hat \eta_s^2  ds \Big\}\exp\Big\{\int_0^t \hat\eta_s^2  ds\Big\} \nonumber
\end{eqnarray}
\noindent where $B_t^M$ is the $\pr^M$-Brownian-motion ($B_t^M = \eta_t dt + d B_t$),
and $\hat \eta_s$ is defined as $\hat\eta_s = \eta_s\ind\{s \leq \tau_k \}$ (note $\hat \eta_s \eta_s = \hat \eta^2_s$);
also note $\hat\eta_s \leq k$ by definition of $\tau_k$ in (\ref{gentau}).
Denote
$$W_t = \exp\Big\{-\int_0^t \hat\eta_s dB_s^M  - \frac{1}{2}\int_0^t \hat \eta_s^2  ds \Big\} \quad\mbox{and}\quad C_t = \exp\Big\{\int_0^t \hat\eta_s^2 ds\Big\} ; $$
\noindent then the expression above for $Z_{t \wedge \tau_k}$ becomes
$$Z_{t \wedge \tau_k} = W_t C_t .$$
\noindent For $C_t$, since each $\hat\eta_s$ is bounded by $k$, we have
$$1 \leq C_t \leq e^{k^2 T}, \; \forall t\in[0,T].$$
\noindent The above immediately implies $W_t \leq Z_{t \wedge \tau_k}$.

Next, clearly $W_t$ is a local martingale undre $\pr^M$, and since $\hat\eta_t \leq k $, we have
$$\ex^M\Big[\frac{1}{2}\exp\Big\{\int_0^T \hat\eta_t^2 dt\Big\}\Big] \leq e^{\frac{1}{2}k^2T} < \infty ; $$
\noindent in other words, Novikov's condition holds, indicating that $W_t$ is a $\pr^M$-martingale.

Combining the above, we have
\begin{eqnarray}
0 &\leq & M_0^{(k)} + \int_0^t \theta^{(k)}_s dX_s = M_t^{(k)} \nonumber\\
&=& \ex^M(M_T^{(k)} \,|\, \calf_t ) = \ex^M(Z_{\tau_k} \,|\, \calf_t ) \nonumber\\
&=& \ex^M(Z_{T \wedge \tau_k} |\calf_t) = \ex^M(W_T C_T \,|\, \calf_t ) \nonumber\\
&\leq & e^{k^2T}W_t \leq  e^{k^2T} Z_{t \wedge \tau_k} \nonumber\\
&\leq & e^{k^2T} \sup_{t \in [0, T]} Z_t
\label{genlongeq1}
\end{eqnarray}
\noindent the first line is just the definition in (\ref{genmk}), and the second line uses $M_T^{(k)} = Z_{\tau_k}$; the first equality on the third line uses the obvious fact $\tau_k \wedge T = \tau_k$, and the second equality makes use of the representation of $Z_{t \wedge \tau_k}$ established above; the fourth line is based on the bound on $C_T$ and the martingale property of $W_t$ established above, as well as $W_t \leq Z_{t \wedge \tau_k}$ as shown above.

Now, combining (\ref{genlongeq1}) and (\ref{gendoob}) implies $\sup_{t \in [0,T]} \int_0^t \theta^{(k)}_sdX_s \in L^2(\pr)$.
Finally, by Cauchy-Schwarz inequality and $d\pr^{\bar M} / d\pr \in L^2(\pr)$
\begin{eqnarray}
\ex^{\bar M}\Big[\Big{|}\sup_{t \in [0,T]} \int_0^t \theta^{(k)}_s dX_s \Big{|}\Big] &=& \ex\Big[\frac{d\pr^{\bar M}}{d\pr}\Big{|}\sup_{t \in [0,T]} \int_0^t \theta^{(k)}_s dX_s \Big{|}\Big] \nonumber
\leq  \ex\Big[\Big(\frac{d\pr^{\bar M}}{d\pr}\Big)^2\Big]\ex\Big[\Big(\sup_{t \in [0,T]} \int_0^t \theta^{(k)}_s dX_s\Big)^2\Big] < \infty .
\end{eqnarray}
Therefore, $\int_0^t \theta^{(k)}_s dX_s = M_t^{(k)} - M_0^k $ is a $\pr^{\bar M}$-martingale
\citep{phambook}.
This establishes the desired result, and proves Lemma~\ref{lem:genztmmg}.

\medskip
Based on Lemma~\ref{lem:genztmmg}, we are now ready to prove Lemma~\ref{lem:keystone}.

For Part (i), suppose $\pr^{\bar M} \in \calm_X$ is given and $\pr^{\bar R}$ follows the stated definition.
First note that clearly $\frac{d\pr^{\bar R}}{d\pr} > 0$ almost surely since $\frac{d\pr^{\bar M}}{d\pr} > 0$ and $Z_T^M = Z_T$ takes exponential form (see (\ref{genZt})); hence $\frac{d\pr^{\bar R}}{d\pr} \sim \pr$. Next, we have
$$\ex\Big[\frac{d\pr^{\bar R}}{d\pr}\Big] = \ex\Big[\frac{d\pr^{\bar M}}{d\pr} \frac{Z_T^M}{Z_0^M}\Big] = \ex^{\bar M}\Big[\frac{Z_T^M}{Z_0^M}\Big] = 1.$$
\noindent The last equality follows from Lemma~\ref{lem:genztmmg}, which indicates that $Z_t^M$ is a $\pr^{\bar M}$-martingale since $\pr^{\bar M} \in \calm_X$ (see Lemma~\ref{lem:genztmmg}).
Next, derive
$$\ex\Big[\Big(\frac{1}{Z_T^M}\frac{d\pr^{\bar R}}{d\pr}\Big)^2\Big] = \ex\Big[\Big(\frac{1}{Z_T^M}\frac{d\pr^{\bar M}}{d\pr}\frac{Z_T^M}{Z_0^M}\Big)^2\Big] = \Big(\frac{1}{Z_0^M}\Big)^2 \ex\Big[\Big(\frac{d\pr^{\bar M}}{d\pr}\Big)^2\Big] < \infty.$$
\noindent The $<$ follows from the fact that $\pr^{\bar M} \in \calm_X$; hence the above implies $\frac{1}{Z_T^M}\frac{d\pr^{\bar R}}{d\pr} \in L^2(\pr)$.

\noindent What remains is to show that $N_t^0$ and $N_t^1$ are $\pr^{\bar R}$-martingales. First note
$$\ex^{\bar R}(N_T^0) = \ex\Big(\frac{d\pr^{\bar M}}{d\pr}\frac{Z_T^M}{Z_0^M}\frac{1}{Z_T^M}\Big) = \frac{1}{Z_0^M} < \infty;$$
\noindent hence we can apply conditional expectation and compute
\begin{eqnarray}
\ex^{\bar R}(N_T^0 \,|\, \calf_t) &=& \ex\Big[\frac{d\pr^{\bar M}}{d\pr}\frac{Z_T^M}{Z_0^M}\frac{1}{Z_T^M} \Big{\vert} \calf_t\Big]\frac{1}{\ex\Big(\frac{d\pr^{\bar M}}{d\pr}\frac{Z_T^M}{Z_0^M} \,\Big{\vert}\, \calf_t \Big)} \nonumber\\
&=& \ex\Big[\frac{d\pr^{\bar M}}{d\pr}\Big{\vert} \calf_t\Big]\frac{1}{\ex^{\bar M}(Z_T^M \,|\, \calf_t)\ex\Big[\frac{d\pr^{\bar M}}{d\pr}\,\Big{\vert}\, \calf_t\Big]} \nonumber\\
&=&\frac{1}{Z_t^M} = N_t^0; \nonumber
\end{eqnarray}
\noindent the first equality applies change of measure from $\pr^{\bar R}$ to $\pr$, and the second equality follows from changing measure from $\pr$ to $\pr^{\bar M}$ on the term $\ex\Big(\frac{d\pr^{\bar M}}{d\pr}\frac{Z_T^M}{Z_0^M} \,\Big{\vert}\, \calf_t \Big)$; the third equality again uses the fact that $Z_t^M$ is $\pr^{\bar M}$-martingale based on Lemma~\ref{lem:genztmmg}.
From above, we can conclude that $N_t^0$ is a martingale under $\pr^{\bar R}$. Similar derivation applies to $N_t^1$ as follows. First note that
$$\ex^{\bar R}(N_T^1) = \ex\Big(\frac{d\pr^{\bar M}}{d\pr}\frac{Z_T^M}{Z_0^M}\frac{X_T}{Z_T^M}\Big) = \frac{1}{Z_0^M}\ex^{\bar M}(X_T) = \frac{X_0}{Z_0^M} < \infty;$$
the last equality accounts for the fact that $\pr^{\bar M}$ is a martingale measure with respect to $X_t$.
Now we can compute the conditional expectation:
\begin{eqnarray}
\ex^{\bar R}(N_T^1 \,|\, \calf_t) &=& \ex\Big(\frac{d\pr^{\bar M}}{d\pr}\frac{Z_T^M}{Z_0^M} \frac{X_T}{Z_T^M} \,\Big{\vert}\, \calf_t \Big)\frac{1}{\ex\Big(\frac{d\pr^{\bar M}}{d\pr}\frac{Z_T^M}{Z_0^M} \Big{\vert} \calf_t \Big)} \nonumber\\
&=& \ex^{\bar M}(X_T \,|\, \calf_t)\ex\Big(\frac{d\pr^{\bar M}}{d\pr} \,\Big{\vert}\, \calf_t \Big) \frac{1}{\ex^{\bar M}(Z_T^M  \,\vert\, \calf_t)\ex\Big(\frac{d\pr^{\bar M}}{d\pr} \,\Big{\vert}\, \calf_t \Big)} \nonumber\\
&=& \frac{X_t}{Z_t^M} = N_t^1;
\end{eqnarray}
\noindent the first equality applies change of measure formula on $\ex^{\bar R}(N_T^1 \,|\, \calf_t)$;
the second equality does the same for $\ex^{\bar M}(X_T \,|\, \calf_t)$ and $\ex^{\bar M}(Z_T^M  \,\vert\, \calf_t)$, respectively;
the third equality uses the fact that $Z_t^M$ is a $\pr^{\bar M}$-martingale (by Lemma~\ref{lem:genztmmg}).
Hence, $N_t^1$ is also a martingale under $\pr^{\bar M}$ by the derivation above. To this point, we have checked that $\pr^{\bar R}$ satisfies conditions specified in $\calm_N$, hence belongs to this set; this proves (i).

\medskip
For Part (ii), the proof is analogous. Suppose $\pr^{\bar R} \in \calm_N$ is given, and define $\pr^{\bar M}$ as stated. First note by the same argument as that for Part (i),
$\pr^{\bar M} > 0$ almost surely, hence equivalent to $\pr$. And,
$$\ex\Big[\frac{d\pr^{\bar M}}{d\pr}\Big] = \ex\Big[\frac{d\pr^{\bar R}}{d\pr} \frac{Z_0^M}{Z_T^M}\Big] = Z_0^M \ex^{\bar R}(N_T^0) = Z_0^M N_0^0 = 1;$$
\noindent the third equality uses the fact that $\pr^{\bar R}$ is a martingale measure for $N_t^0$ and $N_t^1$. Next, check
$$\ex\Big[\Big(\frac{d\pr^{\bar M}}{d\pr}\Big)^2\Big] = (Z_0^M)^2\ex\Big[\Big(\frac{d\pr^{\bar R}}{d\pr}\cdot \frac{1}{Z_T^M}\Big)^2\Big] < \infty;$$
\noindent $<$ follows from $\frac{1}{Z_T^M}\frac{d\pr^{\bar R}}{d\pr} \in L^2(\pr)$; hence the above implies $\frac{d\pr^{\bar M}}{d\pr} \in L^(\pr)$.

Then, the rest is to show $X_t$ is a martingale under $\pr^{\bar M}$.
We start with checking the integrability condition:
$$\ex^{\bar M}(X_T) = \ex\Big(\frac{d\pr^{\bar R}}{d\pr}\frac{Z_0^M}{Z_T^M}X_T\Big) = Z_0^M\ex^{\bar R}(N_T^1) = Z_0^M N_0^1 = X_0 < \infty; $$
the third equality follows from that $\pr^{\bar R} \in \calm_N$ is a martingale measure for $N_t^0$ and $N_t^1$. Next, compute the conditional expectation
\begin{eqnarray}
\ex^{\bar M}(X_T | \calf_t) &=& \ex\Big[\frac{d\pr^{\bar R}}{d\pr}\frac{Z_0^M}{Z_T^M} X_T\Big{\vert}\calf_t \Big]\frac{1}{\ex\Big[\frac{d\pr^{\bar R}}{d\pr}\frac{Z_0^M}{Z_T^M}\Big{\vert}\calf_t \Big]} \nonumber\\
&=&  \ex\Big[\frac{d\pr^{\bar R}}{d\pr} N_T^1\Big{\vert}\calf_t \Big]\frac{1}{\ex\Big[\frac{d\pr^{\bar R}}{d\pr}N_T^0\Big{\vert}\calf_t \Big]} \nonumber\\
&=& \ex^{\bar R}(N_T^1 | \calf_t)\ex\Big[\frac{d\pr^{\bar R}}{d\pr} \Big{\vert}\calf_t \Big] \frac{1}{\ex\Big[\frac{d\pr^{\bar R}}{d\pr} \Big{\vert}\calf_t \Big]\ex^{\bar R}(N_T^0 | \calf_t) } \nonumber\\
&=& \frac{N_t^1}{N_t^0} = X_t;
\end{eqnarray}
\noindent the first equality applies change of measure formula on $\ex^{\bar M}(X_T | \calf_t)$;
the second equality does the same, respectively, for $ \ex^{\bar R}(N_T^1 | \calf_t)$ and $\ex^{\bar R}(N_T^0 | \calf_t) $; the third equality recognizes that $N_t^0$ and $N_t^1$ are martingales under $\pr^{\bar R}$. This concludes that $X_t$ is a $\pr^{\bar M}$-martingale and proves Part (ii). $\hfill\Box$

\subsubsection{Proof of Theorem~\ref{thm:hedgingsol}}
 \label{appendix:thmhedgingsol1}
In this section, we derive the optimal hedging strategy, $\theta_t^*$, of (\ref{opttheta}) as well as the associated minimum variance function $B(m,P,R)$ in (\ref{bmpr}), and thereby provide proofs to Theorem~\ref{thm:hedgingsol} ;
all derivations are based on results established in Section~\ref{appendix:mvgeneraltech}.

We will first transform the equality-constrained problem in (\ref{mvprob}) to an equivalent unconstrained {\it quadratic hedging} problem, and then solve the latter by applying the numeraire-based technique.
Define
\begin{eqnarray}
\label{genA}
A(\lambda) := \inf_{\vartheta \in {\cal A}_X} \ex\Big[ \Big(\lambda - H_T(P,R) - \chi_T (\vartheta) \Big)^2\Big],
\end{eqnarray}
where $A(\g)$ relates to $B(m,P,R)$ by
(see Proposition~6.6.5 in \cite{phambook}):
\begin{eqnarray}
\label{genBA}
B(m,P,R) =\max_\lambda \big[A(\lambda) -(m-\lambda)^2\big].
\end{eqnarray}
\noindent

In addition, the optimal hedging strategy induced by the problem in (\ref{genA}), with $\g$ being set as the optimal solution to the maximization problem (\ref{genBA}), is also optimal to the problem (\ref{mvprob}).
We will show $A(\g)$ takes the following expression:
\begin{eqnarray}
\label{genAexpression}
A(\lambda) = \frac{[\lambda - V_0(P,R)]^2}{Z_0^M} + \int_0^T\ex\Big[\frac{Z_t}{Z_t^M}\delta_t^2(P,R)\Big]dt ,
\end{eqnarray}
where $V_0(P,R)$ and $\delta_t(P,R)$ are terms involved in the martingale representation of $V_t(P,R)$ in (\ref{Vt}); in particular, $V_0(P,R)=\ex^M [H_T(P,R)]$, and $\delta_t(P,R)$ is defined in (\ref{delta}) . $Z_t$ and $Z_t^M$ follow (\ref{genZt}) and (\ref{genZtM}), respectively.

In (\ref{genAexpression}), $\g$ only enters the first component as a quadratic term; the second component is independent {of} $\g$. Thus, the minimization problem in (\ref{genBA}) has a quadratic objective function, since both $A(\g)$ expressed in (\ref{genAexpression}) and $(m-\lambda)^2$ are quadratic functions in $\g$, and it is straightforward to verify that the $\g$ specified in (\ref{opttheta}) solves the right hand side of (\ref{genBA}) and gives the expression of $B(m,P,R)$ in (\ref{bmpr}).

\medskip
Starting from here, we begin to prove (\ref{genAexpression}) by deriving solution to the hedging problem in (\ref{genA}). Write $\hat H_T(\lambda) := \lambda - H_T$, and start with definition of $A(\g)$ in (\ref{genA}):
\begin{eqnarray}
\label{genAR}
A(\lambda) &=& \inf_{\vartheta \in \cala_X} \ex \Big[\Big(\hat H_T(\lambda) - \chi_T (\vartheta) \Big)^2\Big]
\nonumber \\
  &=& \inf_{\vartheta \in \cala_X} Z_0^M  \ex\Big[\frac{(Z_T^M)^2}{Z_0^M}\Big(\frac{\hat H_T(\lambda)}{Z_T^M}-
  \frac{\chi_T(\vartheta)}{Z_T^M} \Big)^2\Big]
\nonumber \\
    &=&  Z_0^M \inf_{\vartheta \in \cala_X} \ex^R \Big[\Big(\frac{\hat H_T(\lambda)}{Z_T^M}-
  \frac{\chi_T(\vartheta)}{Z_T^M} \Big)^2\Big];
\end{eqnarray}
\noindent recall the probability measure $\pr$ is defined in (\ref{genpr}).

Continue with (\ref{genAR}), by Lemma~\ref{lem:genmvkeytech} Part (iii),
\begin{eqnarray}
\label{genAR1}
A(\lambda) = Z_0^M \inf_{\varphi\in{\cal A}_N}\ex^R\bigg[\bigg(\frac{\hat H_T(\lambda)}{Z_T^M} - \int_0^T \phi_t\cdot dN_t\bigg)^2\bigg].
\end{eqnarray}

A natural next step is to express $\frac{\hat H_T(\lambda)}{Z_T^M}$ as a stochastic integral with respect $N_t$, so that we can choose the optimal $\phi_t$ based on this representation. By the fact that $\hat H_T(\g)$ is bounded hence has finite second moment in $\pr$, it is easy to check that $\hat H_T(\g) / Z_T^M$ has finite second moment in $\pr^R$. Hence, we are able to define the Doob's martingale:
\begin{equation}
\label{genhatM}
\hat M_t := \ex^R\bigg[\frac{\hat H_T(\lambda)}{Z_T^M}\,\bigg{|}\,\calf_t\bigg];
\end{equation}
note $\frac{\hat H_T(\lambda)}{Z_T^M} = \hat M_T$. Furthermore, $\hat M_t$ is a square-integrable martingale under $\pr^R$.
Applying Bayes formula, we have
\begin{eqnarray*}
\hat M_t = \ex^M\bigg[\frac{\hat H_T(\lambda)}{Z_T^M}\frac{Z_T^M}{Z_t^M}\,\bigg{|}\,\calf_t\bigg]
= N_t^0 \ex^M[\hat H_T(\lambda) \,|\, \calf_t].
\end{eqnarray*}
Let $M_t:= \ex^M[\hat H_T(\lambda) \,|\, \calf_t]$,
which is clearly a martingale under $\pr^M$.
By Proposition~\ref{pro:mrt} and (\ref{XtM}), $M_t$ admits the martingale representation:
$$dM_t = a_t dB_t^M + b_t d\tilde B_t$$
for some adapted processes $\{a_t\}$ and $\{b_t\}$.
Next, apply It\^{o}'s Lemma on $\hat M_t$, accounting for the dynamics of $N_t^0$ and $N_t^1$ in (\ref{gendNt}) and the definition of $\eta_t^M$ in (\ref{etaM}):
\begin{eqnarray*}
d\hat M_t &=& dN_t^0M_t = N_t^0dM_t + M_tdN_t^0 + dM_tdN_t^0 \\
          &=& N_t^0a_t dB_t^M + N_t^0 b_t d\tilde B_t + M_tdN_t^0 - (N_t^0)^2a_t\psi_tdt \\
          &=& N_t^0a_t[dB_t^M - N_t^0\psi_tdt] + N_t^0b_td\tilde B_t + M_t dN_t^0 \\
          &=& N_t^0a_t[dB_t^M + \eta_t^Mdt] + N_t^0b_td\tilde B_t + M_t dN_t^0 \\
          &=& N_t^0a_tdB_t^R + N_t^0b_td\tilde B_t + M_t dN_t^0.
\end{eqnarray*}
From the dynamics of $N_t^0$ and $N_t^1$ in (\ref{gendNt}) and accounting for the fact that
$N_t^1 = N_t^0X_t$, it is straightforward to verify the following:
$$\sigma_tN_t^1dB_t^R = dN_t^1 - X_t dN_t^0.$$
In particular, note that $\sigma_t >0$ and $N_t^1 = X_t/Z_t^M > 0$.
Substitute this equation to the dynamics of $\hat M_t$ above, we have:
$$d\hat M_t = N_t^0b_td\tilde B_t + M_t dN_t^0 + \frac{N_t^0a_t}{\sigma_tN_t^1}(dN_t^1 - X_tdN_t^0)
= \Big(M_t - \frac{a_t}{\sigma_t}\Big)dN_t^0 + \frac{a_t}{\sigma_t X_t}dN_t^1 + N_t^0b_td\tilde B_t.$$
Therefore, $\hat M_t$ admits the following representation:
\begin{equation}
\label{genhatMGKW}
\hat M_t = \ex^R\bigg[\frac{\hat H_T(\lambda)}{Z_T^M}\bigg] + \int_0^t \phi_s^H \cdot dN_s + \int_0^t \gamma_s d\tilde B_s,
\end{equation}
where $\phi_t^H$ and $\gamma_t$ are some adapted processes.
In particular, note that $\tilde B_t$ is a Brownian motion under $\pr^R$ since the market risk of price process associated with it is $0$.

Substitute (\ref{genhatMGKW}) to (\ref{genAR1}) and take into the consideration that $N_t$ and $\tilde B_t$ are independent,
we can further expand $A(\lambda)$:
\begin{eqnarray}
\label{genAR2}
&&A(\lambda)%
=Z_0^M \inf_{\varphi \in {\cal A}_N} \ex^R \Big[\Big(\ex^R\big[\frac{\hat H_T(\lambda)}{Z_T^M}\big] + \int_0^T \gamma_td\tilde B_t
+\int_0^T \phi^H_t\cdot dN_t  - \int_0^T \phi_t \cdot dN_t \Big)^2\Big]
  \nonumber\\
&=& Z_0^M\inf_{\varphi \in {\cal A}_N} \ex^R \Big[\Big(\ex^R\big[\frac{\hat H_T(\lambda)}{Z_T^M}\big] + \int_0^T \gamma_td\tilde B_t \Big)^2\Big]
+\ex^R \Big[\Big(\int_0^T \phi^H_t \cdot dN_t  - \int_0^T \phi_t\cdot dN_t \Big)^2\Big]. \nonumber\\
\end{eqnarray}
\noindent To reach the second line of above, note the cross term is zero, following the fact that $\int_0^t \gamma_s d\tilde B_s$, $\int_0^t \phi^H_s \cdot dN_s$ and $\int_0^t \phi_s\cdot dN_s$ are all square-integrable martingales under $\pr^R$. Specifically, for $\int_0^t \gamma_s d\tilde B_s$ and $\int_0^t \phi^H_s \cdot dN_s$, this follows from (\ref{genhatMGKW}) and that $\hat M_t$ is a $\pr^R$-square-integrable martingale. For $\int_0^t \phi_s\cdot dN_s$, this follows from the $N_t$-admissibility of $\phi_t$: $\int_0^t \phi_s\cdot dN_s$ is $\pr^R$-martingale and $\int_0^T \phi_t\cdot dN_t \in L^2(\pr^R)$ (since $Z_T^M \int_0^T \phi_t\cdot dN_t \in L^2(\pr)$). Hence, $\int_0^t \gamma_s d\tilde B_s$ and $\int_0^t (\phi^H_s - \phi_s)\cdot dN_s$ are two square-integrable martingales under $\pr^R$, and they are independent since $N_t$ is adapted to $\sigma(B_t)$, which is independent {of} $\tilde B_t$. This
makes the cross term vanish.

Two results follow (\ref{genAR2}). First, it is obvious that the optimal $\phi_t$ of this problem, denoted by $\phi_t^* = (\phi_t^{*0}, \phi_t^{*1})$, should be set as:
\begin{equation}
\label{genoptPhi}
\phi_t^* = \phi_t^H;
\end{equation}
\noindent substituting this to (\ref{genAR2}), apply It\^{o}'s isometry and switch measure from $\pr^R$ to $\pr$, we have:
\begin{eqnarray}
\label{genAR3}
A(\lambda) &=& Z_0^M \ex^R\Big[\Big(\ex^R\big[\frac{\hat H_T(\lambda)}{Z_T^M}\big] + \int_0^T \gamma_td\tilde B_t \Big)^2\Big] \nonumber\\
&=&Z_0^M\Big(\ex^M\Big[\frac{\hat H_T(\lambda)}{Z_T^M} \frac{Z_T^M}{Z_0^M}\Big]\Big)^2 + Z_0^M\int_0^T \ex^M\Big[\gamma_t^2\cdot\frac{Z_t^M}{Z_0^M}\Big]dt \nonumber\\
&=& \frac{(\lambda - V_0)^2}{Z_0^M} + \int_0^T \ex\big(\gamma_t^2 Z_t Z_t^M\big)dt.
\end{eqnarray}
\noindent The expression above for $A(\g)$ does not coincide with (\ref{genAexpression}) yet
(the second term takes a different form);
we will come back to this later.

The other result from (\ref{genAR2}) is the expression for the optimal hedging strategy $\theta_t^*$. With (\ref{genoptPhi}), invoking Part (ii) of Lemma~\ref{lem:genmvkeytech}, we have the expression for $\theta_t^*$:
\begin{equation}
\label{gentheta1}
\theta_t^* = \zeta_t(\int_0^t \phi_s^* \cdot dN_s - \phi_t^*\cdot N_t) + \phi_t^{*1},
\end{equation}
and recall, $\zeta_t$ is defined in (\ref{genZtM}).

\medskip
To this end, we have obtained the expressions for both $A(\lambda)$ and $\theta_t^*$, in (\ref{genAR3}) and (\ref{gentheta1}) respectively, and both expressions involve terms related to $N_t$. Next, we will replace such terms by terms associated with $X_t$ and $V_t$ only.
To do so, the crux is to compare (\ref{Vt}) and (\ref{genhatMGKW}) and match integrands for $dt$, $dB_t^M$ and $d\tilde B_t$. Start with
\begin{equation}
\label{genmatch1}
\hat M_t = \frac{\lambda - V_0}{Z_0^M} + \int_0^t \phi_s^* \cdot dN_s + \int_0^t \gamma_s d\tilde B_s \nonumber,
\end{equation}
and this equation comes directly from (\ref{genhatMGKW}), changing measure from $\pr^R$ to $\pr^M$ for the first term, and accounting for $\phi_t^* = \phi^H$. Alternatively, $\hat M_t$ can also be represented as the following by applying change of measure formula:
\begin{eqnarray}
\label{genmatch2}
\hat M_t &=& \ex^M\Big[\frac{\lambda - H_T}{Z_T^M}\cdot \frac{Z_T^M}{Z_0^M} \, \Big{\vert} \, \calf_t\Big]\cdot \frac{1}{\ex^M\Big[\frac{Z_T^M}{Z_0^M} \, | \, \calf_t\Big]}
= (\lambda - V_t)N_t^0 \nonumber\\
&=& \big[\lambda - V_0 - \int_0^t \xi_s dX_s- \int_0^t\delta_sd\tilde B_s\big]N_t^0,
\end{eqnarray}
where the first equality switches the measure from $\pr^R$ to $\pr^M$; the second equality accounts for the fact $Z_T = Z_T^M$ and definition of $N_t^0$ in (\ref{gendNt}); the third equality makes use of the martingale representation of $V_t$ in (\ref{Vt}). Now, apply It\^{o}'s lemma on both (\ref{genmatch1}) and (\ref{genmatch2}), use the dynamics for $N_t$ in (\ref{gendNt})), and match the $d\tilde B_t$ term, we have:
\begin{equation}
\label{gengammat}
\gamma_t = -\frac{\delta_t}{Z_t^M}.
\end{equation}
Substituting (\ref{gengammat}) in (\ref{genAR3}) gives (\ref{genAexpression}), as desired.

Next, match $dB_t^M$ (matching for $dt$ term gives the same result) and obtain:
\begin{equation}
\label{genthetamatch}
\xi_t + (\lambda - V_t)N_t^0\zeta_t = \zeta_t(\phi_t^*\cdot N_t) - \phi^{*1}_t .
\end{equation}
\noindent Recall, $\zeta_t$ and $\psi_t$ are defined in (\ref{genZtM}) and (\ref{genpsi}) respectively, and to reach (\ref{genthetamatch}) the relation in (\ref{genzetapsi}) is used.
Substituting (\ref{genthetamatch}) to (\ref{gentheta1}) gives (\ref{opttheta}), taking into the following fact implied by Lemma~\ref{lem:genmvkeytech} Part (ii):
$$\frac{\int_0^t\theta^*_sdX_s}{Z_t^M} = \int_0^t\phi_s^*\cdot dN_s.$$

Now, we show that $\theta_t^*$ above is an admissible trading strategy. By Part (ii) of Lemma~\ref{lem:genmvkeytech}, it is sufficient to show $\phi_t^*$ in (\ref{genoptPhi}) is in $\cala_N$. Note $\phi_t^* = \phi_t^H$. Denote $\hat M_t^\circ = \int_0^t \phi^H_s \cdot dN_s $; we already showed that (see the arguments below (\ref{genAR2})) $\hat M_t^\circ$ is a square-integrable martingale under $\pr^R$, and this implies it has finite expected quadratic variation under $\pr^R$:
\begin{equation}
\label{genhatMfiniteQV}
\ex^R([\hat M^\circ, \hat M^\circ]_t) < \infty.
\end{equation}
\noindent Then, for any $\pr^{\bar R} \in \calm_N$, we have
\begin{eqnarray}
\ex^{\bar R}(\sqrt{[\hat M^\circ, \hat M^\circ]_t}) &=& \ex^R\Big[\frac{d\pr^{\bar R}}{d\pr^R}\sqrt{[\hat M^\circ, \hat M^\circ]_t}\Big]
\leq  \ex^R\Big[\Big(\frac{d\pr^{\bar R}}{d\pr^R}\Big)^2\Big]\ex^R([\hat M^\circ, \hat M^\circ]_t)
< \infty ;
\end{eqnarray}
where the $\leq$ follows Cauchy–Schwarz inequality; and the $<$ follows (\ref{genhatMfiniteQV}) and the following
\begin{eqnarray}
\ex^R\Big[\Big(\frac{d\pr^{\bar R}}{d\pr^R}\Big)^2\Big] = \ex\Big[\Big(\frac{d\pr^{\bar R}}{d\pr}\frac{d\pr}{d\pr^R}\Big)^2\frac{\big(Z_T^M\big)^2}{Z_0^M}\Big]
= \ex\Big[\Big(\frac{d\pr^{\bar R}}{d\pr}\Big)^2 \frac{\big(Z_0^M\big)^2}{\big(Z_T^M\big)^4}\frac{\big(Z_T^M\big)^2}{Z_0^M}\Big]
= Z_0^M \ex\Big[\Big(\frac{1}{Z_T^M}\frac{d\pr^{\bar R}}{d\pr}\Big)^2\Big]
< \infty ; \nonumber\\
\end{eqnarray}
where the $<$ follows from the definition of $\calm_N$; see (\ref{genlocNM}). Now by Burkholder-Davis-Gundy inequality, $\hat M^\circ_t$ is an martingale under $\pr^{\bar R}$. Last, $Z_T^M \hat M_T^\circ \in L^2(P)$ easily follows from that $\hat M_t^\circ$ is square-integrable under $\pr^R$. Combining the arguments above, we have $\phi_t^* = \phi_t^H \in \cala_N$, hence $\theta_t^*$ is admissible as argued.
This proves Part (i) of Theorem~\ref{thm:hedgingsol}.

Now, we show that $\var(H_T + \chi_T^*)$ has the expression specified in (\ref{bmpr}) of Part (ii) of this theorem.
First note that from (\ref{genAR3}), $A(\lambda)$ is a quadratic function in $\lambda$ and its second term (the integral) does not involve $\lambda$.
Then, by (\ref{genBA}), clearly the optimal $\lambda$ should solve
$$\max_\lambda \frac{(\lambda - V_0(P,R))^2}{Z_0^M} - (m-\lambda)^2,$$
and it is easy to verify that the solution has the expression in (\ref{opttheta}).
Substituting (\ref{gengammat}) to the second term of $A(\lambda)$ in (\ref{genAR3})
equates this expression to (\ref{genAexpression}).
Then, using the expression of $\lambda_m$ in (\ref{opttheta}) and applying (\ref{genBA}),
the expression of $B(m, P,R)$ in (\ref{bmpr}) immediately follows.

{What remains is to show Part (iii).
Recall, the set of all wealth attainable by admissible trading strategies is defined in (\ref{gentermwealth}).
For simpler notation, $\chi_T(\cala_X)$ is written as $\chi_\cala$ below.
As noted following (\ref{gentermwealth}), $\chi_\cala$ is {\it closed}.
So, being a subspace of $L^2(\pr)$
\footnote{{It is straightforward to verify that any linear combination of admissible strategies is also admissible}},
$\chi_\cala$ is a Hilbert space (with the inner product being that of $L^2(\pr)$).}

{
Next, note that the problem in (\ref{genA}),
which is equivalent to the original hedging problem in (\ref{mvprob}) with $\lambda$ taking the value in (\ref{opttheta}),
is equivalent (in the sense of for any $\chi_T \in \chi_\cala$, it is attained by some trading strategy in $\cala_X$):
\begin{equation}
\label{MSE}
\min_{\chi_T \in \chi_\cala} \; \ex[(\lambda - H_T - \chi_T)^2]
\end{equation}
Then, by Projection Theorem, $\forall \chi_T \in \chi_\cala$, the following must hold:
\begin{equation}
\label{projection}
\ex[(\lambda-H_T - \chi_T^*)\chi_T] = 0.
\end{equation}
Note that due to the fact that $H_T$ is bounded,  $\{-\xi_t, 0\leq t \leq T\} \in \cala_X$.
So, $\{\iota_t = \theta_t^* + \xi_t, 0\leq t \leq T\} \in \chi_\cala$ and thus $\chiiv_T \in \chi_\cala$.
Then, let $\chi_T = \chiiv_T$ in (\ref{projection}), rearrange the terms and apply the definition of $\HTh$ in (\ref{hedgedprodpayoff}), we have (accounting for that $\chi_T^* = \chirm_T + \chiiv_T$):
\begin{eqnarray}
\label{chiivbyHth}
\ex(\chiiv_T\HTh) = \lambda \ex(\chiiv_T) - \ex[(\chiiv_T)^2]
= \lambda (m-V_0)- \ex[(\chiiv_T)^2],
\end{eqnarray}
where the second equality is due to the following facts:
$$\ex(H_T + \chi_T^*) = \ex(\HTh + \chiiv_T) = m$$
and
$$\ex(\HTh) = \ex\Big(H_T + \chirm_T\Big)
= \ex\Big(V_0 + \int_0^T \xi_t dX_t + \int_0^T \delta_t d\tilde B_t + \int_0^T (-\xi_t)dX_t\Big)
=V_0 + \ex\Big(\int_0^T \delta_t d\tilde B_t\Big)= V_0.$$
The second equality is due to $H_T = V_T$ and the representation of $V_t$ of (\ref{GKW}) in Lemma \ref{lem:Vt}.
Then, we can derive the left hand side of  (\ref{invrisk}),
using (\ref{chiivbyHth}) and the expression of $\lambda$ in (\ref{opttheta}):
\begin{eqnarray*}
 \var(\chiiv_T) + \cov(\chiiv_T, \HTh)
 &=& \ex[(\chiiv_T)^2] - [\ex(\chiiv_T)]^2 +  \lambda (m-V_0)- \ex[(\chiiv_T)^2] - \ex(\chiiv_T)\ex(\HTh) \\
 &=& -(m - V_0)^2  +  \lambda (m-V_0) - (m - V_0)V_0 \\
 &=& \frac{1}{Z_0^M-1}(m-V_0)^2,
\end{eqnarray*}
and this establishes (\ref{invrisk}).}

{
Now, note that
$$B(m,P,R) = \var(H_T + \chi_T^*) = \var(\HTh + \chiiv_T) = \var(\chiiv_T) + 2\cov(\HTh, \chiiv_T) + \var(\HTh).$$
Then, combining (\ref{invrisk}) and (\ref{bmpr}) immediately gives (\ref{unhedgeablerisk}).
This completes the proof of Theorem \ref{thm:hedgingsol}. $\hfill\Box$
}

\subsection{Proof of Proposition~\ref{pro:hedgingsolEOU}}
\label{appendix:hedgingsolEOU}
\subsubsection{Verification of Assumption~\ref{assumption:Z}}
First, we prove that under the parametric condition in (\ref{kappa}),
Assumption~\ref{assumption:Z} holds.
\begin{lemma}
\label{lem:OUztsqmg}
{
For $X_t$ specified in (\ref{expOUX}), let $Z_t$ follow the definition in (\ref{genZt})
(then $\eta_t$ involved in $Z_t$ is defined in (\ref{etaeOU})).
Under the parametric condition in (\ref{kappa}) holds, i.e.,
$$\kappa T < \frac{\pi}{4}.$$
Assumption~\ref{assumption:Z} holds---that is, $Z_t$ is a square-integrable martingale under $\pr$.
}
\end{lemma}

{\startb Proof.} By definition of $Z_t$ in (\ref{genZt}), we have
\begin{equation}
\label{OUdZt}
dZ_t = -\eta_t Z_t dB_t = -Z_t\frac{\kappa}{\sigma}\Big(\alpha + \frac{\sigma^2}{2\kappa} - Y_t\Big) dB_t;
\end{equation}
\noindent the second equality uses the expression of $\eta_t$ in (\ref{etaeOU}).

Note that (\ref{OUdZt}) implies that $Z_t$ is a local martingale under $\pr$.
We will prove that it is a martingale as follows.
Introduce the following process, which is a candidate for $\ex(Z_T^2 \,|\, \calf_t)$.
\begin{eqnarray}
\label{OUJ}
J_t := Z_t^2 f(t,Y_t)
    = Z_t^2 \exp\{f_0(T-t) + f_1(T-t)Y_t + f_2(T-t)Y_t^2 \} ;
\end{eqnarray}
where $f_i,\, i= 0,1,2$ are functions defined in (\ref{OUf}); and
\begin{equation}
\label{OUfunc}
f(t, y):= \exp\{f_0(T-t) + f_1(T-t)y + f_2(T-t)y^2 \}.
\end{equation}
\noindent
Since $f_i(0)=0$ for all $i=0,1,2$, we have $f(T, Y_t) = 0$ and thus $J_T = Z_T^2$.
Below we show that properties of $J_t$ guarantee $Z_t$ to be a square-integrable martingale.

Applying It\^{o}'s lemma, it is straightforward to obtain the following dynamics:
\begin{eqnarray}
\label{OUdJtdivZt}
d\Big(\frac{J_t}{Z_t}\Big) &=& dZ_t f(t,Y_t)
= f(t, Y_t)dZ_t + Z_tdf(t,Y_t) + dZ_tdf(t,Y_t) \nonumber\\
&=& Z_t\Big(f_t - \frac{1}{2}\sigma^2 f_y + \frac{1}{2}\sigma^2 f_{yy}\Big)dt + Z_t(\sigma f_y - \eta_t f(t,Y_t))dB_t \nonumber\\
&=& Z_t\Big[f_t - \kappa\big(\alpha + \frac{\sigma^2}{\kappa} - Y_t\big)f_y + \frac{1}{2}\sigma^2f_{yy} + \eta_t^2 f(t,Y_t)\Big]dt \nonumber\\
&-& Z_t\Big[\eta_t f(t, Y_t) - \sigma f_ y \Big]d\Big(dB_t + \eta_t dt \Big);
\end{eqnarray}
\noindent $f_t$, $f_y$ and $f_{yy}$ are usual notations for partial derivatives; note they all depend on $(t, Y_t)$, but for simplicity the arguments are dropped.
The second line uses the expression of $\eta_t$ in (\ref{etaeOU}).

Next, applying It\^{o}'s lemma on $J_t$, it is straightforward to obtain:
\begin{eqnarray}
\label{OUdJt}
dJ_t &=& Z_t d\Big(\frac{J_t}{Z_t}\Big) + \frac{J_t}{Z_t} dZ_t + d\Big(\frac{J_t}{Z_t}\Big) dZ_t \nonumber\\
	 &=& Z_t^2\Big[f_t - \kappa\big(\alpha + \frac{\sigma^2}{\kappa} - Y_t\big)f_y + \frac{1}{2}\sigma^2f_{yy} + \eta_t^2 f(t,Y_t)\Big] dt \nonumber\\
	 &+& \Big(\sigma f_y -\eta_t f(t, Y_t) - \eta_t Z_t^2 f(t, Y_t)\Big) dB_t
\end{eqnarray}

\noindent It can be verified that the function $f(t,y)$ defined in (\ref{OUfunc}) solves the following partial differential equation:
\begin{eqnarray}
\label{OUPDE}
f_t - \kappa\big(\alpha + \frac{\sigma^2}{\kappa} - y\big)f_y &+& \frac{1}{2}\sigma^2 f_{yy} +
\big(\frac{\kappa}{\sigma}\big)^2\big(\alpha + \frac{\sigma^2}{2\kappa}-y\big)^2f(t,y) = 0 \nonumber\\
{\rm s.t.} \qquad f(T,y) &=& 1, \quad \forall y \in \mathbb{R}.
\end{eqnarray}
\noindent To check this, take derivatives of $f(t,y)$ using (\ref{OUfunc}). Then, collect the coefficients for $y^2$, $y$ and the term independent of $y$, and set these coefficients to $0$, then (\ref{OUPDE}) reduces to an ordinary differential equation system:
\begin{eqnarray}
\label{OUODE}
-f_2^\prime + (2\kappa)f_2 + (2\sigma^2)f^2_2 + (\frac{\kappa}{\sigma})^2 &=& 0 \nonumber\\
-f_1^\prime + \kappa[f_1 - 2(\alpha + \frac{\sigma^2}{\kappa})f_2] + (2\sigma^2 f_2)f_1 - 2(\frac{\kappa}{\sigma})^2(\alpha + \frac{\sigma^2}{2\kappa}) &=& 0 \nonumber\\
-f_0^\prime - \kappa(\alpha + \frac{\sigma^2}{\kappa})f_1 + \frac{1}{2}\sigma^2(f_1^2 + 2f_2) + (\frac{\kappa}{\sigma})^2(\alpha + \frac{\sigma^2}{2\kappa})^2 &=& 0 \nonumber\\
{\rm s.t.} \qquad f_i(0) = 0, i = 0,1,2.
\end{eqnarray}
\noindent It is straightforward to verify that under the parameter condition in (\ref{kappa}), $f_i,\; i = 0,1,2$ specified in (\ref{OUf}), solve this ODE system. So, $f(t, y)$ solves the PDE above, and this makes the $dt$-term of $dJ_t$ in (\ref{OUdJt}) vanish, reducing (\ref{OUdJt}) to:
\begin{eqnarray}
\label{OUdJt1}
dJ_t =  \Big[\sigma f_y -\eta_t f(t, Y_t) - \eta_t Z_t^2 f(t, Y_t)\Big] dB_t .
\end{eqnarray}
\noindent
And, (\ref{OUdJtdivZt}) reduces to (note the PDE expression is also involved in the $dt$-term on the second line of (\ref{OUdJtdivZt})):
\begin{eqnarray}
\label{OUdJtdivZt1}
d\Big(\frac{J_t}{Z_t}\Big)  &=& - Z_t\Big[\eta_t f(t, Y_t) - \sigma f_y \Big]d\Big(dB_t + \eta_t dt \Big).
\end{eqnarray}
\noindent Clearly, by (\ref{OUdJt1}), $J_t$ is a local martingale under $\pr$.

Now, observe that the term inside the exponential of $f(t, Y_t)$ is a quadratic function in $Y_t$ at each time $t$. Under (\ref{kappa}), $f_2(T - t) \ge 0$ for all $t \in [0, T]$, hence
$$f(t, Y_t) \ge \exp\Big\{f_0(T-t) - \frac{f_1^2(T-t)}{4f_2(T-t)}\Big\}, \quad t\in[0, T); $$
\noindent It is easy to verify that $f_1^2(T-t)/f_2(T-t) \to 0$ as $t \to T$. Clearly, the function $f_0(T-t) - \frac{f_1^2(T-t)}{4f_2(T-t)}$ is continuous, hence admits a minimum on $[0, T]$; so, there exists a positive number, $c > 0$, such that $f(t, Y_t) \ge c > 0$ for all $t \in [0, T]$. Therefore,
\begin{equation}
\label{OUZtbound}
J_t = Z_t^2f(t, Y_t) \ge cZ_t^2 \;\Rightarrow \; Z_t \leq \sqrt{\frac{1}{c}J_t}.
\end{equation}
Next, we will make use of (\ref{OUZtbound}) to prove $Z_t$ is a square-integrable martingale under $\pr$. Define the following sequence of stopping times with respect to $\calg_t$:
\begin{equation}
\label{OUtau}
\tau_k := \inf \{t\in[0,T] \,|\, J_t \ge k \} \wedge T, \quad k\in \mathbb{N}.
\end{equation}
\noindent Clearly, $J_t$ is a continuous process, hence $\tau_k \uparrow T$ as $k\to\infty$.
Since each $\tau_k$ bounds the stopped version of $J_t$, hence also bounds the stopped version of $Z_t$ via (\ref{OUZtbound}). So, both $Z_{t\wedge \tau_k}$ and $J_{t\wedge \tau_k}$ are bounded $\pr$-martingales, and we can apply Doob's inequality:
\begin{eqnarray}
\label{OUDoob}
\ex\Big[\sup_{t\in[0,T]} Z_{t\wedge \tau_k}^2\Big] \leq 2\ex(Z_{T \wedge \tau_k}^2) &=& 2\ex(Z_{\tau_k}^2)
\leq  \frac{2}{c}\ex(J_{\tau_k}) = \frac{2}{c}J_0 .
\end{eqnarray}
\noindent the first $\leq$ is application of Doob's inequality, and the following equality uses the obvious fact $\tau_k \leq T$.
The second $\leq$ uses (\ref{OUZtbound}) and the $=$ is application of optional stopping theorem on the bounded martingale $J_{t\wedge \tau_k}$.
So, (\ref{OUDoob}) implies
$$\ex\Big[\sup_{t\in[0,T]} Z_{t\wedge \tau_k}^2\Big] \leq \frac{2}{c}J_0; $$
\noindent clearly, because $\tau_k$ increases in $k$, so does $\sup_{t\in[0,T]} Z_{t\wedge \tau_k}^2$ (since the $\sup$ is taken on a longer time interval for larger $\tau_k$), so we can let $k\to\infty$ and apply monotone convergence to above to reach the following, accounting for $\tau_k \uparrow T$:
$$\ex\Big[\sup_{t\in[0,T]} Z_{t}^2\Big] \leq \frac{2}{c}J_0; $$
\noindent this is sufficient to establish that $Z_t$ is a square-integrable martingale under $\pr$, and completes the proof of Lemma~\ref{lem:OUztsqmg}. $\hfill\Box$

\subsubsection{Proof of Proposition~\ref{pro:hedgingsolEOU}}
Now that we know $Z_t$ is a $\pr$-martingale, $\pr^M$ specified in (\ref{genZt}) is well-defined.
Then, Girsanov Theorem applies and we have the $\pr^M$-Brownian-motion:
$$dB_t^M = dB_t - \eta_t dt.$$
Then, (\ref{OUdJtdivZt1}) becomes
\begin{eqnarray}
\label{OUdJtdivZt2}
d\Big(\frac{J_t}{Z_t}\Big)  &=& - Z_t\Big[\eta_t f(t, Y_t) - \sigma f_y \Big]dB_t^M;
\end{eqnarray}
\noindent using the expression of $f(t, y)$ in (\ref{OUfunc}), it is straightforward to derive an explicit expression for $f_y$.
Then, arrange the terms to explicitly write (\ref{OUdJtdivZt2}) as:
\begin{eqnarray}
\label{OUdJtdivZt3}
d\Big(\frac{J_t}{Z_t}\Big)  &=& - \sigma\Big(\frac{J_t}{Z_t}\Big)\Big[\frac{\kappa}{\sigma^2}b(T-t)(\alpha - Y_t) + a(T-t) \Big]dB_t^M;
\end{eqnarray}
\noindent with the two deterministic functions $a(\cdot)$ and $b(\cdot)$ specified in (\ref{abtau}); note $b(T-t) > 0$ for all $t \in [0, T]$. Next, observe that $V_t:= \frac{\kappa}{\sigma^2}b(T-t)(\alpha - Y_t) + a(T-t)$ follows a linear stochastic differential equation of the following form:
$$ dV_t = (x_t + y_t V_t)dt + z_t dB_t^M; $$
\noindent where $x_t$, $y_t$ and $z_t$ above are deterministic functions; the linearity comes from $Y_t$, which also follows a linear SDE (see (\ref{expOUX})), as well as from $\eta_t$ in (\ref{etaeOU}), which is linear in $Y_t$.
Now, Lemma~A.4 in \cite{wang2013mean} immediately applies, and we conclude that $J_t/Z_t$ is a martingale under $\pr^M$.
Next,
\begin{eqnarray}
Z_t^M := \ex^M(Z_T \,|\, \calf_t) = \ex^M\Big(\frac{J_T}{Z_T} \,\Big{|}\, \calf_t\Big)
= \frac{J_t}{Z_t} = Z_tf(t, Y_t).
\end{eqnarray}
\noindent
The first $=$ follows from $J_T = Z_T^2$, and the second one uses the established fact that $J_t/Z_t$ is a martingale under $\pr^M$.
The above can be written as:
\begin{equation}
\label{OUZtZtM}
\frac{Z_t}{Z_t^M} = \frac{1}{f(t, Y_t)} = \exp\{-f_0(T-t) - f_1(T-t)Y_t - f_2(T-t)Y_t^2 \}.
\end{equation}

\noindent
Now, (\ref{OUdJtdivZt3}) can be written as:
\begin{eqnarray}
\label{OUdJtdivZt4}
dZ_t^M  &=& - Z_t^M \Big[\frac{\kappa}{\sigma^2}b(T-t)(\alpha - Y_t) + a(T-t) \Big] \sigma dB_t^M \nonumber\\
		&=& -\frac{Z_t^M}{X_t} \Big[\frac{\kappa}{\sigma^2}b(T-t)(\alpha - Y_t) + a(T-t) \Big] dX_t;
\end{eqnarray}
\noindent where the second line uses $dX_t= \sigma X_t dB_t^M$. Hence, here the quantity $\zeta_t$ defined in (\ref{genZtM}) has the expression
\begin{equation}
\label{OUzeta}
\zeta_t = -\frac{Z_t^M}{X_t} \Big[\frac{\kappa}{\sigma^2}b(T-t)(\alpha - Y_t) + a(T-t) \Big].
\end{equation}

Now we can apply Theorem~\ref{thm:hedgingsol} to establish Proposition~\ref{pro:hedgingsolEOU}. In particular, $Z_t / Z_t^M$ involved in $B(m,P,R)$ expressed in (\ref{bmpr}) follows (\ref{OUZtZtM}), and $\zeta_t$ involved in $\theta_t^*$ specified in (\ref{opttheta}) follows (\ref{OUzeta}). This completes the proof. $\hfill\Box$

\section{Proofs for Section \ref{sec:production}}

\subsection{Proof of Propositions~\ref{pro:prx0-positive} and \ref{pro:prx0-negative}}
\label{appendix:prx0}
It is clear that both Propositions~\ref{pro:prx0-positive} and \ref{pro:prx0-negative} immediately follow from the following lemma.
Specifically, for Proposition~\ref{pro:prx0-positive}
(resp., Proposition~\ref{pro:prx0-negative}),
$C_T = C_T^{(2)}$ and $C_T^M = C_T^{(1)}$
(resp., $C_T = C_T^{(1)}$ and $C_T^M = C_T^{(2)}$),
where $C_T^{(1)}$ and $C_T^{(2)}$ are defined in Lemma~\ref{lem:CTorder} below.
\begin{lemma}
\label{lem:CTorder}
 Let $(\nvp, \nvr)$ be the solution to problem~(\ref{nv1}),
 i.e., $(\nvp, \nvr) = \arg\max_{P,R} \ex[H_T(P,R)]$.
Then, as  $C_T$ increases stochastically,
both $\nvp$ and $\nvr$ increase.
Specifically, for any $C_T^{(1)} \preceq C_{T}^{(2)}$,
the corresponding solutions to problem~(\ref{nv1}),
denoted $(\nvp_i, \nvr_i)$ for $i=1,2$,
satisfy $\nvp_1 \leq \nvp_2$ and $\nvr_1 \leq \nvr_2$.
In addition, let $H_T^{(i)}$ denote the production payoff with $C_T = C_T^{(i)}$.
Then, $\ex[H_T^{(1)}(P,R)] \leq \ex[H_T^{(2)}(P,R)]$ for all $(P,R)$.
\end{lemma}
{\startb Proof.}
  Let the distribution functions of $A_T^{(i)}$ be $F_i$.
  For a given $P$, let $R_i^{\rm NV}(P)$  be the corresponding profit-maximizing decision, i.e.,
  \begin{eqnarray*}
  R_i^{\rm NV}(P) &:=& \argmax_R \ex[H_T(P, R)] \\
  &=& \argmax_R \; (P-c) (R-bP) - (P-s)\ex[(R-A^{(i)}_{T})^{+}]
  = F_i^{-1}\Big(\frac{P - c}{P - s}\Big).
  \end{eqnarray*}
  Clearly, $R_1^{\rm NV}(P) \leq R_2^{\rm NV}(P)$ due to $A_T^{(1)} \preceq A_T^{(2)}$
  (refer to Remark~\ref{rem:ATorder}).

  Define
  $$H_i(P) = (P-c)(R-bP)- (P-s) \ex[(R_i^{\rm NV}(P)-A_T^{(i)})^+], \quad i = 1,2.$$
  Then, the first-order derivative with respect to $P$ is:
  $$H'_{i}(P)=R_i^{\rm NV}(P) -2bP+ bc- \ex[(R_i^{\rm NV}(P)-A_T^{(i)})^+], \quad i=1, 2$$
  For any given $P$, we show $H'_{1}(P) \leq H'_{2}(P)$:
  \begin{eqnarray*}
  H'_{1}(P) &=& R_1^{\rm NV}(P) -2bP+ bc-  \ex[(R_1^{\rm NV}(P)-A_T^{(1)})^+] \\
            &\leq& R_1^{\rm NV}(P) -2bP+ bc-  \ex[(R_1^{\rm NV}(P)-A_T^{(2)})^+] \\
            &=& \ex[R_1^{\rm NV}(P) \wedge A_T^{(2)}] - 2bP + bc \\
            &\leq& \ex[R_2^{\rm NV}(P) \wedge A_T^{(2)}] - 2bP + bc \\
            &=& R_2^{\rm NV}(P) -2bP+ bc- \ex[(R_2^{\rm NV}(P)-A_T^{(2)})^+] = H'_{2}(P).
  \end{eqnarray*}
	The first inequality is due to $A_T^{(1)}\preceq A_T^{(2)}$ and
	the second one follows from
 $R_1^{\rm NV}(P) \leq R_2^{\rm NV}(P) $.
	Suppose $P_{i}^{\rm NV}$ is the one of the larger zeros to $H^{'}_{i}(P) = 0$.
	(Recall, $H^{'}_{i}(P) = 0$ has two solutions and the lager one is the newsvendor's solution; see Section~\ref{appendix:nvsol}.)
	Then,
	$$0=H^{'}_1(P_1^{\rm NV}) \leq H^{'}_2(P_1^{\rm NV}),$$
	thus $H^{'}_2(P)$ stays nonnegative before $P$ exceeds $P_1^{\rm NV}$, indicating $P_2^{\rm NV} \ge P_1^{\rm NV}$.
	Furthermore,
	$$R_1^{\rm NV}(P_1^{\rm NV})=F_1^{-1}\Big(\frac{P_1^{\rm NV}-c}{P_1^{\rm NV}-s}\Big)\leq F_2^{-1}\Big(\frac{P_1^{\rm NV}-c}{P_1^{\rm NV}-s}) \leq F_2^{-1}(\frac{P_2^{\rm NV}-c}{P_2^{\rm NV}-s}\Big)= R_2^{\rm NV}(P_2^{\rm NV})=R_2^{\rm NV}.$$

 The ``in addition'' part follows immediately from the fact that $H_T(P,R)$ is an increasing function in $A_T$ for any $(P,R)$, and $A_T^{(1)} \preceq A_T^{(2)}$.
This completes the proof.
	$\hfill\Box$

\subsection{Proof of Propositions~\ref{pro:prbound-part2} and \ref{pro:prbound-part1}}
\label{appendix:prbound}
We first prove Proposition~\ref{pro:prbound-part1} and,
as an intermediate step, establish Lemma~\ref{lem:prboundeachother}.
Based on this lemma, we further prove Proposition~\ref{pro:prbound-part2}.
\subsubsection*{Proof of Proposition~\ref{pro:prbound-part1}}
Denote the second term of $B(m, P, R)$ as $\Psi(P, R)$.
Clearly, $\Psi(P, R)$ strictly increases in both $P$ and $R$.
Then, write
$$B(m,P,R) = C[(m - V_0(P,R))]^2 + \Psi(P,R), $$
where $C = 1/(Z_0^M-1) > 0$.
(Below, we refer to $C[(m - V_0(P,R))]^2$ (resp.,\ $\Psi(P,R)$) as the ``first term'' (resp.,\ ``second term'') of
$B(m, P,R)$.)
Note that $(P_m^h, R_m^h)$ satisfy the optimality equations
\begin{equation*}
2C(V_0 - m) \frac{\partial V_0(P_m^h, R_m^h)}{\partial P} + \frac{\partial \Psi(P_m^h, R_m^h)}{\partial P} = 0, \quad
2C(V_0 - m) \frac{\partial V_0(P_m^h, R_m^h)}{\partial R} + \frac{\partial \Psi(P_m^h, R_m^h)}{\partial R} = 0.
\end{equation*}
Since $\Psi(P,R)$ is strictly increasing in both $P$ and $R$, we must have
\begin{equation}
\label{bmprdersign}
(V_0 - m) \frac{\partial V_0(P_m^h, R_m^h)}{\partial P} < 0,
\quad
(V_0 - m) \frac{\partial V_0(P_m^h, R_m^h)}{\partial R} < 0.
\end{equation}

Recall, $V_0(P, R)$ is concave in $P$ (resp.,\ $R$) for any given $R$ (resp.,\ $P$).
We use the notation in (\ref{pmr}):
$$P^{\rm NV(M)}(R) := \argmax_P \; V_0(P,R), \quad R^{\rm NV(M)}(P) := \argmax_R \; V_0(P,R),$$
and it is straightforward to derive:
\begin{equation}
\label{nvmpr1}
P^{\rm NV(M)}(R) = \frac{bc + \ex^M(R \wedge A_T)}{2b},
\quad R^{\rm NV(M)}(P) = F^{-1}_M\Big(\frac{P-c}{P-s}\Big);
\end{equation}
where $F_M$ is the distribution function of $A_T$ under $\pr^M$.
Clearly, $\nvmp(R)$ (resp., $\nvmr(P)$) increases in $R$ (resp.,\ $P$).
Moreover, by its partial concavity in $P$ (resp.,\ $R$),
for given $R$ (resp.,\ $P$),
$V_0(P,R)$ increases in $P$ (resp.,\ $R$) for $P \leq \nvmp(R)$ (resp.,\ $R \leq \nvmr(P)$) and then decreases.
Thus, for any given $R$, the corresponding $P$ that minimizes $B(m, P, R)$ cannot exceed $P^M(R)$.
Otherwise, a $P$ smaller than $\nvmp(R)$, $P'$, can be found to satisfy $V_0(P, R) = V_0(P', R)$.
Then, $B(m, P', R) < B(m, P, R)$, contradicting with $P$ minimizing $B(m, P, R)$.
Completely analogous argument can be applied to $R$.
Applying the above to optimality of $(P_m^h, R^m_h)$, we must have:
\begin{lemma}
\label{lem:prboundeachother}
$P^h_m \leq \nvmp(R_m^h), \quad R^h_m \leq \nvmr(P_m^h)$.
\end{lemma}
Clearly, Lemma~\ref{lem:prboundeachother} implies:
\begin{equation}
\label{V0derbound}
\frac{\partial V_0(P_m^h, R_m^h)}{\partial P} \ge 0, \quad \frac{\partial V_0(P_m^h, R_m^h)}{\partial R} \ge 0.
\end{equation}
Combining (\ref{V0derbound}) with (\ref{bmprdersign}),
we conclude $V_0(P_m^h, R_m^h) \leq m$.
This proves Proposition~\ref{pro:prbound-part1}.

\subsubsection*{Proof of Proposition~\ref{pro:prbound-part2}}
$P_m^h \leq  P^{\rm NV(M)}$ and $R_m^h \leq R^{\rm NV(M)}$.
First, we consider the case of $m \ge V_0(P^{\rm NV(M)},R^{\rm NV(M)})$.
For this case, $P_m^h$ and $R_m^h$ cannot both exceed $P^{\rm NV(M)}$ and $R^{\rm NV(M)}$, respectively;
otherwise both terms of $B(m, P_m^h, R_m^h)$ will exceed those of $B(m, P^{\rm NV(M)}, R^{\rm NV(M)})$,
contradicting optimality of $(P_m^h, R_m^h)$.
Then, if $P_m^h \leq P^{\rm NV(M)}$, by Lemma~\ref{lem:prboundeachother},
$$R_m^h \leq \nvmr(P_m^h) \leq \nvmr(P^{\rm NV(M)}) = R^{\rm NV(M)}.$$
If $R_m^h \leq R^{\rm NV(M)}$, again by Lemma~\ref{lem:prboundeachother},
$$P_m^h \leq \nvmp(R_m^h) \leq \nvmp(R^{\rm NV(M)}) = P^{\rm NV(M)}.$$
Hence, for both cases, we must have $P_m^h \leq  P^{\rm NV(M)}$ and $R_m^h \leq R^{\rm NV(M)}$.

Now, we consider the other case, $m < V_0(P^{\rm NV(M)},R^{\rm NV(M)})$.
Let $v(P) := V_0(P, \nvmr(P))$.
Clearly, $v(c) \leq 0 \leq m$.
On the other hand,
$$v(P^{\rm NV(M)}) = V_0(P^{\rm NV(M)}, R^M(P^{\rm NV(M)})) = V_0(P^{\rm NV(M)}, R^{\rm NV(M)}) > m.$$
Thus, $\exists P_1 \in [c, P^{\rm NV(M)}]$ such that $v(P_1) = m$;
let $R_1 =\nvmr(P_1)$.
Note, $R_1 =\nvmr(P_1) \leq \nvmr(P^{\rm NV(M)}) = R^{\rm NV(M)}$.
Then, $P_m^h$ and $R_m^h$ cannot both exceed, respectively, $P_1$ and $R_1$, otherwise both terms of $B(m, P_m^h, R_m^h)$ are larger than those of $B(m, P_1, R_1)$ (the first term of which is zero).
Then, if $P_m^h \leq P_1 (\leq P^{\rm NV(M)})$, applying Lemma~\ref{lem:prboundeachother},
$$R_m^h \leq \nvmr(P_m^h) \leq \nvmr(P_1) \leq \nvmr(P^{\rm NV(M)}) = R^{\rm NV(M)}.$$
If $R_m^h \leq R_1 (\leq R^{\rm NV(M)})$, then again by Lemma~\ref{lem:prboundeachother},
$$P_m^h  \leq \nvmp(R_m^h) \leq \nvmp(R_1) \leq \nvmp(R^{\rm NV(M)}) = P^{\rm NV(M)}.$$
Then, for both cases, we must also have $P_m^h \leq  P^{\rm NV(M)}$ and $R_m^h \leq R^{\rm NV(M)}$.
Summarizing all above leads to $P_m^h \leq  P^{\rm NV(M)}$ and $R_m^h \leq R^{\rm NV(M)}$.
This proves Proposition~\ref{pro:prbound-part2}.

\subsection{Proof of Theorem~\ref{thm:hurtpr}}
\label{appendix:hurtpr}
We layout key definitions and technical preparations in Section~\ref{appendix:thm3prep} and then prove Theorem~\ref{thm:hurtpr} in Section~\ref{appendix:hurtpr}.
\subsubsection{Definitions and Technical Preparations}
\label{appendix:thm3prep}
Throughout the proof, $m$ denotes the newsvendor's maximum expected profit, i.e.,
$$m =\ex[H_T(\nvp, \, \nvr)] > 0.$$
(The $>0$ is due to Assumption~\ref{assumption:nvm}.)
Also recall,
\begin{align*}
\nvmp(R) :={}& \arg\max_P V_0(P,R) = \frac{\ex^M(R \wedge A_T) + bc}{2b}, \\
\nvmr(P):={}& \arg\max_R V_0(P,R) = F_M^{-1}\Big(\frac{P - c}{P-s}\Big).
\end{align*}
Related to $m$, below we define three critical values.
\begin{definition}
\label{defn:3pr}
(i) $P_1$ is the smallest solution to $\ex^M[H_T(P_1, \nvmr(P_1))] = m$ and $R_1 =: \nvmr(P_1)$.
(ii) $P_2$ is the solution to $ \nvmr(P_2) = \nvr$.
(iii) $P_3$ is the smallest solution to $V_0(P_3, \nvr) = m$.
\end{definition}

The lemma below collects properties of $P_i$, $i = 1,2,3$.
\begin{lemma}
\label{lem:P123}
All $P_1$, $P_2$ and $P_3$ exist, hence by definition they are unique.
Furthermore,
\begin{equation}
\label{P2}
P_2=  \frac{c-s\pr^M(A_T\leq \nvr)}{1-P^M(A_T\leq \nvr)} = s+ \frac{c-s}{\pr^M(A_T\geq \nvr)}.
\end{equation}
And
\begin{equation}
\label{P3}
P_3 = \nvmp(\nvr) - \sqrt{(\nvmp(\nvr)-\nvp)(\nvmp(\nvr) + \nvp-2s)}.
\end{equation}
Moreover,
\begin{equation}
\label{P123bound}
P_i \leq \nvp, \quad i = 1,2,3.
\end{equation}
\end{lemma}
{\startb Proof.}
We first show $P_1$ exists.
(Existence of $P_2$ and $P_3$ will be clear after (\ref{P2}) and (\ref{P3}) are proved.)
Recall,
$$V_0(P,R) = (P - c)(R - bP) - (P-s)\ex^M[(R - A_T)^+].$$
As $P \to c$, $\nvmr(P) = F_M^{-1}((P-c)/(P-s)) \to -\infty$.
Thus, as $P \to c$, $(P - c)(R - bP)$ is eventually nonpositive (hence $<$ m) and so is $V_0(P,\nvmr(P))$.
Now, we check $V_0(\nvp, \nvmr(\nvp))$:
\begin{eqnarray*}
V_0(\nvp, \nvmr(\nvp)) &=& \max_R \, V_0(\nvp, R) \\
&\ge& V_0(\nvp, \nvr)\\
&=& (\nvp - c)(\nvr - b\nvp) - (\nvp - s)\ex^M[(\nvr - A_T)^+] \\
&\ge&(\nvp - c)(\nvr - b\nvp) - (\nvp - s)\ex[(\nvr - A_T)^+] \\
&=& \ex[H_T(\nvp, \nvr)] = m.
\end{eqnarray*}
The second $\leq$ is due to $A_T \preceq A_T^M$.
Summarizing the above, there exists value(s) of $P$ in $[c, \nvp]$ such that $V_0(P, \nvmr(P)) = m$, hence $P_1$ uniquely exists and in particular, $P_1 \leq \nvp$.

For $P_2$, by definition:
$$\nvmr(P_2) = F_M^{-1}\Big(\frac{P_2 - c}{P_2 -s}\Big) = \nvr, $$
which immediately leads to the expression in (\ref{P2}).
Furthermore,
\begin{eqnarray*}
P_2 = s+ \frac{c-s}{\pr^M(A_T\geq R^{NV})} \leq s+ \frac{c-s}{\pr(A_T\geq R^{NV})} = \nvp(\nvr) = \nvp.
\end{eqnarray*}
The $\leq$ is due to $A_T \preceq A_T^M$.

For $P_3$, note that given $R = \nvr$, $V_0(P, \nvr) = m$ is a quadratic equation in $P$.
Rearranging the terms and accounting for the following:
$$\nvmp(\nvr) = \frac{\ex^M(\nvr \wedge A_T) + bc}{2b},$$
the equation becomes
\begin{equation}
\label{P3eqn}
-bP^2 + 2b\nvmp(\nvr)P - (c-s)\nvr -2bs\nvmp(\nvr) + bcs = m.
\end{equation}
Using the optimality equations specified in Proposition~\ref{pro:nvsol},
it is straightforward to derive the following:
$$m = b(\nvp)^2 - (c-s)\nvr - 2bs\nvp + bcs.$$
Substituting this expression of $m$ to (\ref{P3eqn}), the equation becomes:
\begin{equation}
\label{P3eqn1}
-P^2+2\nvmp(\nvr)P - [(\nvp)^2 + 2s\nvmp(\nvr) - 2s\nvp] = 0.
\end{equation}
Then, it is straightforward to verify that (\ref{P3}) is the smallest solution to (\ref{P3eqn1}).
In particular, by $A_T \preceq A_T^M$,
$$\nvp = \frac{1}{2b}[\ex(\nvr \wedge A_T) + bc] \leq \frac{1}{2b}[\ex^M(\nvr \wedge A_T) + bc] \leq
= \nvmp(\nvr).$$
In addition, by Assumption~\ref{assumption:nvm}, $\nvp > c > s$, hence $\nvmp(\nvr) > c > s$.
Therefore, $P_3$ in (\ref{P3}) is well-defined (i.e., real-valued).
Next, using the fact that the function $x - \sqrt{(x-y)(x + y - 2s)}$ decreases in $x$ for $x \ge y \ge s$,
$P_3 \leq \nvp$ immediately follows from $\nvmp(\nvr) \ge \nvp$.
$\hfill\Box$

\medskip
The next result indicates that $R_1$ upper bounds $R_m^h$.
\begin{lemma}
\label{lem:R1}
$R_m^h \leq R_1.$
\end{lemma}
{\startb Proof.}
If $R_m^h > R_1$ and $P_m^h > P_1$,  then $B(m, P_1, R_1)< B(m, P_m^h, R_m^h)$
since the both the first term (which is zero) and the second term of the former are smaller than the latter.
This contradicts with optimality of $(P_m^h, R_m^h)$.
Therefore $R_m^h>R_1$ and $P_m^h> P_1$ cannot hold at the same time.
If $R_m^h>R_1$, we must have $P_m^h \leq P_1$, but this introduces contradiction
because by Lemma~\ref{lem:prboundeachother}, we must also have
$$R_m^h \leq R^{\rm NV(M)}(P_m^h) \leq  R^{\rm NV(M)}(P_1) = R_1.$$
Therefore, $R_m^h \leq  R_1$ must hold.
$\hfill\Box$

Finally, we need the following definition and lemma.
\begin{eqnarray}
\label{pmr}
P^{\rm NV(M)}(R) := \argmax_P \; V_0(P,R), \quad R^{\rm NV(M)}(P) := \argmax_R \; V_0(P,R).
\end{eqnarray}
Based on (\ref{pmr}), we define a critical point, $(P^*, \, R^*)$:
\begin{equation}
\label{pr*}
R^* \;\;\mbox{is the smallest root of} \;\; V_0(P^{\rm NV(M)}(R), R) = \ex[H_T(\nvp, \nvr)], \quad P^*:= P^{\rm NV(M)}(R^*).
\end{equation}
\begin{lemma}
\label{lem:r*}
Suppose Assumptions \ref{assumption:r}--\ref{assumption:Abc} hold and
$C_T \preceq C_T^M$.
Then, $(P^*, R^*)$  defined in (\ref{pr*}) {exists} and satisfies
\begin{equation*}
\label{pr*bnd}
P^* \leq \nvp \quad \mbox{and} \quad R^* \leq \nvr.
\end{equation*}
\end{lemma}

{\startb Proof.}
Let $f(R) := V_0(P^{\rm NV(M)}(R), R)$.
For notational simplicity, let $P(R) = P^{\rm NV(M)}(R)$
and $m =\ex[H_T(\nvp, \nvr)]$.
Given $R$, $V_0(P,R)$ is concave in $P$,
hence by setting $\partial V_0 / \partial P$ to zero,
$P(R)$ satisfies the following optimality equation:
\begin{equation}
\label{propteqn}
2bP(R) = \ex^M[R \wedge A_T] + bc.
\end{equation}
Then, it is straightforward to verify:
$$f(R) = b(P(R))^2 - 2bsP(R) - (c-s)R + bcs.$$
Next, we will show $f(\nvr) \ge m$ and $f(bc) \leq m$,
which in turn indicates that $f(R)$ has at least one root within $[bc, \nvr]$
and thus proves the result.
First, examine $f(\nvr)$.
With $C_T \preceq C_T^M$, by independence of $\tilde B_T$ from $\{B_t, 0\leq t\leq T\}$,
by Lemma~\ref{lem:CTorder} we have
$A_T \preceq A_T^M$ ($A_T^M$ is the version of $A_T$ under $\pr^M$: $\pr(A_T^M \leq a) = \pr^M(A_T \leq a)$ for all $a$.).
Then,
$$2bP(\nvr) = \ex^M[\nvr \wedge A_T] + bc \ge \ex[\nvr \wedge A_T] + bc = 2b\nvp, $$
where the last equality follows from (\ref{nvopteqns}),
leading to $P(\nvr) \ge \nvp$.
Then,
by the first optimality equation in (\ref{nvopteqns}),
it is straightforward to verify the following:
$$m = b(\nvp)^2 - 2bs\nvp - (c-s)\nvr + bcs.$$
Analogously, it can be verified that
$$f(\nvr) = b(P(\nvr))^2 - 2bsP(\nvr) - (c-s)\nvr + bcs.$$
Since $P(\nvr) \ge \nvp \ge s$, we have $f(\nvr) \ge m$.

Next, we check $f(bc)$.
Let $\epsilon = \ex^M[(bc - A_T)^+]$,
then clearly $\epsilon \leq [\ex[(bc - A_T)^+]].$
From Assumption~\ref{assumption:Abc}, we have
$$\frac{[\ex[(bc - A_T)^+]]^2}{4bm} \leq 1.$$
Note $P(bc) = [\ex^M(bc \wedge A_T) + bc] / 2b = c - \epsilon / 2b$,
then it is straightforward to verify that:
$$f(bc) <\frac{\epsilon^2}{4b}.$$
Thus,
$$\frac{f(bc)}{m} < \frac{\epsilon^2}{4bm} \leq \frac{[\ex[(bc - A_T)^+]]^2}{4bm}  \leq 1
\quad \Rightarrow \quad f(bc) < m.$$
Combining the above, $f(R)$ must have root(s) within $[bc, \nvr]$, hence by definition of $R^*$,
$R^* \leq \nvr$.

Next, we proceed with showing $P^* \leq \nvp$.
Using (\ref{propteqn}), it is easy to verify that
$$(m=) \quad f(R^*) = b(P^*)^2 - 2bsP^* - (c-s)R^* + bcs.$$
Comparing with the expression of $m$ above and taking into account $R^* \leq \nvr$,
clearly, $P^* \leq \nvp$ must hold.
This completes the proof. 	$\hfill\Box$

\subsubsection{Proof of Theorem~\ref{thm:hurtpr}}
Now we are ready to prove Theorem~\ref{thm:hurtpr}. \\
\noindent
\textbf{Proof of Part (i):}
By Lemma~\ref{lem:r*}, $P^* \leq \nvp$, so it is sufficient to prove $P^h_m \leq P^*$.
To this end, we consider two cases, $R_m^h \leq R^*$ or $R_m^h \ge R^*$.
For the first case, apply Lemma~\ref{lem:prboundeachother}, we have
$$P_m^h \leq  P^{\rm NV(M)}(R_m^h)\leq P^{\rm NV(M)}(R^*) = P^*.$$
Now, consider the other case, $R_m^h \ge R^*$.
Note that $P_m^h$ and $R_m^h$ cannot exceed $P^*$ and $R^*$ at the same time:
the first term of $B(m, P^*, R^*)$ is zero (since $V_0(P^*, R^*) = m$) and thus
increasing both $P$ and $R$ beyond $P^*$ and $R^*$ increases both terms $B(m, P, R)$.
Therefore, $R_m^h \ge R^*$ must lead to $P_m^h \leq P^*$.

\medskip
\noindent
\textbf{Proof of Part (ii):}
Applying Lemma~\ref{lem:P123},
it is straightforward to verify that the stated inequality in (\ref{hurtrcond}) is equivalent to $P_3 \leq P_2$.
By Lemma~\ref{lem:R1}, it is sufficient to prove $R_1 \leq \nvr$ under this condition.
To this end, first note that
$$\nvmp(\nvr) = \frac{1}{2b}[\ex^M(\nvr \wedge A_T) + bc] \ge \frac{1}{2b}[\ex(\nvr \wedge A_T) + bc] = \nvp. $$
The $\ge$ is due to $A_T \preceq A_T^M$.
Combining this with (\ref{P123bound}) and accounting for the concavity of $V_0(P, R)$ in $P$ with a given $R$,
$P_3 \leq P_2$ leads to
$$V_0(P_1, R_1) = m = V_0(P_3, \nvr) \leq V_0(P_2, \nvr).$$
By $V_0(P_1, R_1) = \max_R V_0(P_1, R) \ge V_0(P_1, \nvr)$, we have:
$$V_0(P_1, \nvr) \leq V_0(P_2, \nvr).$$
Applying $P_1, P_2 \leq \nvmp(\nvr)$ and concavity of $V_0(P,R)$ in $P$ with given $R$ again, we have
$$P_1 \leq P_2.$$
Therefore,
$$R_1 = \nvmr(P_1) \leq \nvmr(P_2) = \nvr, $$
which completes the proof of Part (ii).

\medskip
\noindent
\textbf{Proof of Part (iii):}
Violation of (\ref{hurtrcond}) is equivalent to $P_2 < P_3$.
Reversing the argument in proof of Part (ii) above, it is easy to obtain $P_1 > P_2$,
hence $R_1 = \nvmr(P_1) > \nvmr(P_2)  = \nvr$.
Next,
$$V_0(P_1, R_1) = \max_R V_0(P_1, R) = m = V_0(P_3, \nvr) \quad \Rightarrow \quad V_0(P_1, \nvr) \leq V_0(P_3, \nvr).$$
Again, by concavity of $V_0(P, \nvr)$ in $P$ and $P_1, P_3 \leq \nvp \leq \nvmp(\nvr)$,
we have $P_1 \leq P_3$.
In summary, we have
\begin{equation}
\label{P123relation}
c < P_2 < P_1 \leq P_3, \quad R_1 > \nvr.
\end{equation}
($P_2 > c$ is obvious from (\ref{P2}.)

Next, rearranging the terms of $V_0(P,R)$ leads to:
$$V_0(P,R) = (-bP^2 + bcP) + (P-s)\ex^M(R \wedge A_T) - (c-s)R.$$
Then, $m = V_0(P_1, R_1) = V_0(P_3, \nvr)$ leads to:
\begin{align*}
    & (-bP_1^2 + bcP_1) - (-bP_3^2 + bcP_3) + (P_1-s)\ex^M[R_1\wedge A_T]  - (c-s) R_1 \\
    ={}&  (P_3-s)\ex^M[\nvr \wedge A_T]  - (c-s) \nvr.
\end{align*}
Since the function $-bP^2 + bcP$ decreases in $P \ge c/2$ and $c < P_1 \leq P_3$,
we have $(-bP_1^2 + bcP_1) - (-bP_3^2 + bcP_3) \ge 0$.
Then, the equality above implies:
\begin{equation}
\label{keyinthepart3}
    (P_1-s)\ex^M[R_1\wedge A_T]  - (c-s) R_1 \leq (P_3-s)\ex^M[\nvr \wedge A_T]  - (c-s) \nvr.
\end{equation}
The LHS of (\ref{keyinthepart3}) is:
$$\ex^M(R_1\wedge A_T) = R_1 \pr^M(A_T\geq R_1) + \ex^M[A_T \ind\{A_T\leq R_1\}]
 = R_1 \frac{c-s}{P_1-s} + \ex^M[A_T \ind{\{A_T\leq R_1}\}],$$
 where the equality is due to definition of $R_1$:
 $\pr^M(A_T\leq R_1) =(P_1-c)/(P_1-s)$,
substituting the above expression of $\ex^M[R_1\wedge A_T]$ to the LHS of (\ref{keyinthepart3}), we have:
\begin{eqnarray*}
\text{LHS of (\ref{keyinthepart3})} &=& R_1(c-s) +  (P_1-s)\ex^M[A_T \ind\{A_T\leq R_1\}] - (c-s)R_1 \\  &=& (P_1-s)\ex^M[A_T \ind\{A_T\leq R_1\}] \\
&=& \ex^M[A_T \ind\{A_T\leq R_1\}] \frac{c-s}{\pr^M(A_T\geq R_1)},
\end{eqnarray*}
where the last equality is due to:
$\pr^M(A_T\geq R_1) = (c-s)/(P_1-s)$.

Now, we focus on the RHS of (\ref{keyinthepart3}).
We first show that it is positive.
Using $V_0(P_3, \nvr) = m > 0$ and rearranging terms of $V_0(P_3, \nvr)$,
we have:
\begin{eqnarray*}
0 < m = V_0(P_3, \nvr) = (-bP_3)(P_3 -c) + (P_3-s)\ex^M(\nvr \wedge A_T) - (c-s)\nvr.
\end{eqnarray*}
Note $P_3 > P_2 > c$ and thus $(-bP_3)(P_3 -c) < 0$,
which implies
$$\text{RHS of (\ref{keyinthepart3})} = (P_3-s)\ex^M(\nvr \wedge A_T) - (c-s)\nvr > 0.$$
Next, we have:
$$\ex^M[\nvr \wedge A_T] = 2bP^{\rm NV(M)}(\nvr)-bc = 2b\bar{P}^{\rm NV(M)}(\nvr) + 2bs- bc
=2b\bar P^\circ + 2bs - bc,$$
where the last equality follows from the definition of $\bar P^\circ $ in (\ref{PCirc}).
By (\ref{P3}) and definition of $r^\circ$ in (\ref{PCirc}), it is straightforward to verify the following:
$$\bar{P_3} \equiv P_3- s = \frac{\bar P^{\rm NV}}{r^\circ + \sqrt{(r^\circ)^2 - 1}} .$$
Therefore, the RHS of (\ref{keyinthepart3}) is rewritten as:
$$ \text{RHS of (\ref{keyinthepart3})} = \frac{\bar P^{\rm NV}( 2b\bar P^\circ+ 2bs- bc ) }{r^\circ + \sqrt{(r^\circ)^2 - 1}} -(c-s)\nvr  \;(>0).$$
Combining the above and dividing both sides by $c-s$, the inequality (\ref{keyinthepart3}) is equivalent to:
\begin{equation}
\label{keyinthepart3-2}
     \frac{\ex^M[A_T \ind\{A_T\leq R_1\}]}{\pr^M(A_T\geq R_1)} \leq \frac{\bar P^{\rm NV}( 2b\bar P^\circ+ 2bs- bc ) }{(c-s)(r^\circ + \sqrt{(r^\circ)^2 - 1})} -\nvr .
\end{equation}
In particular, the RHS of (\ref{keyinthepart3-2}) is positive.

\medskip
Next, we show that (\ref{keyinthepart3-2}) produces an upper bound on $R_1$, $R^\circ$,
hence by Lemma~\ref{lem:R1} also bounds $R_m^h$.
Define the following function in $R$:
$$ T(R) = \frac{\ex^M[A_T \ind\{A_T\leq R\}]}{\pr^M(A_T\geq R)}.$$
Taking derivative of $T(R)$ leads to:
$$T'(R) = \frac{f^M(R)}{[\pr^M(A_T \ge R)]^2} \ex^M(R \wedge A_T),$$
where $f^M(r)$ is the probability density function of $A_T$ under $\pr^M$.
Note, by $A_T \succeq A_T^M$ and Assumption~\ref{assumption:nvm},
$$\ex^M(\nvr \wedge A_T) \ge \ex(\nvr \wedge A_T ) = 2b\nvp - bc > 2bc - bc  = bc >0.$$
So, clearly $T'(R) > 0$ for $R \ge \nvr $.
Thus, $T(R)$ strictly increases in $R$ for $R \in [\nvr, \infty)$.

Next, we show $R(\nvr)$ is smaller than RHS of (\ref{keyinthepart3-2}).
There are two cases: $\ex^M[A_T \ind\{A_T\leq \nvr\}] < 0$ or $\ex^M[A_T \ind\{A_T\leq \nvr\}] \ge 0$.
For the first case, clearly $T(\nvr) < 0$ and thus smaller than RHS of (\ref{keyinthepart3-2}).
Now, suppose the other case holds, i.e., $\ex^M[A_T \ind\{A_T\leq \nvr\}] \ge 0$,
then we have:
\begin{eqnarray*}
T(\nvr) &=& \frac{\ex^M[A_T \ind\{A_T\leq \nvr\}]}{\pr^M(A_T\geq \nvr)}
\leq \frac{P_3-s}{c-s}\ex^M[A_T \ind\{A_T\leq \nvr\}] \\
&=&\frac{P_3-s}{c-s}\Big(\ex^M[A_T \ind\{A_T\leq \nvr\}] + \nvr\frac{c-s}{P_3-s} \Big) - \nvr \\
&\leq& \frac{P_3-s}{c-s}\Big(\ex^M[A_T \ind\{A_T\leq \nvr\}] + \pr^M(A_T \geq \nvr)\nvr \Big) - \nvr \\
&=& \frac{P_3-s}{c-s}\ex^M[\nvr \wedge A_T]  - \nvr
= \frac{\bar P^{\rm NV}(2b\bar P^\circ + 2bs - bc)}{(c-s)(r^\circ + \sqrt{(r^\circ)^2-1})} - \nvr \\
&=& \text{RHS of (\ref{keyinthepart3-2})}.
\end{eqnarray*}
The two $\leq$ involved in the derivations above uses the fact $\pr^M(A_T \leq \nvr) \leq (P_3 - c)/(P_3-s)$,
which follows from (based on the proved fact that $R_1 \ge \nvr$ and $P_3 \ge P_1$):
$$\pr^M(A_T \leq \nvr ) \leq \pr^M(A_T \leq R_1) = \frac{P_1 -c}{P_1 - s} \leq \frac{P_3 - c}{P_3 -s},
\quad \mbox{and thus}\quad \pr^M(A_T \ge \nvr) \ge \frac{c-s}{P_3 - s}.$$
Combining both cases, we have:
$$T(\nvr) \leq \text{RHS of (\ref{keyinthepart3-2})}.$$
Furthermore, since $\ex^M(A_T ) \ge \ex^M(\nvr \wedge A_T) \ge \ex(\nvr \wedge A_T) = 2b\nvp - bc > 0$,
clearly $T(R) \to \infty$ as $R \to \infty$.
Recall,  $T(R)$ strictly increases in $R$ for $R \ge \nvr$,
the analysis above indicates that there exist a unique $R^\circ \ge \nvr$ such that:
$$ T(R^\circ) = \frac{\ex^M\big[A_T \ind\{A_T \leq R^\circ\}\big]}{\pr^M(A_T \ge R^\circ)}
    = \text{RHS of (\ref{keyinthepart3-2})}
    = \frac{\bar \nvp(2b\bar P^\circ + 2bs - bc)}{(c-s)(r^\circ + \sqrt{(r^\circ)^2-1})} - \nvr.$$
Combining the proved fact that $R_1 \ge \nvr$ with $T(R_1) \leq \text{RHS of (\ref{keyinthepart3-2})}$, $R_1 \leq R^\circ$.
Applying Lemma~\ref{lem:R1}, $R^h_m \leq R_1 \leq R^\circ$ immediately follows.

What remains is to show $R^\circ \leq \nvmr$.
Let $R = \nvmr$ and use $\pr^M(A_T \geq  \nvmr) = (c-s)/(\nvmp-s)$:
\begin{equation} \label{Tnvmr}
 T(\nvmr) = \frac{\ex^M[A_T \ind\{A_T\leq \nvmr\}]}{\pr^M(A_T\geq \nvmr)}
 = \frac{\nvmp-s}{c-s}\ex^M[A_T \ind\{A_T\leq \nvmr\}].
\end{equation}
Here we show $\ex^M[A_T \ind\{A_T\leq \nvmr\}] > 0$ (hence $T(\nvmr) > 0$) as follows.
First note that
$$0 < m = \ex[H_T(\nvp, \nvr)] \leq \ex^M[H_T(\nvp, \nvr)] \leq \max_{P,R} V_0(P,R) = V_0(\nvp, \nvr)$$
(the first $\leq$ is due to $A_T \succeq A_T^M$).
Next,
\begin{eqnarray*}
 V_0(\nvmp, \nvmr) &=& (\nvmp - c)(\nvmr - b\nvmp) - (\nvmp - s)\ex^M[(\nvmr - A_T)^+] \\
 &=& (\nvmp - c)(\nvmr - b\nvmp) - (\nvmp - s)\nvmr\pr^M(A_T \leq \nvmr) \\
  &+& (\nvmp-s)\ex^M[A_T \ind\{A_T\leq \nvmr\}] \\
 &=& (- b\nvmp)(\nvmp - c) + (\nvmp-s)\ex^M[A_T \ind\{A_T\leq \nvmr\}] > 0.
\end{eqnarray*}
The third equality uses the optimality equation $\pr^M(A_T \leq \nvmr) = (\nvmp - c)/(\nvmp-s)$.
Since $(- b\nvmp)(\nvmp - c) < 0$ (by $\nvmp > c > s$),
we must have $\ex^M[A_T \ind\{A_T\leq \nvmr\}] > 0$.

 Let $R_3:=F_M^{-1}((P_3-c)/(P_3-s))$, i.e., $R_3 = \nvmr(P_3)$.
 As $P_2<P_3$,
 $$\nvr = \nvmr(P_2) \leq \nvmr(P_3) = R_3.$$
By $P_3 \leq \nvp$,
  $$R_3 = \nvmr(P_3) \leq \nvmr(\nvp) \leq \nvmr .$$
  Combining the above, we have $\nvr \leq R_3 \leq \nvmr(\nvp) \leq \nvmr$.
  Let
  $$g(R): = \frac{P_3-s}{c-s}\ex^M[R \wedge A_T]  - R.$$
  Clearly, $g(\nvr)=\text{RHS of (\ref{keyinthepart3-2})}$.
  Taking derivative of $g(R)$,  we have
  $$g^{\prime}(R)= \frac{P_3-s}{c-s}\pr^M(A_T \ge R)-1\geq 0, \quad\mbox{for any}\quad R \leq R_3,$$
  since $\pr^M(A_T \ge R)\geq (c-s)/(P_3-s)$ for $R \leq R_3$.
  Thus, $g(\nvr)\leq g(R_3)$.
  Now we check $g(R_3)$:
  \begin{eqnarray*}
 g(R_3)& = &\frac{P_3-s}{c-s}\ex^M[R_3 \wedge A_T]  - R_3\\
       & =& \frac{P_3-s}{c-s}\ex^M[A_T \ind\{A_T\leq R_3\}] + \frac{P_3-s}{c-s}R_3\pr^M(R_3\leq A_T) - R_3\\
       & = & \frac{P_3-s}{c-s}\ex^M[A_T \ind\{A_T\leq R_3\}]\\
       &\leq & \frac{\nvmp-s}{c-s}\ex^M[A_T \ind\{A_T\leq \nvmr\}] \\
       &= &T(\nvmr).
  \end{eqnarray*}
  The third $=$ is due to the definition of $R_3$: $\pr^M(R_3\leq A_T) = (c-s)/(P_3-s)$ and the last $=$ is due to (\ref{Tnvmr}).
  The $\leq$ is due to $\nvmp \ge \nvp \ge P_3 > c > s$, $\ex^M[A_T \ind\{A_T\leq \nvmr\}] > 0$ and $\nvmr \ge R_3$.
  Therefore,
  $$T(R^\circ) = \text{RHS of (\ref{keyinthepart3-2})} = g(\nvr) \leq g(R_3)\leq T(\nvmr), $$
  which leads to $R^\circ\leq \nvmr$.
  This completes the proof. $\hfill\Box$

\subsection{Proof of Proposition~\ref{pro:optfrontier}}
\label{appendix:optfrontier}

To establish the efficient frontier, direct differentiation using the expression of $B(m, P, R)$ in (\ref{bmpr}) yields (with $C = 1/(Z_0^M - 1) > 0$):
\begin{eqnarray*}
\frac{dB(m, P^h_m, R^h_m)}{dm} &=& 2C[m-V_0(P^h_m, R^h_m)] + \frac{\partial B(m, P^h_m, R^h_m)}{\partial P}\frac{dP^h_m}{dm} + \frac{\partial B(m, P^h_m, R^h_m)}{\partial R}\frac{dR^h_m}{dm} \\
&=& 2C[m-V_0(P^h_m, R^h_m)] \ge 0,
\end{eqnarray*}
where the second equality follows from optimality of $P^h_m$ and $R^h_m$
(which makes the partial derivatives vanish) and the $\ge 0$ follows from Proposition~\ref{pro:prbound-part1}.
$\hfill\Box$

\section{Proofs for  Section~\ref{appendix:supplementaryS6}}\label{ECsec:proofinsupplyS6}
\subsection{Proof of Proposition \ref{pro:rcvx}}
\label{appendix: prorcvx}
{
For any given $R$, $V_0(P,R)$ is concave in $P$;
in particular, it increases in $P$ up to $P^{\rm NV(M)}(R)$ and then decreases.
First, suppose $m > \max_P V_0(P, R)$.
Then, increasing $P$ beyond $P^{\rm NV(M)}(R)$ increases both terms of $B(m,P,R)$,
thus is not optimal.
Therefore, for this case the optimal $P$ is bounded by $P^{\rm NV(M)}(R)$.
Now consider the other case, $m \leq  \max_P V_0(P, R)$.
Then, increasing $P$ beyond $\bar P(R)$
(i.e.\ the smaller root of $V_0(P, R) = m$) is not optimal
as it also increases both terms of $B(m,P,R)$. }

{
In summary, for a given $R$, the optimal $P$ satisfies $V_0(P,R) \leq m$ over
$P \leq \bar P(R)$, and thus, by concavity of $V_0(P,R)$, the first term of $B(m, P, R)$ is convex in $P$.
The second term of $B(m, P, R)$ is a convex quadratic function in $P$, and this concludes the proof.
}
$\hfill\Box$

\section{Proofs for  Section~\ref{sec:generalization}}\label{ECsec:general-payoff}

Here we spell out the solution to the hedging problem in (\ref{hedstraprobgenepayfunc}).
Similar to the price-setting newsvendor model, the solution approach relies on the {\it projected} production payoff process:
\begin{equation}
\label{Vt_genefunc}
V_t(\mathbf{q}) := \ex^M[H_T(\mathbf{q}, \mathbf{A}_T) \,|\, \calf_t].
\end{equation}
Clearly, $V_t$ is a martingale under $\pr^M$.
Then, we apply the martingale representation theorem to decompose $V_t$ and we summarize the decomposition in the following lemma.
The proof is collected in Section~\ref{appendix:lem:Vt_genefunc}.
\begin{lemma}
\label{lem:Vt_genefunc}
$V_t(\mathbf{q})$ defined in (\ref{Vt_genefunc}) has the following representation:
\begin{equation}
\label{Vt_repregeneral}
V_t(\mathbf{q}) = V_0(\mathbf{q}) + \int_0^t \xi_s(\mathbf{q}) dX_s + \int_0^t \bdelta_s(\mathbf{q})d\tilde{\mathbf{B}}_{s},
\end{equation}
with $V_0(\mathbf{q}) = \ex^M[H_T(\mathbf{q}, A_T)]$.
$\xi_t$ and $\bdelta_t$ are processes adapted to $\calf_t$. In particular,
\begin{equation}
\label{xidelta_genefunc}
\bdelta_t(\mathbf{q}) = \nabla_{\mathbf{a}}^{\prime} f(t, X_t, \bA_t, \sigma_t)\tilde\bSigma,
\end{equation}
where
$f(t, x, \mathbf{a},\sigma) := \ex^M[H_T(\mathbf{q}, \bA_T) \,|\, X_t = x, \bA_t = \mathbf{a}, \sigma_t =\sigma].$
\end{lemma}
The expression of $ \xi_t(\mathbf{q}) $ needs further to specify the general processes $\bC_t$ and $\sigma_t$.

The optimal solution to the hedging problem (\ref{hedstraprobgenepayfunc}) is summarized by following theorem, with proof detailed in Section~\ref{appendix:thm:genpayhedsol}.
\begin{theorem}
\label{thm:genpayhedsol}
\begin{enumerate}[label=(\roman*)]
\item The optimal trading strategy is
\begin{equation}
\label{opttheta_genefunc}
\theta^*_t(\mathbf{q}) = \underbrace{-\xi_t(\mathbf{q})\phantom{\frac{{}_{}}{{}^{}_{}}}}_{\text{risk-mitigation position}}  \underbrace{-\frac{\zeta_t}{Z_t^M}[\lambda_m - V_t(\mathbf{q}) - \chi_t^*]}_{\text{investment position}},
\end{equation}
\begin{equation}
\label{lambda_genefunc}
\lambda_m = \frac{mZ_0^M - V_0(\mathbf{q})}{Z_0^M - 1}.
\end{equation}
(Note: $Z_0^M = \ex(Z_T^2) > 1$.)
\item
The optimal objective function value  has the following expression:
\begin{equation}
\label{bmq_genefunc}
B(m, \mathbf{q}) = \underbrace{\frac{[m - V_0(\mathbf{q})]^2}{Z_0^M-1}}_{\text{squared investment risk}} + \underbrace{\int_0^T \ex\Big[\frac{Z_t}{Z_t^M}||\bdelta_t(\mathbf{q})||^2\Big]dt}_{\text{squared unhedgeable risk}},
\end{equation}
In particular, $Z_0^M = \ex(Z_T^2) > 1$, and $0\leq Z_t / Z_t^M \leq 1$.
\end{enumerate}
\end{theorem}

\subsection{Proof of Lemma \ref{lem:Vt_genefunc}}
\label{appendix:lem:Vt_genefunc}
Recall that  $X_t$ is a (local) martingale under $\pr^M$ and admits the following dynamics:
$$
dX_t^M = \sigma_tX_t dB_t^M.
$$
According to the definition of $\bA_{t}$, for all $u \ge t \ge 0$,
\begin{eqnarray}
\label{Audecomp_general}
\bA_{u} = \bC_{u} + \tilde\bSigma \tilde{\mathbf{B}}_{u}
=  \bA_t +  \bC_{u} -\bC_t +  \tilde\bSigma (\tilde{\mathbf{B}}_{u} -\tilde{\mathbf{B}}_t).%
\end{eqnarray}
By the assumption that $\bC_u - \bC_t \in \sigma(\{X_s, \, t\leq s \leq u \})$ and $(X_t, \sigma_t)$ is Markovian under $\pr^M$,
the distribution of $\bC_u - \bC_t$ (conditional on $\calf_t$) is in turn determined by $(X_t, \sigma_t)$.
Then, clearly
$(\bA_t, X_t, \sigma_t)$ is also Markovian under $\pr^M$.
Next,
express $V_t(\mathbf{q})$ as:
\begin{eqnarray*}
V_t(\mathbf{q})  &:=& \ex^M[H_T(\mathbf{q}, \bA_T) \,|\, \calf_t]
=  \ex^M[H_T(\mathbf{q}, \bA_T) \,|\, X_t, \bA_t, \sigma_t]
= f(t,\mathbf{q}, X_t, \bA_t, \sigma_t),
\end{eqnarray*}
where the second equality follows from the Markovian property of $(\bA_t, X_t, \sigma_t)$
and the function $f(\cdot)$ in the last line is defined as follows:
$$f(t, \bq, x,\ba, \sigma) := \ex^M[H_T(\mathbf{q}, \bA_T) \,|\, X_t = x, \bA_t = \ba, \sigma_t =\sigma].$$
Clearly, $V_t$ is a martingale under $\pr^M$.
By Proposition~\ref{pro:mrt} (see Remark~\ref{rem:mrthd}) and (\ref{XtM}),
$V_t(\bq)$  has the following representation
with $V_0(\bq) = \ex^M[H_T(\mathbf{q}, \bA_T)]$ ($\bq$ is dropped from the notation since it is given):
\begin{equation*}
V_t = V_0 + \int_0^t \psi_s d B_t^M + \int_0^t \bdelta_s \cdot d\tilde{\mathbf{B}}_s.
\end{equation*}
$\psi_t$ and $\bdelta_t$ are processes adapted to $\calf_t$.
To show that it leads to (\ref{Vt_repregeneral}),
apply It\^{o}'s lemma:
\begin{eqnarray*}
dV_t & = & d f(t,\mathbf{q}, X_t, \bA_t,  \sigma_t) \\
 & = &  f_x dX_t + \sum_{j=1}^J f_{a_j}dA_{jt} + f_{\sigma}d\sigma_t  + \Big(f_t + \frac{1}{2}f_{xx} \sigma_t^2X_t^2 +  \frac{1}{2}\sum_{i = 1}^J \sum_{j=1}^J \bsigma_i\cdot\bsigma_j\frac{\partial^2f}{\partial a_i\partial a_j}  \Big)dt\\
 & & + \sum_{j=1}^J f_{xa_j}(dX_tdA_{jt}) + \sum_{j=1}^J f_{\sigma a_j}(d\sigma_tdA_{jt}) +  f_{x\sigma}(dX_td\sigma_t)\\
 & = & f_xdX_t + \nabla_{\mathbf{a}}^{\prime} f\tilde\bSigma d\mathbf{\tilde B}_t+ \sum_{j=1}^J f_{a_j} dC_{jt} + f_{\sigma} d\sigma_t  + \Big(f_t  + \frac{1}{2} f_{xx} \sigma_t^2x^2
 +\frac{1}{2}\sum_{i = 1}^J \sum_{j=1}^J \bsigma_i\cdot\bsigma_j\frac{\partial^2f}{\partial a_i\partial a_j} \Big)dt \\
 & & + \sum_{j=1}^J f_{xa_j}(dX_tdA_{jt}) + \sum_{j=1}^J f_{\sigma a_j}(d\sigma_tdA_{jt}) +  f_{x\sigma}(dX_td\sigma_t)\\
 & = & [...]dt + \psi_t dB_t^M + \nabla_{\mathbf{a}}^{\prime} f\tilde\bSigma d\mathbf{\tilde B}_t \\
 & = & \frac{\psi_t}{\sigma_t X_t}dX_t  + \nabla_{\mathbf{a}}^{\prime} f\bSigma d\mathbf{\tilde B}_t
\end{eqnarray*}
In the last line, the $dt$-term vanishes due to martingale property of $V_t$.
Furthermore, $dB_t^M$ is related to $dX_t$ based on the dynamics of $X_t$ in $\pr^M$; see (\ref{XtM}).
Comparing the last line with the martingale representation of $V_t(\bq)$ above and matching the coefficient of the term $d \tilde{\bB}_t$ leads to:
\begin{equation*}
\bdelta_t(\bq) = \nabla_{\mathbf{a}}^{\prime} f(t,\mathbf{q}, X_t, \bA_t, \sigma_t)\tilde\bSigma.
\end{equation*}
(Matching the $dX_t$-term leads to $\xi_t = \psi_t/(\sigma_t X_t)$).
$\hfill\Box$

\subsection{Proof of Theorem \ref{thm:genpayhedsol}}
\label{appendix:thm:genpayhedsol}
The methodology specified in Section~\ref{appendix:thmhedgingsol} leading to Theorem \ref{thm:hedgingsol} directly applies to prove Theorem \ref{thm:genpayhedsol},
because it does not assume any specific form of the payoff function and the asset price model is the same.
The only difference here is that the $\delta_t^2$ involved in $B(\cdot)$ in (\ref{bmpr}) is replaced by $\|\bdelta_t\|^2$ in (\ref{bmq_genefunc}).
Specifically, the only modification needed is to replace the one-dimensional $\delta_t$ in (\ref{GKW}) with the vector $\bdelta_t$ in (\ref{xidelta_genefunc}),
hence the one-dimensional $\gamma_t$ in (\ref{gengammat}) is replaced by a vector
$\bgamma_t = -\bdelta_t/Z_t^M$, accordingly.
Then, when applying It\^{o}'s isometry in (\ref{genAR3}), we have
$$Z_0^M\ex^R\Big[\Big(\int_0^T \bgamma_t \cdot d\tilde \bB_t\Big)^2\Big]
= Z_0^M\int_0^T\ex^M\Big[\|\bgamma_t\|^2\frac{Z_t^M}{Z_0^M}\Big]dt
=\int_0^T\ex\Big[\|\bgamma_t\|^2 Z_t^M Z_t\Big]dt
=\int_0^T\ex\Big[\frac{Z_t}{Z_t^M}\|\bdelta_t\|^2\Big]dt.$$
The rest of the proof is the same as that in Section~\ref{appendix:thmhedgingsol1},
which leads to the expression of $B(m, \bq)$ in (\ref{bmq_genefunc}).
This completes the proof.
$\hfill\Box$

\subsection{Proof of Theorem \ref{thm:genpay}}
\label{appendix:thm:genpay}
\subsubsection{Technical Preparation}

In this section, we present a collection of technical results that will be used later in the proof of Theorem \ref{thm:genpay}.
\begin{lemma} \label{lem:increasingdifference}
For any given $\mathbf{A}_1 \preceq \mathbf{A}_2$(defined as $\pr(\mathbf{A}_1 \in U) \leq \pr(\mathbf{A}_2 \in U)$ for any upper set
$U \subseteq \mathbb{R}^J$), suppose $H_T(\mathbf{q}, \mathbf{a})$ has increasing differences in $(\mathbf{q}, \mathbf{a})$.  Then, $\ex[H_T(\mathbf{q}, \mathbf{A}_2)] - \ex[H_T(\mathbf{q}, \mathbf{A}_1)]$ is entry-wise increasing in $\mathbf{q}$.
\end{lemma}
{\startb Proof.}
For any $\mathbf{q}_2\geq \mathbf{q}_1 \ge \bzero$,  we want to prove:
$$\ex[H_T(\mathbf{q}_2, \mathbf{A}_2) -H_T(\mathbf{q}_2, \mathbf{A}_1)] \geq \ex[H_T(\mathbf{q}_1, \mathbf{A}_2) -H_T(\mathbf{q}_1, \mathbf{A}_1)]  $$
which is equivalent to
$$\ex[H_T(\mathbf{q}_2, \mathbf{A}_2)- H_T(\mathbf{q}_1, \mathbf{A}_2)]\geq \ex[H_T(\mathbf{q}_2, \mathbf{A}_1) -H_T(\mathbf{q}_1, \mathbf{A}_1)].$$
To this end, let $\phi(\mathbf{a}) = H_T(\mathbf{q}_2, \mathbf{a})-H_T(\mathbf{q}_1, \mathbf{a})$, then $\phi(\mathbf{a})$ is entry-wise increasing in $\mathbf{a}$ as $H_T(\mathbf{q}, \mathbf{a})$ has increasing differences in $(\mathbf{q}, \mathbf{a})$.
Then, by $\bA_1 \preceq \bA_2$, we have:
$$\ex [\phi(\mathbf{A}_2)] \geq \ex [\phi(\mathbf{A}_1)], $$
which is equivalent to $\ex[H_T(\mathbf{q}_2, \mathbf{A}_2)]- H_T(\mathbf{q}_1, \mathbf{A}_2)] \geq \ex[H_T(\mathbf{q}_2, \mathbf{A}_1) -H_T(\mathbf{q}_1, \mathbf{A}_1)] $.
This completes the proof.
$\hfill\Box$

Note that Lemma \ref{lem:increasingdifference} implies:
\begin{equation*}
\frac{\partial{\ex[H_T(\mathbf{q}, \mathbf{A}_2)}]}{\partial q_{i}} \geq \frac{\partial{\ex[H_T(\mathbf{q}, \mathbf{A}_1)}]}{\partial q_{i}} \quad \text{for all} \,\,i.
\end{equation*}

Next, we present an extension of Topkis Theorem (\cite{topkis1978}) to the payoff function $H_T(\bq, \bA_T)$ described in Section~\ref{sec:generalization}.
\begin{theorem}[Extension of Topkis Theorem]
\label{thm:extendedtopkis}
Suppose $H_T(\mathbf{q}, \mathbf{a})$ satisfies Assumption \ref{assumption:genepayfunc}.
Let
$$\mathbf{q}_{i}^* = \min\arg \max_{\bq \ge \bzero} \ex[H_T(\mathbf{q}, \mathbf{A}_i)], \quad i = 1,2.$$
If $\mathbf{A}_1 \preceq \mathbf{A}_2$, then $\mathbf{q}^*_{1}\leq  \mathbf{q}^*_{2}$ .
\end{theorem}
\noindent
{\startb Proof.}
We prove this result by mathematical induction on the dimension of $\bq$,
denoted by $N$.
    For $N = 1$,
    let
    $$q_{i}^* = \min \arg \max_{q \ge 0} \ex[H_T(q, \mathbf{A}_i)], \quad i=1, 2. $$
    The optimality condition yields:
    $$\frac{\partial \ex[H_T(q_i^*, \mathbf{A}_i)] }{\partial q}  \leq 0, \quad i=1, 2. $$
    (Specifically, the above holds as equation for $q^*_i > 0$ and may hold as a strict inequality of $q^*_i = 0$.)
    Then, by Lemma \ref{lem:increasingdifference},
    $$\frac{\partial \ex[H_T(q_2^*, \mathbf{A}_1)] }{\partial q} \leq  \frac{\partial \ex[H_T(q_2^*, \mathbf{A}_2)] }{\partial q}  \leq 0. $$
    If $q^*_1 = 0$, then $q_1^* \leq q_2^*$ clearly holds.
    Otherwise, $q^*_1 > 0$ and we have
    $$\frac{\partial \ex[H_T(q_2^*, \mathbf{A}_1)] }{\partial q} \leq 0 = \frac{\partial \ex[H_T(q_1^*, \mathbf{A}_1)] }{\partial q}.$$
    By concavity of $\ex[H_T(q, \mathbf{A}_1)]$ in $q$, we have  $q_1^* \leq q_2^*$.
    Combining both cases, $q_1^* \leq q_2^*$ holds for $N = 1$

    Now, suppose that the result holds for $N$ and below we prove that it also holds for $N+1$.
    Let
    $$\mathbf{q}_{i}^* = \min \arg \max_{\bq \ge 0} \ex[H_T(\bq, \bA_i)],
    \quad \text{where} \,\, \bq= (q_1, q_2, ..., q_{N+1}), \quad i=1, 2.$$
    For any given $q_j \ge 0$, $j \in \{ 1, 2,...,N+1\}$, let
      $$\bq^*_{i, -j}(q_j) = \min \arg \max_{\bq_{-j} \ge \bzero} \ex[H_T(\bq, \bA_i)], \quad i=1,2, $$
      where $ \bq_{-j} = (q_1, q_2, ..., q_{j-1}, q_{j+1}, ...,  q_{N+1})$.
    Clearly, for any $j \in \{ 1, 2,...,N+1\}$, $\bq^*_{i, -j}(\bq^*_{i,j})$ is just $\bq^*_i$ with the $j$-th entry excluded.
    The argument $\bq_{-j}$ is an N-dimensional vector.
    Then, by induction hypothesis,
    \begin{equation}\label{inductionN}
        \mathbf{q}^*_{1, -j}(q_j) \leq \mathbf{q}^*_{2, -j}(q_j)
    \end{equation}
   for any given $q_j \ge 0$, $j \in \{ 1, 2,...,N+1\}$.
   Furthermore, by supermodularity of $H_T$ in $\bq$ and Topkis Theorem,
   $ \mathbf{q}^*_{i, -j}(q_j) $ is increasing in $q_j$ for both $i=1,2$.

   Suppose there exists a $j\in \{ 1, 2,...,N+1\}$, such that $\mathbf{q}^*_{1, j} \leq  \mathbf{q}^*_{2, j}$.
   From
   (\ref{inductionN})
   and that $\mathbf{q}^*_{i, -j}(q_j) $
   increases in
   $q_j$,  we must have
   $$\mathbf{q}^*_{1, -j}(\mathbf{q}^*_{1, j}) \leq \mathbf{q}^*_{2, -j}(\mathbf{q}^*_{1, j})  \leq \mathbf{q}^*_{2, -j}(\mathbf{q}^*_{2, j}) $$
   which leads to
   $$\mathbf{q}^*_{1} \leq  \mathbf{q}^*_{2}.$$
   In other words, if there exists some $j$ such that $\mathbf{q}^*_{1, j} \leq  \mathbf{q}^*_{2, j}$, then $\mathbf{q}^*_{1} \leq  \mathbf{q}^*_{2}$ must hold.
  Otherwise, suppose there does not exist such $j$,
  i.e., $\mathbf{q}^*_{1, j} > \mathbf{q}^*_{2, j}$ for all $j\in \{ 1, 2,...,N+1\}$;
  in other words,
  $\mathbf{q}^*_{1} > \mathbf{q}^*_{2}$.
   By definition of $\mathbf{q}^*_{1}$, we have
   $$\ex[H_T(\mathbf{q}^*_{1}, \mathbf{A}_1)] -  \ex[H_T(\mathbf{q}^*_{2}, \mathbf{A}_1)]>0. $$
   This is because that by the assumed $\mathbf{q}^*_{1} > \mathbf{q}^*_{2}$,
   $\mathbf{q}^*_{2}$ can not be the optimal solution to $\max_{\bq \ge \bzero} \ex[H_T(\mathbf{q}, \mathbf{A}_1)]$ due to $\bq^*_1$ being defined as the smallest optimal solution.
   Then, by Lemma \ref{lem:increasingdifference},
    $$\ex[H_T(\mathbf{q}^*_{1}, \mathbf{A}_2)] -  \ex[H_T(\mathbf{q}^*_{2}, \mathbf{A}_2)] \geq \ex[H_T(\mathbf{q}^*_{1}, \mathbf{A}_1)] -  \ex[H_T(\mathbf{q}^*_{2}, \mathbf{A}_1)]>0$$
   and this contradicts the optimality of $\mathbf{q}^*_{2}$ to $\max_{\bq \ge \bzero} \ex[H_T(\mathbf{q}, \mathbf{A}_2)]$.
   By contradiction, there must exists $j\in \{ 1, 2,...,N+1\}$ such that $\mathbf{q}^*_{1, j} \leq  \mathbf{q}^*_{2, j}$ and as already proved above, this leads to $\mathbf{q}^*_{1} \leq  \mathbf{q}^*_{2}$.
   This completes the proof.
$\hfill\Box$

\begin{lemma}
\label{lem:qbound}
Let
\begin{equation} \label{maxiV0}
\mathbf{q}^{*M} := \min \arg \max_{\bq \ge \bzero} \ex^M[H_T(\bq, \bA_T)] = \min \arg \max_{\bq \ge \bzero} \ex[H_T(\bq, \bA^M_T)],
\end{equation}
 and
\begin{equation} \label{miniBmq}
\cals_m^h = \argmin_{\bq \ge \bzero} B(m, \bq).
\end{equation}
Then, for any $m \ge  \ex^M[H_T(\bzero, \bA_T)]$, it holds that:
$$\mathbf{q}_m^h \leq \mathbf{q}^{*M}, \quad \forall \bq_m^h \in \cals_m^h.$$
\end{lemma}

\noindent
{\startb Proof.}
Recall the notation $V_0 := \ex^M(H_T)$.

We prove the stated result by mathematical induction on the dimension of $\bq$, denoted by $N$.
First, consider $N = 1$.
If $q_m^h = 0$, clearly $q_m^h \leq q^{*M}$ holds.
Otherwise, $q_m^h > 0$ and the following optimality equation $\partial B(m,q_m^h) / \partial q = 0$ must hold, which is expressed as:
$$2C(V_0(q_m^h) - m) \frac{\partial V_0(q_m^h)}{\partial q} + \frac{\partial \Psi(q_m^h )}{\partial q} = 0.$$
By Assumption \ref{assumption:Psi}, $\partial \Psi(q_m^h ) / \partial q \ge 0$,
hence $(V_0(q_m^h) - m) (\partial V_0(q_m^h) / \partial q) \leq 0$ must hold.
If $m \ge V_0(q^{*M})$, then clearly $V_0(q_m^h) - m \leq 0$ and $\partial V_0(q_m^h) / \partial q \ge 0$.
By concavity of $V_0$ in $q$, this indicates $q_m^h \leq q^{*M}$.
Now consider the other case, $m < V_0(q^{*M})$.
Since $V_0(0) = 0 \leq m$, there exists $\tilde q \in [0, q^{*M}]$ such that $V_0(\tilde q) = m$, and thus the first term of $B(m, \tilde q)$, $C[m - V_0(\tilde q)]^2$, is zero.
Therefore, any $q$ with $q \ge \tilde q$ can not be optimal since for such $q$,
$B(m, q)$ has both terms larger than those of $B(m, \tilde q)$.
Hence, $q_m^h \leq \tilde q \leq q^{*M}$ must hold.
In summary, the result stated in this theorem holds for $N = 1$.

Now, suppose the stated result holds for $N \ge 1$.
For an $(N+1)$-dimensional decision vector $\bq$,
write $\bq = (q_j, \bq_{-j})$,
where $\bq_{-j}$ is $\bq$ excluding $q_j$ for some $j \in \{ 1, 2,...,N+1\}$;
clearly, $\bq_{-j}$ is of dimension $N$.
Following this notation,
for the problem of dimension-$(N+1)$,
express $\bq_m^h $ as $\bq_m^h = (q^h_{m,j}, \bq^h_{m, -j})$.
In the following, we will prove two results.
First, we prove that if there exists a $j \in \{ 1, 2,...,N+1\}$ such that $q^h_{m,j} \leq  q_j^{*M} $, then $\mathbf{q}^h_{m, -j} \leq  \mathbf{q}_{-j}^{*M} $.
Second, we prove that it is impossible that $q^h_{m,j} \geq  q_j^{*M} $ for all $j \in \{ 1, 2,...,N+1\}$;
in other words, $\bq^h_m \ge \bq^{*M}$ can not hold.
Combining these two results immediately leads to $\bq_m^h \leq \bq^{*M}$ for the dimension-$(N+1)$ problem, thereby proves the stated result stated by this lemma.

Next, we prove the two results stated above.

    First, we prove that if there exists a $j \in \{ 1, 2,...,N+1\}$ such that $q^h_{m,j} \leq  q_j^{*M} $, then $\bq^h_{m, -j} \leq  \bq_{-j}^{*M} $.
    Suppose for some $j \in \{ 1, 2,...,N+1\}$,  $q^h_{m,j} \leq  q_j^{*M} $ holds.
    Next, fix $q_j =q^h_{m,j} $ in the optimization problems (\ref{maxiV0}) and (\ref{miniBmq}) so that the decision variable is just $\bq_{-j}$ for both problems.
    Then, for any $\tilde\bq^h_{m,-j}  \in \arg\min_{\bq_{-j} \ge \bzero} B(m, \bq_{-j},\,  q^h_{m, j} )$, it holds that
   $$
    \tilde\bq^h_{m,-j}  \leq  \min \arg \max_{\bq_{-j} \ge \bzero} V_0(\bq_{-j},\,  q^h_{m, j} )
  \leq \min \arg \max_{\bq_{-j} \ge \bzero} V_0(\bq_{-j},\,  q_j^{*M} ) = \bq_{-j}^{*M}.
$$
The first $\leq$ is by induction hypothesis
(note that $\bq_{-j}$ is of dimension $N$)
and the second $\leq$ is by Topkis Theorem
(recall, $V_0$ is supermodular in all $q_j$'s).
Recognizing that $\bq^h_{m,-j} \in \arg\min_{\bq_{-j}} B(m, \bq_{-j},\,  q^h_{m, j} )$,
$\bq^h_{m, -j} \leq  \bq_{-j}^{*M} $ immediately follows.

Now, we prove the second result:
it is impossible that $q^h_{m,j} \geq  q_j^{*M} $ for all $j \in \{ 1, 2,...,N+1\}$.
First, consider the case of $m\geq V_0(\bq^{*M})$.
Then, any $\bq$ with $\bq \geq  \bq^{*M} $ can not be optimal since it induces a higher variance than $\bq^{*M}$ does:
the latter has both terms of $B(m, \bq)$ smaller than the former.
Hence, for this case the stated result holds.
Now we focus on the other case,  $m < V_0(\bq^{*M})$.
 Suppose $\bq_m^h \ge \bq^{*M}$.
 Let $V_{00} = V_0(\bzero)$;
 note $V_{00} = \ex^M[H_T(\bzero, \bA_T)] \leq m$.
 In addition, let $\bq_{[n]} = (q_j)_{j=1}^n$ for $n = 1,2,\cdots, N+1$,
 i.e., $\bq_{[n]}$ is the vector of the first $n$ entries of $\bq$.
For each $n \in \{1,2,\cdots, N+1\}$, define problem $n$ as:
\begin{equation}
V_{0n} := \max_{\bq_{[n]} \ge \bzero} V_0(\bq_{[n]}, 0,\cdots,0),
\quad \bq^{(n)}:= \min \arg\max_{\bq_{[n]} \ge \bzero} V_0(\bq_{[n]}, 0,\cdots,0).
\end{equation}
Clearly, $\bq^{(N+1)} = \bq^{*M}$ and thus $V_{0(N+1)} = V_0(\bq^{*M}) > m$.
So,
$V_{0n}$ increases from $V_{00} \leq m$ to $V_{0(N+1)} > m$ as $n$ increases from $0$ to $N+1$.
Therefore, $\exists \tilde n \leq N-1$ such that
$$V_{0\tilde n} \leq m \leq V_{0(\tilde n+1)}.$$
Now consider the problem $\tilde n+1$, which produces $\bq^{(\tilde n+1)}$.
Consider the vector
$\bq(q) = (\bq_{[\tilde n]}^{(\tilde n+1)}, q, 0, \cdots, 0)$.
First, $V_0(\bq(q^{(\tilde n+1)}_{\tilde n+1})) = V_0((\bq^{(\tilde n+1)}, 0, \cdots,0))
= V_{0(\tilde n+1)}\ge m$.
Next, note
$$V_0(\bq(0)) = V_0((\bq_{[\tilde n]}^{(\tilde n+1)}, 0, \cdots,0))
\leq \max_{\bq_{[\tilde n] \ge \bzero}} V_0(\bq_{[\tilde n]},0,\cdots,0) = V_{0\tilde n} \leq m.$$
Therefore, there exists $q^\circ \in [0,q^{(\tilde n+1)}_{\tilde n+1}]$ such that
$$V_0(\bq(q^\circ)) = V_0((\bq_{[\tilde n]}^{(\tilde n+1)}, q^\circ, 0,\cdots, 0)) = m.$$
Clearly, $\bq(q^\circ) \leq \bq(q^{(\tilde n+1)}_{\tilde n+1})$.
Now, with $\bq^{*M} \ge \mathbf{0}$ and Topkis Theorem:
$$\bq^{(\tilde n+1)} = \min \arg\max_{\bq_{[\tilde n+1]} \ge \bzero} V_0(\bq_{[\tilde n+1]},0,\cdots, 0)
\leq \min \arg\max_{\bq_{[\tilde n+1] \ge \bzero}} V_0(\bq_{[\tilde n+1]},q^{*M}_{\tilde n+2}, \cdots, q^{*M}_{N+1}) = \bq^{*M}_{[\tilde n+1]}, $$
so,
$$\bq^{(\tilde n+1)}_{[\tilde n]} \leq \bq^{*M}_{[\tilde n]}
\quad\mbox{and}\quad
q^\circ \leq q_{\tilde n+1}^{(\tilde n+1)} \leq q^{*M}_{\tilde n+1}. $$
Therefore,
$$\bq(q^\circ) \leq \bq(q^{(\tilde n+1)}_{\tilde n+1}) = (\bq^{(\tilde n+1)}, 0, ..., 0) \leq \bq^{*M}$$
In summary, there exists $\bq(q^\circ) \leq \bq^{*M} \leq \bq^h_m$ such that the first term of $B(m, \bq(q^\circ))$ (which is zero) is smaller than that of $B(m, \bq_m^h)$ and so is its second term,
this leads to $B(m, \bq(q^\circ)) \leq B(m, \bq_m^h)$.
Thus, $\bq^h_m$ cannot be optimal.
By contradiction, it is impossible that $\bq^h_m \ge \bq^{*M}$.
$\hfill\Box$

\subsubsection{Proof of Theorem \ref{thm:genpay}.}
When the asset price trend positively impacts the stochastic operational factors,
i.e., $\bC_T^M \preceq \bC_T$,
we have $\bA_T^M \preceq \bA_T$.
From Theorem \ref{thm:extendedtopkis}, we have:
$$\mathbf{q}^{*M}\leq \mathbf{q}^*.$$
By $\bA_T^M \preceq \bA_T$, $\ex[H_T(\bzero, \bA_T)] \ge \ex^M[H_T(\bzero, \bA_T)]$.
So, $m = \max_{\bq \ge \bzero} \ex[H_T(\bq, \bA_T)] \ge \ex[H_T(\bzero, \bA_T)]  \ge \ex^M[H_T(\bzero, \bA_T)]$
and thus Lemma \ref{lem:qbound} immediately leads to
$$\mathbf{q}_m^h \leq \mathbf{q}^{*M} \leq  \mathbf{q}^* .$$
This completes the proof.
$\hfill\Box$

\subsection{Proof of Corollary \ref{cor:multi}}
\label{appendix:cor:multi}

We begin with verifying that $H_T$ satisfies all conditions specified in Assumption \ref{assumption:genepayfunc}.
Recall,
$$ H_{T}((\bP, \bR), \bA_T) = \sum_{j=1}^{J}\left[P_j \Big(R_{j}-\sum_{i=1}^J \gamma_{ji} P_i \Big)-\left(P_j + c_j\right)\left(R_{j}  -A_{j T} \right)^{+}\right].$$
For each $R_j$, the term in $H_T$ that involves this variable is
$P_j(R_j - \sum_{i=1}^J\gamma_{ij}P_i) - (P_j+c_j)(R_j-A_{jT})^+$,
which is a linear function minus a convex function of $R_j$, hence concave in $R_j$.
For each $P_j$, the nonlinear term in $H_T$ involving this variable is:
$-\gamma_{jj}P_j^2$, which is concave in $P_j$.
Hence, $H_T$ is concave in each $R_j$ and $P_j$.
Next, clearly, $H_T$ is increasing in $A_{jT}$.
Now, we check the supermodularity of $H_T$ in the decision variables,
and this is equivalent to verifying this property for each pair of variables.
For any $i \neq j$, there is no interaction term involving $R_i$ and $R_j$,
or involving $P_i$ and $R_j$.
So, we only need to consider pairs of $(P_i, P_j)$ for $i \neq j$ and $(P_j, R_j)$.
It is straightforward to obtain:
$$\frac{\partial^2 H_T}{\partial P_i \partial P_j} = -(\gamma_{ji} + \gamma_{ij}) > 0,$$
hence $H_T$'s supermodularity in $(P_i, P_j)$ holds.
For any $P_j$, it is easy to derive:
$$\frac{\partial H_T}{\partial P_j} = R_j \wedge A_{jT} - 2\sum_{i=1}^J \gamma_{ji}P_i, $$
which increases in $R_j$ and leads to supermodularity of $H_T$ in $P_j$ and $R_j$.
Hence, $H_T$ is supermodular in all the decision variables.
Last, we verify the increasing difference property.
For $\bA_T \leq \bA'_T$, it is straight forward to derive
$$H_T((\bP, \bR), \bA_T) - H_T((\bP, \bR), \bA'_T) = \sum_{j=1}^J (P_j+c_j)(R_j \wedge A'_{jT} - A_{jT})^+,$$
which clearly increases in all $P_j$ and $R_j$ and this confirms the increasing difference property.
In summary, $H_T$ satisfies Assumption \ref{assumption:genepayfunc}.

Now we turn to verify Assumption \ref{assumption:Psi}.
Recall, by Theorem \ref{thm:genpayhedsol},
$$\Psi(\bP, \bR) = \int_0^T \ex\Big[\frac{Z_t}{Z_t^M} \|\bdelta^2_t(\bP, \bR)\|\Big]dt, $$
where
$$\bdelta_t(\bP, \bR) = \nabla'f\tilde\bSigma = \sum_{j=1}^J f_{a_j}\tilde\bsigma'_j,
\quad
f(t,x,\ba,\sigma) = \ex^M(H_T \,|\, X_t = x, \bA_t = \ba, \sigma_t = \sigma).$$
To obtain the expression of $\bdelta_t$, we derive $f_{a_j}$ as follows.
Note
\begin{eqnarray*}
f(t,x,\ba,\sigma) &=& \ex^M(H_T \,|\, X_t = x, \bA_t = \ba, \sigma_t = \sigma) \\
                  &=& \sum_{j=1}^J P_j(R_j - \sum_{i=1}^J \gamma_{ji}P_i)
                  - \sum_{j=1}^J (P_j + c_j)\ex^M[(R_j - A_{jT})^+ \,|\,  X_t = x, \bA_t = \ba, \sigma_t = \sigma] \\
&=& \sum_{j=1}^J P_j(R_j - \sum_{i=1}^J \gamma_{ji}P_i) \\
    &-& \sum_{j=1}^J (P_j + c_j)\ex^M[(R_j - A_{jt} - (C_{jT} - C_{jt}) - \tilde\bsigma_j\cdot (\tilde \bB_T - \tilde \bB_t))^+ \,|\,  X_t = x, \bA_t = \ba, \sigma_t = \sigma]  \\
&=& \sum_{j=1}^J P_j(R_j - \sum_{i=1}^J \gamma_{ji}P_i) \\
    &-& \sum_{j=1}^J (P_j + c_j)\ex^M[(R_j - a_j - (C_{jT} - C_{jt}) - \tilde\bsigma_j\cdot (\tilde \bB_T - \tilde \bB_t))^+ \,|\,  X_t = x, \sigma_t = \sigma],
\end{eqnarray*}
where the last $=$ is follows from $C_{jT} - C_{jt} \in \sigma(\{X_u, t\leq u \leq T\})$, Markovian property of $(X_t, \sigma_t)$ and independence of $\tilde \bB_T - \tilde \bB_t$ from $\calf_t$.
Then, for each $j = 1,\cdots, J$, we have:
\begin{eqnarray*}
f_{a_j}(t, x,\ba, \sigma) = \sum_{i=1}^J(P_i + c_i)\pr^M(a_i + (C_{iT} - C_{it}) + \tilde\bsigma_i\cdot (\tilde \bB_T - \tilde \bB_t) \leq R_i \,|\, X_t = x, \sigma_t = \sigma),
\end{eqnarray*}
hence
\begin{eqnarray*}
f_{a_j}(t, X_t,\bA_t, \sigma_t) &=& \sum_{i=1}^J(P_i + c_i)\pr^M(A_{it} + (C_{iT} - C_{it}) + \tilde\bsigma_i\cdot (\tilde \bB_T - \tilde \bB_t) \leq R_i \,|\, X_t, \sigma_t) \\
&=&\sum_{i=1}^J(P_i + c_i)\pr^M(A_{it} + (C_{iT} - C_{it}) + \tilde\bsigma_i\cdot (\tilde \bB_T - \tilde \bB_t) \leq R_i \,|\, \calf_t) \\
&=& \sum_{i=1}^J(P_i + c_i)\pr^M(A_{iT} \leq R_i \,|\, \calf_t),
\end{eqnarray*}
the second $=$ follows from, again, $C_{jT} - C_{jt} \in \sigma(\{X_u, t\leq u \leq T\})$, Markovian property of $(X_t, \sigma_t)$ and independence of $\tilde \bB_T - \tilde \bB_t$ from $\calf_t$.
Then, we can derive:
\begin{eqnarray*}
\|\bdelta_t(\bP, \bR)\|^2 = \| \sum_{j=1}^J f_{a_j}\tilde\bsigma'_j\|^2
= \sum_{i=1}^J\sum_{j=1}^J (P_i+c_i)(P_j+c_j)\pr^M(A_{iT} \leq R_i \,|\, \calf_t)
\pr^M(A_{jT} \leq R_j \,|\, \calf_t)(\tilde\bsigma_i \cdot \tilde\bsigma_j).
\end{eqnarray*}
Clearly, when $\tilde\bsigma_i \cdot \tilde\bsigma_j \ge 0$ for all $i,j$,
$\|\bdelta_t^2(\bP, \bR)\|^2$ increases in all $P_j$ and $R_j$,
hence so does $\Psi(\bP, \bR)$.
This completes the proof.
$\hfill\Box$

\subsection{Proof of Corollary \ref{cor:pgamma}}
\label{appendix:cor:pgamma}

Recall that the  profit function is
$$H_T((P, \gamma), A_T) = Pd(A_T, P, \gamma) - k(\gamma).$$
To prove this corollary, it suffices to check that the operational payoff function $H_T((P, \gamma), A_T)$ above satisfies Assumptions \ref{assumption:genepayfunc} and \ref{assumption:Psi},
then Theorem \ref{thm:genpay} applies.
To this point, we verify each of the conditions specified in these two assumptions.

    We start with Assumption \ref{assumption:genepayfunc}.
    First, concavity of $H_T$ in $P$ and $\gamma$.
    For $\gamma$, the concavity immediately follows from the concavity of $d(\cdot)$ and convexity of $k(\cdot)$ in this variable.
    For $P$, it is straightforward to obtain:
    $$\frac{\partial^2 H_T}{\partial P^2} = 2\frac{\partial d}{\partial P} + P \frac{\partial^2 d}{\partial P^2}.$$
    Since $d(\cdot)$ is decreasing  and concave in $P$, the derivative above is negative,
    which leads to the concavity in $P$.
    Next, since $d(\cdot)$ is increasing in the market size $A_T$,
    clearly so does $H_T$.
    The supermodularity of $H_T$ in $(P, \gamma)$ follows from:
    $$\frac{\partial^2 H_T}{\partial P \partial \gamma} = \frac{\partial d}{\partial \gamma}+ P\frac{\partial^2 d}{\partial P \partial \gamma} \ge 0.$$
    The $\ge 0$ follows from $d(\cdot)$ increasing in $\gamma$ and being supermodular in $(P, \gamma)$.
    Last, for the increasing difference property of $H_T$ in $((P, \gamma), A_T)$, we show that $H_T$ is supermodular in $(P, \gamma, A_T)$, which is a sufficient condition for this property.
    To this end, it is sufficient to show that $H_T$ is supermodular in all three pairs:
    $(P, \gamma)$, $(\gamma, A_T)$ and $(P, A_T)$.
    Supermodularity in the first pair is already shown above and that in the second pair follows immediately from supermodularity of $d(\cdot)$ in this pair.
    Now, supermodularity of $H_T$ in $(P, A_T)$ is a result of:
    $$\frac{\partial^2 H_T}{\partial P \partial A_T} = \frac{\partial d}{\partial A_T} + P\frac{\partial^2 d}{\partial P \partial A_T} \ge 0, $$
    where the $\ge 0$ follows from $d(\cdot)$ being increasing in the market size $A_T$ and supermodular in $(P, A_T)$.
    In summary, $H_T$ specified in this corollary satisfies Assumption \ref{assumption:genepayfunc}.

Next, we verify that Assumption \ref{assumption:Psi} holds.
Recall, by Theorem \ref{thm:genpay}, the squared unhedgeable risk has the following expression:
$$\Psi(P, \gamma) = \int_0^T \ex\Big[\frac{Z_t}{Z_t^M} \delta^2_t(P, \gamma)\Big]dt, $$
where
$$\delta_t(P, \gamma) = \tilde\sigma\frac{\partial f(t, X_t, A_t, \sigma_t)}{\partial a},
\quad \mbox{with}\quad
f(t,x,a,\sigma) = \ex^M(H_T \,|\, X_t = x, A_t = a, \sigma_t = \sigma).$$
With $H_T = Pd(A_T, P, \gamma) - k(\gamma)$, we have
\begin{eqnarray*}
f(t, x, a, \sigma)
& = & \ex^M[Pd(A_T, P, \gamma) - k(\gamma) \,|\, X_t = x, A_t = a, \sigma_t =\sigma] \\
& = & \ex^M[Pd(a + C_T -C_t +\tilde{\sigma} (\tilde{B}_T -\tilde{B}_t) , P, \gamma) - k(\gamma) \,|\, X_t = x, \sigma_t =\sigma],
\end{eqnarray*}
where the second $=$ follows from $C_T - C_t \in \sigma(\{X_u, t \leq u \leq T\})$,
Markovian property of $(X_t, \sigma_t)$ and independence of $\tilde{B}_T -\tilde{B}_t$ from $\calf_t$.
This gives
$$\frac{\partial f}{\partial a} = P\ex^M\Big[\frac{\partial d(a + C_T -C_t +\tilde{\sigma} (\tilde{B}_T -\tilde{B}_t), P, \gamma)}{\partial A_T} \,\Big{|}\, X_t = x, \sigma_t =\sigma\Big].$$
Hence, $\partial f / \partial a$ is nonnegative (since $d(\cdot)$ in increases in $A_T$) and
is increasing in both $P$ and $\gamma$ (by supermodularity of $d(\cdot)$ in $(A_T, P, \gamma)$),
and this leads to the fact that $\delta_t^2(P, \gamma)$ increases in $P$ and $\gamma$.
Since $Z_t / Z_t^M > 0$, this further implies that $\Psi(P, \gamma)$ increases in $P$ and $\gamma$.
This completes the proof.
$\hfill\Box$

\end{document}